\documentclass[a4paper,11pt]{article}
\pdfoutput=1

\usepackage{amsmath}
\usepackage{a4wide}
\usepackage{slashed}
\usepackage{graphicx}
\usepackage[utf8]{inputenc}
\usepackage{amsfonts}
\usepackage{lscape}
\usepackage{booktabs}
\usepackage{array}
\usepackage{rotating}
\usepackage{physics}
\usepackage{siunitx}
\usepackage{slashed}
\usepackage[dvipsnames]{xcolor} 
\usepackage[normalem]{ulem}

\usepackage{jheppub}

\hypersetup{%
  colorlinks=true, linktocpage=true, pdfstartpage=1, pdfstartview=FitV,
  breaklinks=true, pageanchor=true,
  pdfpagemode=UseNone,
  plainpages=false, bookmarksnumbered, bookmarksopen=true, bookmarksopenlevel=1,
  hypertexnames=true, pdfhighlight=/O,
  urlcolor=RoyalBlue, linkcolor=ForestGreen, citecolor=RoyalBlue,
  pdftitle={Leptophilic Axion-Like Particles in Beam-Dump Experiments},
  pdfauthor={Alexander M. Eberhart, Marco Fedele, Felix Kahlhoefer, Eike Ravensburg, Robert Ziegler},
  pdfsubject={},
  pdfkeywords={},
  pdfcreator={pdfLaTeX},
  pdfproducer={LaTeX with hyperref}
}

\usepackage{subcaption}

\newcommand{\MeV}{{\, \rm MeV}}
\newcommand{\GeV}{{\, \rm GeV}}

\newcommand{\eps}{\epsilon}

\newcommand{\be}{\begin{equation}}
\newcommand{\ee}{\end{equation}}

\newcommand{\hi}[1]{\ensuremath{^{\textrm{#1}}}}
\newcommand{\lo}[1]{\ensuremath{_{\textrm{#1}}}} 

\makeatletter
\newcommand*{\rom}[1]{\expandafter\@slowromancap\romannumeral #1@} 
\makeatother

\preprint{TTP25-011, P3H-25-026, MITP-25-027}

\author[a]{Alexander Eberhart,}
\author[b,c]{Marco Fedele,}
\author[a,d]{Felix Kahlhoefer,}
\author[e,f]{Eike Ravensburg,}
\author[a]{Robert Ziegler}

\affiliation{$^a$ Institute for Theoretical Particle Physics (TTP), Karlsruhe Institute of Technology (KIT), Wolfgang-Gaede-Str. 1, D-76131 Karlsruhe, Germany\\
$^b$ Departament de Física Teòrica, IFIC, Universitat de València – CSIC,
Parque Científico, Catedrático José Beltrán 2, E-46980 Paterna, Spain\\
$^c$ PRISMA+ Cluster of Excellence \& Mainz Institute for Theoretical Physics,
Johannes Gutenberg University, D-55099 Mainz, Germany\\
$^d$ Institute for Astroparticle Physics (IAP), Karlsruhe Institute of Technology (KIT), Wolfgang-Gaede-Str. 1, D-76131 Karlsruhe, Germany\\
$^e$ The Oskar Klein Centre, Department of Physics, Stockholm University, SE-10691 Stockholm, Sweden\\
$^f$ CP3-Origins, University of Southern Denmark, Campusvej 55, DK-5230 Odense M, Denmark}

\emailAdd{alexander.m.eberhart@durham.ac.uk}
\emailAdd{mfedele@uni-mainz.de}
\emailAdd{kahlhoefer@kit.edu}
\emailAdd{ravensburg@cp3.sdu.dk}
\emailAdd{robert.ziegler@kit.edu}

\title{Leptophilic ALPs in Laboratory Experiments}

\abstract{We study the collider phenomenology of leptophilic axion-like particles (ALPs), i.e. pseudoscalar particles that couple only to charged leptons. Loops of charged leptons induce effective interactions of the ALPs with photons, which depend on the momenta of the interacting particles and differ between pseudoscalar and derivative lepton couplings. We systematically discuss the form of the interaction with photons for general external momenta and identify the regimes when it can be safely approximated by an effective coupling constant. We use these results to derive novel constraints from LEP and calculate state-of-the-art limits from E137 and NA64 for four different scenarios, in which the ALPs couple either to a single lepton generation or universally to all, for both pseudoscalar and derivative lepton couplings. We collect complementary bounds from astrophysics, flavour, and other laboratory experiments to chart the allowed parameter space of leptophilic ALPs in the MeV-GeV mass range.}

\begin{document}

\maketitle

\section{Introduction}

Some of the simplest modifications of the Standard Model (SM) involve a single axion-like particle (ALP), which is a pseudo-Goldstone boson of a suitable, spontaneously broken global symmetry. In contrast to the QCD axion~\cite{Wilczek:1977pj,Weinberg:1977ma, Peccei:1977hh,Peccei:1977ur}, this symmetry is subject to a large explicit breaking, which induces an ALP mass that is simply taken as a free parameter. Of particular interest are masses in the MeV--GeV range, which lead to a very rich phenomenology at particle colliders~\cite{Bauer:2017ris}, depending on the precise form of ALP interactions with SM particles. Though the explanation of the dark matter abundance is one of the main motivations of these scenarios (see, e.g. refs.~\cite{Nomura:2008ru, Freytsis:2010ne, Dolan:2014ska, Berlin:2015wwa, Fan:2015sza, No:2015xqa, Bauer:2017ota, Baek:2017vzd, Kamada:2017tsq, Kaneta:2017wfh, Banerjee:2017wxi, Arcadi:2017wqi, Hochberg:2018rjs, Berlin:2018bsc, deNiverville:2019xsx, Buttazzo:2020vfs,  Darme:2020sjf, Ge:2021cjz, Gola:2021abm, Domcke:2021yuz, Zhevlakov:2022vio, Bauer:2022rwf, Panci:2022wlc, Bharucha:2022lty,  Fitzpatrick:2023xks, Ghosh:2023tyz, Dror:2023fyd, Capozzi:2023ffu, Armando:2023zwz}), it is important to map current limits from laboratory experiments, astrophysics and cosmology into the general parameter space of the ALP Lagrangian.   

While this task has been performed in much detail for ALP couplings to photons, couplings to leptons have received far less attention (notable exceptions include refs.~\cite{Altmannshofer:2022ckw, Bauer:2021mvw, Ferreira:2022xlw, Armando:2023zwz, Cornella:2019uxs, Calibbi:2020jvd,Bertuzzo:2022fcm}). In the present work, we aim to systematically study the model-independent laboratory constraints on ``leptophilic ALPs", which, by definition, couple dominantly to charged leptons\footnote{The case of ALPs that couple dominantly to neutrinos was recently discussed in ref.~\cite{Bonilla:2023dtf}.}. As a result, we provide an exhaustive overview of the relevant constraints from beam-dump, flavour and collider experiments on leptophilic ALPs, which complement limits from astrophysics~\cite{Caputo:2024oqc, Carenza:2021pcm, Ferreira:2022xlw, Fiorillo:2025sln}. 

We restrict ourselves to flavour-diagonal couplings and consider four simple scenarios, where either the coupling to a single lepton family dominates ($e,\mu,\tau$) or the ALP couples in a flavour-universal way to leptons. In contrast to previous analyses, we explicitly consider two cases of possible leptonic ALP interactions, which take the form of either pseudoscalar or derivative couplings. It is well-known that these interactions are largely equivalent for tree-level processes involving a single ALP but differ for multi-ALP interactions and, in particular, by an effective ALP-photon coupling induced by the chiral anomaly. Therefore, the two leptophilic interactions lead to different predictions for observables that involve an effective ALP-photon-photon interaction, which is induced by a leptonic loop. The associated vertex function depends on the external momenta and, thus, on the typical energy scales of the relevant process (apart from a constant part depending on whether pseudoscalar or derivative couplings are considered). This energy dependence implies that the phenomenology of the leptophilic ALP cannot be simply obtained from existing limits on the  ALP-photon coupling but has to be re-calculated using the full momentum dependence for all processes involving ALP-photon interactions~\cite{Bharucha:2022lty, Ferreira:2022xlw}. The complete analysis of the resulting limits from E137~\cite{Bjorken:1988as}, NA64~\cite{Gninenko:2719646,NA64:2019auh} and LEP~\cite{Jaeckel:2015jla} represents the main result of this work. 

Although such loop-induced processes are suppressed with respect to tree-level diagrams involving only leptons, we will show that beam dump experiments like E137 can give highly relevant constraints on leptophilic ALP, even when both the production and decay of the ALPs proceed exclusively via photon couplings. When the relevant energy scales are much larger than the ALP and lepton masses, the loop function becomes suppressed. The resulting vertex inherits this suppression for pseudoscalar interactions, while the derivative vertex is essentially given by the constant part induced by the chiral anomaly. This implies that, e.g. limits from high-energy colliders like LEP that have been derived for effective photon couplings~\cite{Jaeckel:2015jla} are relevant only for leptophilic ALPs with derivative couplings, and similar conclusions hold for highly energetic leptonic beam dumps. More generally, a large separation of scales often permits the approximation of the fully off-shell vertex function by the on-shell coupling, which largely simplifies calculations, as we discuss in detail. We also point out the great potential of muon beam-dump experiments, such as NA64$\mu$~\cite{NA64:2024klw}, to constrain models of leptophilic ALPs~\cite{Li:2025yzb}. Indeed, NA64$\mu$ already sets relevant constraints on several of the scenarios that we consider, which will substantially improve in coming years with larger data sets.

The rest of this work is organised as follows. In Section~\ref{setup}, we define the basic setup of the leptophilic ALP models and discuss the difference between the pseudoscalar and derivative basis and the resulting ALP-photon-photon vertex function. In Section~\ref{sec:Beam_Dumps}, we use these results in order to calculate ALP production at leptonic beam dumps in all relevant channels, focusing on E137 and NA64. Constraints on leptophilic ALPs from other experiments are discussed in Section~\ref{sec:other_constraints}, and in particular, we derive novel limits by recasting LEP1 data on multi-photon events (largely following ref.~\cite{Jaeckel:2015jla}). Here, we also collect existing bounds in the literature from $B$ factories, $W^+$ boson and rare meson decays, SN 1987A, leptonic anomalous magnetic moments and cosmology. Our results are presented in Section~\ref{results}, which mainly consist of novel limits on leptophilic ALPs from  E137 and NA64 that have been carefully derived including the track-length distribution of the beam electrons as well as secondary electrons and positrons, which we combine with other existing limits in order to provide a concise overview of all relevant constraints on leptophilic ALPs. We summarise our conclusions in Section~\ref{conclusions} and include various appendices with details on the experimental setup in appendix~\ref{app:experiments} and the cross-section calculations in appendix~\ref{Xsecs}. We also compare our results for the exclusion limit from E137 with previous analyses in appendix~\ref{comparison}.

\section{Setup}
\label{setup}

In this section, we introduce the effective Lagrangians for leptophilic ALP models, discuss in detail the phenomenological and theoretical differences between pseudoscalar and derivative couplings and study the general induced one-loop ALP couplings to photons.

\subsection{ALP Couplings to Fermions}
\label{sec:Theory:Der_vs_PS}

The ALP is assumed to be a pseudoscalar that is not charged under the SM gauge group, arising as a pseudo-Goldstone boson from the spontaneous breaking of a global Peccei-Quinn (PQ) symmetry at some UV scale $\Lambda$. At scales much smaller than $\Lambda$, the effective Lagrangian is governed by the non-linearly realised  PQ symmetry under which the ALP shifts, $a \to a + {\rm const}$. While for the QCD axion, this symmetry is only broken by QCD instantons, for the ALP that we consider here, there is also a soft breaking by an explicit ALP mass term. Apart from this mass term, the most general Lagrangian then involves shift-invariant couplings to fermions and anomalous couplings to gauge bosons, just as the QCD axion~\cite{Georgi:1986df}.

Here, we focus on leptophilic ALPs, which are described by an effective Lagrangian only involving charged leptons. Up to higher-dimensional operators and restricting to flavour-diagonal couplings, the general ALP Lagrangian reads
\begin{align}
    \mathcal{L}_{\rm D} &= \frac{1}{2} \partial_\mu a  \, \partial^\mu a - \frac{1}{2} m_a^2 a^2 +  \frac{\partial_\mu a}{2 \Lambda}   \sum_{\ell=e,\mu,\tau} C_\ell  \, \overline \psi_\ell \gamma^\mu \gamma_5 \psi_\ell \,,
    \label{eq:Theory:Lagrangian_Der}
\end{align}
with real couplings $C_\ell$. Equivalently, one can change the field basis and write this Lagrangian in terms of pseudoscalar couplings to fermions. Under the chiral ALP-dependent rotation of the lepton fields
\begin{align}
\psi_\ell & \to e^{i \frac{C_\ell}{2 \Lambda} a \gamma_5} \psi_\ell \, , &     \overline \psi_\ell & \to \overline{\psi}_\ell \, e^{i \frac{C_\ell}{2 \Lambda} a \gamma_5} \, ,
\label{redef}
\end{align}
the derivative couplings above are cancelled when plugging the transformed fields into the lepton kinetic terms, and the ALP appears in the lepton mass term. Since this transformation is anomalous, i.e.\ it does not leave the fermion path integral measure invariant, it also gives rise to a contribution to the ALP couplings to photons, so that in the new field basis, the ALP Lagrangian reads (see e.g. refs.~\cite{Bauer:2017ris, Armando:2023zwz} for details)
\begin{align}
    \mathcal{L}_{\rm D} = \frac{1}{2} \partial_\mu a  \, \partial^\mu a - \frac{1}{2} m_a^2 a^2  - i a \sum_{\ell=e,\mu,\tau}  \frac{m_\ell}{\Lambda}  C_\ell \overline \psi_\ell \gamma_5 \psi_\ell + \frac{\alpha}{4 \pi}  \frac{a}{\Lambda}  F^{\mu\nu} \Tilde{F}_{\mu\nu} \sum_{\ell=e,\mu,\tau}  C_\ell   \,,
    \label{eq:Theory:Chiral_Rotation_Expanded}
\end{align}
up to higher-order terms in $a/\Lambda$, and  $\tilde{F}^{\mu \nu} =  \epsilon^{\mu \nu \rho \sigma} F_{\rho \sigma}/2$ with $\epsilon^{0123} = 1$. Note that the truncation explicitly breaks the shift symmetry of lepton couplings, which is restored by higher-order terms where the shift symmetry just corresponds to a re-phasing of the leptons.

It is well-known~\cite{Chisholm:1961tha, Kamefuchi:1961sb} that all physical observables remain unchanged under non-linear field redefinitions of the form in eq.~\eqref{redef}, which is just a change of variables in the path integral formulation of QFT. In the context of effective field theories, it has been shown in ref.~\cite{Arzt:1993gz} that such field redefinitions are largely equivalent to using the equations of motion for the involved fields (even when off-shell), although care has to be taken beyond the leading order~\cite{Criado:2018sdb}. Working consistently at leading order in $1/\Lambda$, i.e. only using vertices involving a single ALP, the Lagrangians in eq.~\eqref{eq:Theory:Lagrangian_Der} and eq.~\eqref{eq:Theory:Chiral_Rotation_Expanded} therefore make identical predictions, even though the latter features ALP couplings to \emph{both} leptons and photons. 

This equivalence implies that a leptophilic Lagrangian in the pseudoscalar basis defined by 
\begin{align}
    \mathcal{L}_{\rm P} = \frac{1}{2} \partial_\mu a  \, \partial^\mu a - \frac{1}{2} m_a^2 a^2  - i a \sum_{\ell=e,\mu,\tau}  \frac{m_\ell}{\Lambda}  C_\ell  \overline \psi_\ell \gamma_5 \psi_\ell   \,,
    \label{eq:LagP}
\end{align}
does not give the same results as~\eqref{eq:Theory:Lagrangian_Der} when the corresponding process involves effective vertices to photons. For this reason, there are two ``leptophilic" ALP models with distinct phenomenology, which in the following we refer to as the ``derivative" (eq.~\eqref{eq:Theory:Lagrangian_Der}) and ``pseudoscalar" (eq.~\eqref{eq:LagP}) model. To follow the conventional notation in the literature, we also introduce the dimensionless coupling $g_{a \ell}$ via the relation
\begin{align}
    \frac{C_\ell}{\Lambda} = \frac{g_{a \ell}}{m_\ell} \, , 
\label{eq:Theory:Coupling_Conversion}
\end{align}
such that the pseudoscalar interactions can be written as $
    \mathcal{L}_{\text{P}} \supset - i g_{a \ell} a \overline{\psi}_\ell \gamma_5 \psi_\ell
$. In order to clearly distinguish between derivative and pseudoscalar leptophilic models, we use the notation $C_\ell/\Lambda$ exclusively for the derivative interaction and the coupling $g_{a \ell}/m_\ell$ exclusively for the pseudoscalar interaction. Moreover, we will discuss below four different scenarios, depending on the values of $C_\ell$ for the three families. Either the ALP couples to a single lepton species, e.g.\ $C_e \neq 0$, while $C_\mu = C_\tau = 0$, or it has flavour-universal couplings in the derivative basis, i.e.\ $C_e = C_\mu = C_\tau$, corresponding to universal PQ charges.

Finally, we briefly discuss the question of which type of leptophilic interaction (eq.~\eqref{eq:Theory:Lagrangian_Der} or eq.~\eqref{eq:LagP}) is best motivated from a UV perspective. Simple UV completions seem to suggest the pseudoscalar form, which can arise from a renormalisable Yukawa coupling with a non-linearly realised Goldstone boson in a straightforward way. For example, one can imagine a standard QCD axion scenario like the DFSZ model~\cite{Zhitnitsky:1980tq,Dine:1981rt}, in addition to some small explicitly PQ breaking mass term, for example induced by $M_{\rm Planck}$-suppressed operators that generally spoil the solution to the Strong CP Problem in QCD axion models (the so-called PQ Quality Problem~\cite{Kamionkowski:1992mf}). In DFSZ scenarios, SM fermions are charged under the PQ symmetry, which is spontaneously broken by the Higgs fields (at least two) and a SM singlet scalar. The axion is then a linear combination of the pseudoscalar components in the Higgs fields and the singlet and couples to fermions through the Yukawa interactions (see e.g. refs.~\cite{DiLuzio:2020wdo, Badziak:2021apn} for the explicit construction of generalised DFSZ models)
\begin{align}
    \mathcal{L}\lo{DFSZ} \supset -  \bar{u}\lo{L} m_U u\lo{R} e^{i \chi_{H_u} \frac{a}{v_{\rm PQ}}} -  \bar{d}\lo{L} m_D d\lo{R} e^{i \chi_{H_d} \frac{a}{v_{\rm PQ}}} -  \bar{e}\lo{L} m_E e\lo{R} e^{i \chi_{H_d} \frac{a}{v_{\rm PQ}}} + \text{h.c.} \,,
    \label{eq:Theory:DFSZ_Lagrangian}
\end{align}
where $\chi_{H_{u,d}}$ are the PQ charges of the Higgs doublets $H_{u,d}$ and $v_{\rm PQ} \propto \Lambda$ is the PQ breaking scale. 

Expanding the exponential functions, one obtains mass terms for the fermions as well as pseudoscalar ALP-fermion interaction terms, while a tree-level coupling between ALPs and photons is absent in this basis. A chiral redefinition of the fermion fields would trade the pseudoscalar couplings of eq.~\eqref{eq:Theory:DFSZ_Lagrangian} for derivative couplings and generally non-vanishing couplings to gauge bosons, unless the PQ charges are such that the electromagnetic (EM) anomaly coefficient vanishes. In the four leptophilic scenarios discussed above, this anomaly is indeed non-vanishing since, by construction, only leptons contribute, and the leptonic PQ charge matrix is not traceless. Thus, in order to obtain one of these scenarios with derivative lepton couplings and a vanishing photon coupling, one would need to extend the simple DFSZ UV model with new heavy fermions that have suitable PQ charges to cancel the EM anomaly, along the lines of common KSVZ models~\cite{Kim:1979if,Shifman:1979if}. 

On the other hand, since the derivative coupling is the only dimension-5 operator that preserves the shift-symmetry of the ALP, other shift-symmetric couplings (such as those to gauge bosons, e.g. in a KSVZ model) will generally induce this form of coupling through RGE-evolution~\cite{Chala:2020wvs}. An example is the photophobic ALP discussed in ref.~\cite{Craig:2018kne}, which, however, certainly requires more model-building effort than the above-mentioned DFSZ-like constructions. Moreover, as discussed in ref.~\cite{Craig:2018kne}, it comes with additional interactions with the electroweak gauge bosons (which we do not consider here). 

In the remainder of this article, we choose to remain agnostic about the nature of the UV completions of both bases and simply derive constraints for both scenarios in eq.~\eqref{eq:Theory:Lagrangian_Der} and eq.~\eqref{eq:LagP}, which will deviate when probing ALP couplings to photons. These couplings are effectively generated from lepton loops and differ for the two interactions by a constant, as we discuss in more detail now.

\subsection{Effective ALP Coupling to Photons}
\label{sec:Theory:Effective_Coupling}

\begin{figure}[tbp]
    \centering
    \includegraphics[width=0.3\linewidth]{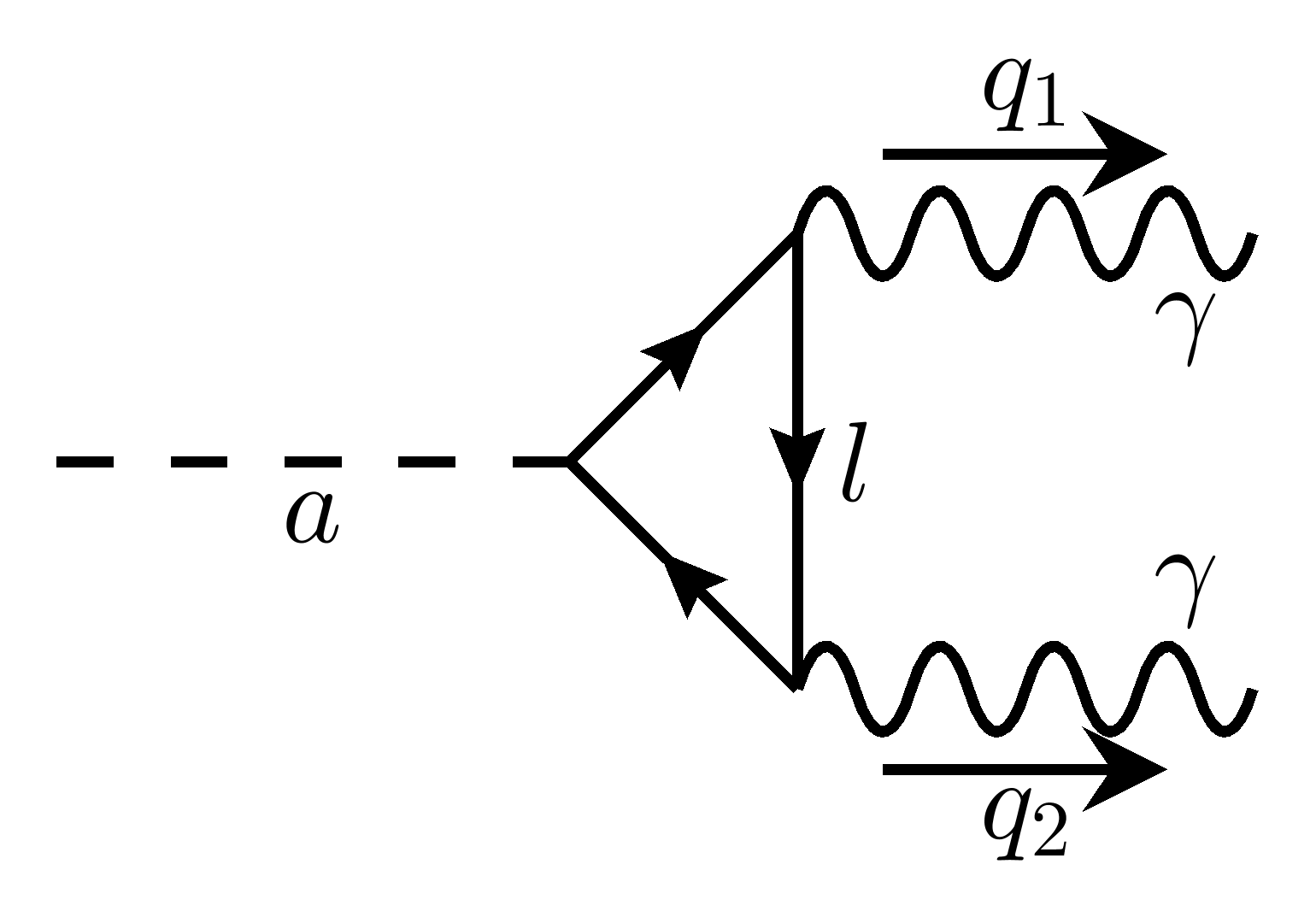}
    \caption{Triangle diagram inducing the effective ALP-photon interaction.}
    \label{fig:Theory:Triangle_Diagram}
\end{figure}

Even though leptophilic ALPs do not couple to photons at tree level, an effective ALP-photon interaction is induced by the triangle diagram shown in figure~\ref{fig:Theory:Triangle_Diagram}, which corresponds to an off-shell ALP-photon-photon vertex function $\Gamma_{a^* \gamma^* \gamma^*}^{\mu \nu} (q_1, q_2)$ that depends on the outgoing photon momenta $q_{1,2}$, with incoming ALP momentum $q_a = q_1 + q_2$. It is convenient to introduce an effective coupling as a scalar form factor by  
\begin{align}
\Gamma_{a^* \gamma^* \gamma^*}^{\mu \nu} (q_1, q_2) =  i g_{a^* \gamma^* \gamma^*}^{\rm eff}  (q_a^2, q_1^2, q_2^2) q_{1 \alpha} q_{2 \beta} \epsilon^{\mu \nu \alpha \beta} \, , 
\end{align}
such that  $g_{a^* \gamma^* \gamma^*}^{\rm eff}  (q_a^2, q_1^2, q_2^2) $ is given by a scalar one-loop integral with three external momenta $q_a, q_1, q_2$, which can be expressed in terms of  Passarino-Veltman functions. For ALPs with a derivative coupling to a single lepton species of the form in eq.~\eqref{eq:Theory:Lagrangian_Der}, the  effective off-shell coupling reads~\cite{Ferreira:2022xlw} (the superscript ``D" denotes the derivative basis)
\begin{align}
    g_{a^* \gamma^* \gamma^*}^{\rm eff,D}  (q_a^2, q_1^2, q_2^2) &=  \frac{\alpha}{\pi} \frac{C_\ell}{\Lambda} \qty[1+2 m_\ell^2 C_0\qty(q_1^2,q_2^2,q_a^2,m_\ell^2,m_\ell^2,m_\ell^2)]  \, , 
    \label{GammaD}
    \end{align}
where $C_0$ is the Passarino-Veldman three-point function defined as
\begin{align}
    \begin{split}
        C_0&\qty(q_1^2,q_2^2,(q_1+q_2)^2,m_1^2,m_2^2,m_3^2) 
        \equiv \int \frac{\dd^4 k}{i \pi^2} \frac{1}{\qty(k^2-m_1^2)\qty (k_1^2-m_2^2)\qty( k_{12}^2-m_3^2)} \, , 
    \end{split}
\end{align}
with $k_1 \equiv k - q_1$ and $k_{12} = k-q_1 - q_2$, following the conventions in ref.~\cite{Shtabovenko:2023idz}. If the ALP couples to several leptons, the effective coupling is given by the sum over all contributions.

Since the Lagrangian in eq.~\eqref{eq:Theory:Lagrangian_Der} is equivalent to the one in eq.~\eqref{eq:Theory:Chiral_Rotation_Expanded} for single ALP vertices, the scalar vertex function for the pseudoscalar coupling in eq.~\eqref{eq:LagP} differs from eq.~\eqref{GammaD} by a constant, which is given by the coefficient of the $\tfrac{1}{4} a F \tilde{F}$ operator in eq.~\eqref{eq:Theory:Chiral_Rotation_Expanded}, $\alpha C_\ell /  (\pi \Lambda)$. Indeed, for the effective coupling in the pseudoscalar basis (denoted by the superscript ``P"), one has
\begin{align}
    g_{a^* \gamma^* \gamma^*}^{\rm eff,P}  (q_a^2, q_1^2, q_2^2)  &= \frac{\alpha}{\pi} \frac{g_{a \ell}}{m_\ell} \qty[2 m_\ell^2 C_0\qty(q_1^2,q_2^2,q_a^2,m_\ell^2,m_\ell^2,m_\ell^2)] \, ,
    \end{align}
where we used $g_{a \ell}/m_\ell = C_\ell/\Lambda$.

It is instructive to evaluate the fully off-shell couplings for the simplifying cases where one or more external particles are on-shell. If a single photon is on-shell, we find (in agreement with ref.~\cite{Bauer:2021mvw}) 
\begin{align}
\label{2off1onshell}
g_{a^* \gamma^* \gamma }^{\rm eff,D}  (q_a^2, q_1^2) & =  \frac{ \alpha}{\pi} \frac{C_\ell}{\Lambda}   \left(  1 - \int_0^1 dx \int_0^1 dy \frac{m_\ell^2}{m_\ell^2 - x (1-x) (q_a^2 - 2 y q_a q_1)  } \right) \, ,
\end{align}
which is valid in the derivative basis, i.e.\ using the Lagrangian in eq.~\eqref{eq:Theory:Lagrangian_Der}. For the analogous result in the pseudoscalar basis, one simply needs to drop the constant part. 

Taking the ALP on-shell\footnote{The alternative case that both photons are on-shell but the ALP is off-shell is discussed in ref.~\cite{Bonilla:2021ufe}.}, $q_a^2 = m_a^2$, the vertex function depends only on a single off-shell photon squared momentum $q_1^2 \equiv t$, and reads
\begin{align}
g^{\rm eff, D}_{a  \gamma^* \gamma}  (t)  & =   \frac{ \alpha}{\pi} \frac{C_\ell}{\Lambda}    \left( 1 - \int_0^1 dx \int_0^1 dy \frac{m_\ell^2}{m_\ell^2 - x (1-x) (m_a^2 - y (m_a^2 + t) ) } \right)  \, ,
\label{off1X}
\end{align}
where we have taken eq.~\eqref{2off1onshell} and set $2 q_a q_1 = m_a^2 + t$. This equality becomes evident by rewriting the momentum $q_2$ of the on-shell photon as $q_2^2 = (q_a - q_1)^2 = m_a^2 - 2 q_a q_1 + t = 0$.

Finally, if all particles are on-shell, i.e. if $t = 0$, the vertex function becomes a constant 
 \begin{align}
g^{\rm eff, D}_{a \gamma \gamma}   & =  \frac{ \alpha}{\pi} \frac{C_\ell}{\Lambda}     \left( 1 - \int_0^1 dx \int_0^1 dy \frac{m_\ell^2}{m_\ell^2 - x (1-x) (1-y) m_a^2 } \right)  =  \frac{ \alpha}{\pi} \frac{C_\ell}{\Lambda}  \left( 1 - \tau_\ell f(\tau_\ell)^2   \right)   \, ,
     \label{eq:Theory:C_0_Decay}
\end{align}
with  $\tau_\ell = 4 m_\ell^2 / m_a^2$ and  the function
\begin{align}
    f(\tau) = \begin{cases}
        \arcsin(\frac{1}{\sqrt{\tau}}) ,&\qif \tau \geq 1\\
        \frac{\pi}{2} + \frac{i}{2}\log(\frac{1+\sqrt{1-\tau}}{1-\sqrt{1-\tau}}) ,&\qif 0 < \tau < 1
    \end{cases} \, ,
\end{align}
consistent with the expression derived in refs.~\cite{Ferreira:2022xlw,Bauer:2017ris}. This constant can be identified with the effective Lagrangian coupling
\begin{align}
    \mathcal{L}_\gamma &= \frac{g_{a \gamma \gamma}^{\rm eff}}{4} a F^{\mu\nu} \tilde{F}_{\mu\nu} \,.
    \label{eq:Theory:Effective_Photon_Interaction}
\end{align}
and controls the ALP decay into two photons. Since, asymptotically,
\begin{align}
\tau f^2 (\tau) = \begin{cases}  1 + \frac{1}{3 \tau} & \tau \gg 1 \\ \frac{\tau}{4} \left( \pi + i \ln \frac{4}{\tau} \right)^2 & \tau \ll 1 \end{cases} \, ,
\end{align}
the effective photon coupling defined in eq.~\eqref{eq:Theory:Effective_Photon_Interaction} for a derivatively coupled ALP in eq.~\eqref{eq:Theory:Lagrangian_Der} has the  limits
\begin{align}
    \label{eq:Theory:C_0_Decay_Der_Cases}
    g_{a \gamma \gamma}^{\rm eff,D} & \propto \ 1 - \tau_\ell f(\tau_\ell)^2 \propto
    \begin{cases}
-\tfrac{1}{12} \, (m_a/m_\ell)^2 &\qif m_a \ll m_\ell \\
    1 &\qif m_a \gg m_\ell   
      \end{cases} \, .
\end{align}
Instead, the effective photon coupling for an ALP coupled to the pseudoscalar current in eq.~\eqref{eq:LagP} has the asymptotic behaviour \begin{align}
    \label{eq:Theory:C_0_Decay_PS_Cases}
  g_{a\gamma \gamma}^{\rm eff, P}&  \propto - \tau_\ell f(\tau_\ell)^2  \propto \begin{cases}  - 1 - \tfrac{1}{12} \, (m_a/m_\ell)^2& m_a \ll m_\ell \\    \order{\log[2]( m_a / m_\ell) \left( m_\ell / m_a \right)^2} & m_a \gg m_\ell \end{cases} \, .
\end{align}
in agreement with ref.~\cite{Ferreira:2022xlw}.
 
Therefore, for couplings to leptons significantly lighter than the ALP ($\tau_\ell \ll 1$), the contribution to the effective photon coupling coming from $C_0$ itself is negligible. In the case of derivative couplings, this leaves only a constant that results in an effective coupling of $g_{a \gamma \gamma}^{\rm eff, D} \approx \alpha C_\ell /  (\pi \Lambda)$. Instead, for pseudoscalar couplings, the ALP-photon coupling is strongly suppressed in this limit. Conversely, in the case of couplings to leptons significantly heavier than the ALP ($\tau_\ell \gg 1$), the limits are effectively reversed. The derivative coupling leads to a strong suppression in $g_{a \gamma \gamma}^{\rm eff, D}$ due to cancellation between the constant and $C_0$, while the pseudoscalar coupling leads to a similar result as the tree-level coupling $g_{a \gamma \gamma}^{\rm eff, P} \approx - \alpha g_{a \ell}/(\pi m_\ell)$, albeit with the opposite sign. The ALP decay rate to two photons will feature the same limiting behaviour, see section~\ref{decayrates}. To summarise, the effective on-shell photon couplings are given by
\begin{align}
g_{a \gamma \gamma}^{\rm eff, D}   & =  \frac{ \alpha}{\pi} \frac{C_\ell}{\Lambda}  \left( 1 - \tau_\ell f(\tau_\ell)^2   \right)   \, , &
g_{a \gamma \gamma}^{\rm eff, P}  & = -   \frac{ \alpha}{\pi} \frac{g_{a \ell}}{m_\ell}    \tau_\ell f(\tau_\ell)^2     \, ,
\label{effcouplings}
\end{align}
for derivative and pseudoscalar couplings, respectively. 

The loop function can also be calculated analytically for the case in which one of the photons is off-shell, i.e.\ eq.~\eqref{off1X}. In this case, one finds
\begin{align}
    2 m_\ell^2 C_0(0,t,m_a^2,m_\ell^2,m_\ell^2,m_\ell^2) = B\qty(\frac{4 m_\ell^2}{m_a^2},\frac{4 m_\ell^2}{t}) \,,
    \label{eq:Theory:C_0_t}
\end{align}
where the function $B (\tau_1, \tau_2)$ is defined as
\begin{align}\label{eq:Bt1t2}
    B(\tau_1,\tau_2) = \frac{\tau_1 \tau_2}{\tau_1 - \tau_2} \qty(f(\tau_1)^2 - f(\tau_2)^2) \,.
\end{align}
Since $t$ can also be negative, we extend the definition of $f$ to
\begin{align}
    f(\tau) = \begin{cases}
        \arcsin(\frac{1}{\sqrt{\tau}}) ,&\qif \tau \geq 1 \\
        \frac{\pi}{2} + \frac{i}{2}\log(\frac{1+\sqrt{1-\tau}}{1-\sqrt{1-\tau}}) ,&\qif 0 < \tau < 1 \\
        \frac{i}{2} \log(\frac{\sqrt{1-\tau}+1}{\sqrt{1-\tau}-1}) ,&\qif \tau < 0
    \end{cases} \, .
    \label{eq:Theory:f_Full_Definition}
\end{align}
Note that eq.~\eqref{eq:Theory:C_0_t} is consistent with eq.~\eqref{eq:Theory:C_0_Decay}  since in the limit of $t \to 0$
\begin{align}
 B(\tau_1,\tau_2)\xrightarrow{\tau_2 \to \infty} - \tau_1 f(\tau_1)^2  \, , 
\end{align}
as $\tau_2 f(\tau_2)^2$ becomes constant. Instead, in the limit of $t \gg m_\ell^2$, one has 
\begin{align}
 B(\tau_1,\tau_2)\xrightarrow{\tau_2 \to 0} - \tau_2 f(\tau_2)^2 \propto \tau_2 \ln^2 {\tau_2} \, , 
\end{align} 
which is particularly relevant for ALP searches with large momentum transfer.

The limit for $C_0$ in eq.~\eqref{eq:Theory:C_0_t} is sufficient for our phenomenological discussion since at least one of the photons will be on-shell, or at least approximately so, see section~\ref{sec:Implementation:DB_P}. To summarise the resulting effective coupling, one has
\begin{align}
g^{\rm eff, D}_{a  \gamma^* \gamma}  (t)  &  = 
        \frac{\alpha}{\pi} \frac{C_\ell}{\Lambda} \qty[1 + B\qty(\frac{4 m_\ell^2}{m_a^2},\frac{4 m_\ell^2}{t})] \, , &       
g^{\rm eff, P}_{a  \gamma^* \gamma}  (t)  &  =                  \frac{\alpha}{\pi} \frac{g_{a \ell}}{m_\ell} B\qty(\frac{4 m_\ell^2}{m_a^2},\frac{4 m_\ell^2}{t}) \, , 
      \label{eq:Theory:Effective_Coupling}
\end{align}
for derivative and pseudoscalar coupling, respectively. For $t \gg m_\ell^2$, the vertex function for the pseudoscalar couplings is suppressed, while it approaches a constant for the derivative coupling. In the following, we will often suppress the argument, as it is clear from the subscript whether we refer to the on- or off-shell coupling. 

\subsection{ALP decay rates}
\label{decayrates}

The decay rates for the decay of an ALP into a pair of photons or leptons are~\cite{Bauer:2017ris}
\begin{align}
    \Gamma_{a \rightarrow \gamma\gamma} &= \frac{|g_{a \gamma \gamma} \hi{eff}|^2 m_a^3}{64 \pi} \,, \\
    \Gamma_{a \rightarrow \ell\ell} &= \frac{g_{a \ell}^2 m_a}{8 \pi} \sqrt{1-\frac{4 m_\ell^2}{m_a^2}} = \frac{C_\ell^2 m_\ell^2 m_a}{8 \pi \Lambda^2} \sqrt{1-\frac{4 m_\ell^2}{m_a^2}} \,,
\end{align}
where the effective on-shell photon coupling for derivative and pseudoscalar couplings is given in eq.~\eqref{effcouplings} with the limiting values in eqs.~\eqref{eq:Theory:C_0_Decay_Der_Cases} and~\eqref{eq:Theory:C_0_Decay_PS_Cases}, respectively. For leptophilic ALPs with arbitrary couplings to the three lepton species, the total decay rate is given by
\begin{align}
    \Gamma_a &= \Gamma_{a \rightarrow \gamma\gamma} + \sum_{l=e,\mu,\tau} \Gamma_{a \rightarrow \ell \ell} \,.
\end{align}

\section{ALP Production at Lepton Beam Dumps}
\label{sec:Beam_Dumps}

\begin{figure}[t]
    \centering
    \includegraphics[width=\linewidth]{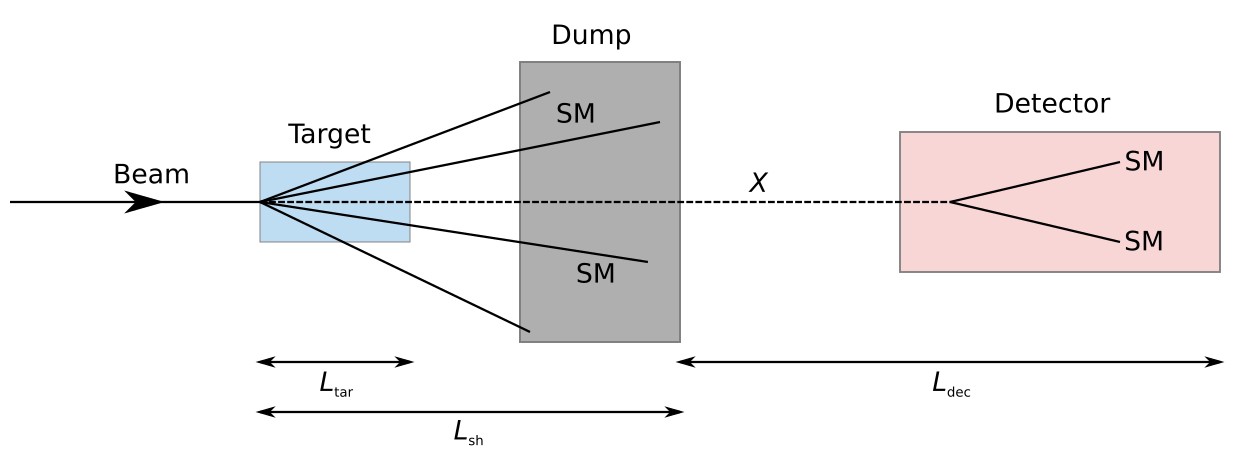}
    \caption{Schematic illustration of a beam dump. A BSM particle $X$ is produced at the target and passes through the dump. It can be detected if it decays into SM particles in the decay volume.}
    \label{fig:Beam_Dumps:Beam_Dump_Diagram}
\end{figure}

Beam-dump experiments are a valuable asset in the search for physics beyond the Standard Model (BSM). The general setup of a beam-dump experiment is depicted schematically in figure~\ref{fig:Beam_Dumps:Beam_Dump_Diagram}.
A high-intensity beam, usually consisting of protons, electrons or muons, impinges upon a target material. Inside this target, countless Standard Model processes take place, creating particles such as pions or kaons~\cite{Ertas:2021mkt}. However, a small fraction of the interactions might produce a weakly interacting BSM particle $X$, for example, an ALP. In order to shield the detectors from the SM background, the beam dump is positioned downstream of the target. The dump absorbs the shower of particles, and only weakly interacting particles are able to pass. Once past the dump, the particle $X$ may decay back into Standard Model particles, which leads to a signature in the detectors placed downstream of the dump~\cite{Cesarotti:2023sje}.

Beam-dump experiments are well-suited to searches for particles in the mass range $\order{\text{MeV}} - \order{\text{GeV}}$ with tiny couplings to Standard Model particles. These feebly interacting particles have long lifetimes and decay lengths and are consequently difficult to detect in collider experiments. In contrast, the long decay volumes of beam-dump experiments allow for greater sensitivity to these types of particles since they are able to decay inside the experiment and be detected~\cite{Cesarotti:2023sje,Kim:2024vxg}.

The most sensitive beam-dump experiments for leptophilic ALPs are the ones employing lepton beams. Concretely, we consider the E137 beam-dump experiment performed at SLAC~\cite{Bjorken:1988as} to search for neutral metastable particles, as well as the NA64 beam-dump experiment running at the CERN Super Proton Synchrotron. The latter performs three different types of searches:
\begin{itemize}
\item The standard \emph{invisible mode} of the NA64 experiment differs from E137 and figure~\ref{fig:Beam_Dumps:Beam_Dump_Diagram} in that the experiment employs an active beam dump. Instead of the detector being downstream of the target and detecting potential decay products, the target itself is the detector, and the experiment searches for missing energy~\cite{Gninenko:2719646}.
\item In contrast to the invisible setup, NA64 run in \emph{visible mode} uses a similar search strategy to the one employed by E137~\cite{NA64:2019auh}.
\item The NA64$\mu$ experiment~\cite{NA64:2024klw} employs a muon beam to search for new particles that couple primarily to muons, such as the gauge boson of a new $\text{U(1)}_{L_\mu-L_\tau}$ gauge symmetry. Clearly, this \emph{muon mode} also possesses great sensitivity to muonphilic ALP models.
\end{itemize}
The details of the different experiments are summarised in appendix~\ref{app:experiments}.

In the remainder of this section, we explain how the constraints for these experiments are calculated. Complementary constraints from other types of experiments will be discussed in section~\ref{sec:other_constraints}.

\subsection{ALP Yield in E137}
\label{sec:Implementation:Yield}

For electron beam-dump experiments like E137 or NA64 run in visible mode, the general formula for the expected number of ALPs that can be detected by the experiment is given by the expression~\cite{Liu:2023bby}
\begin{align}
    N_a = \frac{N_e X}{M\lo{target} / N\lo{avo}} \int_{E\lo{min}}^{E\lo{max}}\dd E \int_{x\lo{min}}^{x\lo{max}}\dd x \int_0^T \dd t \, I_e(E_0,E,t) \dv{\sigma}{x} e^{-L\lo{sh}\qty(\frac{1}{l_a}+\frac{1}{l_\lambda})} \qty(1-e^{-\frac{L\lo{dec}}{l_a}}) \,,
    \label{eq:Implementation:N_a_original}
\end{align}
with the number of electrons on target (EOT) $N_e$, the unit length of the radiation target $X$, the molar mass of the target atom $M\lo{target}$, and the Avogadro constant $N\lo{avo}$. Furthermore, $L\lo{sh}$ denotes the length of the beam dump, while $L\lo{dec}$ is the length of the decay volume, i.e. the distance from the end of the beam dump to the detector. A summary of the experimental parameters of E137 and NA64 is given in appendix~\ref{app:experimental_parameters}.

The variable $x = E_a/E$ defines the ratio of the energy transferred from an electron (or muon for muon beam dumps) of energy $E$ to an ALP with energy $E_a$. The lower integration bound for $x$ can be determined from the requirement that the ALP energy must be greater than the energy threshold of the detector $E_\text{cut}$, as well as the ALP rest mass. Assuming the ALP and electron/muon initial and final states are collinear~\cite{Liu:2017htz}, the upper bound can be approximated by requiring energy conservation in the case that the outgoing electron/muon is at rest, with the entire energy (excluding the lepton's rest mass) being transferred to the ALP. This means the bounds for $x$ are
\begin{align}
    x\lo{min} = \frac{\max(m_a \,,\, E\lo{cut})}{E} \,\qc x\lo{max} = \frac{E-m_\ell}{E} = 1 - \frac{m_\ell}{E} \,.
    \label{eq:Implementation:Boundaries_x}
\end{align}
In this work, we only consider ALPs with mass below \SI{1}{GeV}, i.e.\ below the energy threshold, such that $x_\text{min} = E_\text{cut} / E$. 

The upper integration boundary for the electron energy is given by the incident beam energy $E\lo{max} = E_0$. The energy of the electrons decreases after impacting the beam dump due to shower creation and scattering processes~\cite{Tsai:1966js}. This is not the case for muon beams (see section~\ref{sec:Implementation:Modifications} below). The lower energy bound is given for the electron energy $E\lo{min}$ at which the bounds $x\lo{min,max}$ are equal. This integration range for the energy $E$ is, therefore, in the interval between
\begin{align}
    E\lo{min} = E\lo{cut}+m_\ell \,\qc E\lo{max} = E_0 \,.
    \label{eq:Implementation:Boundaries_E}
\end{align}
The function $I_e(E_0,E,t)$ denotes the energy distribution function for electrons with an energy $E$ at a depth $t$ inside the beam dump. The maximum depth for $t$ is given by $T = \rho L\lo{tar} / X$, where $\rho$ is the density of the target and $L\lo{tar}$ is the thickness of the target. 

\begin{figure}[tbp]
    \centering
    \includegraphics[width=0.6\linewidth]{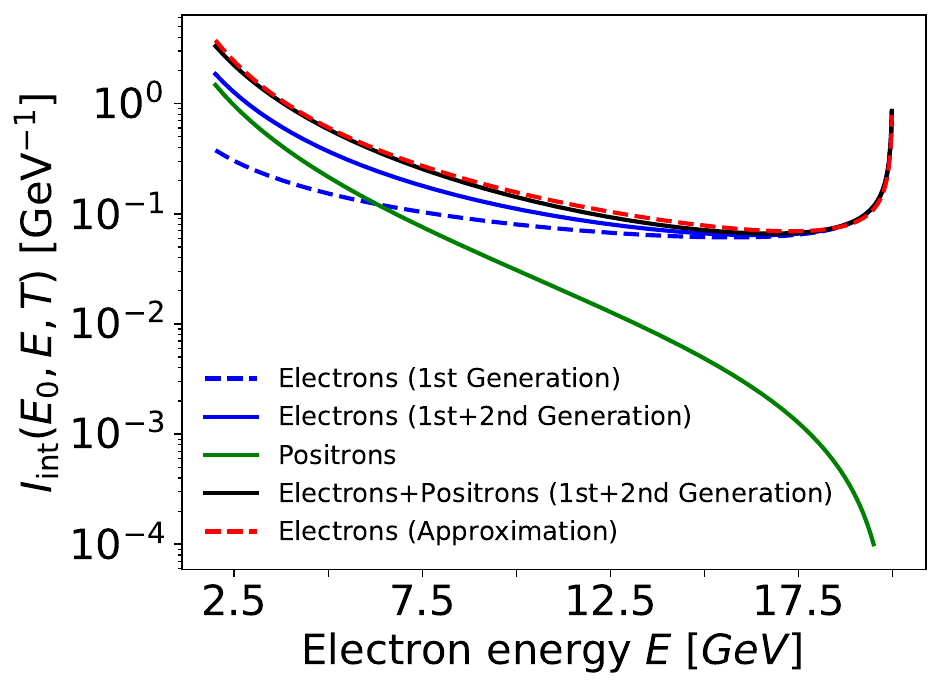}
    \caption{Integrated electron/positron energy distributions for E137.}
    \label{fig:Beam_Dumps:Iint}
\end{figure}

In this work, we consider the primary shower of electrons $I_e^{(1)}$, as well as the secondary shower of electrons and positrons $I_{e,p}^{(2)}$, which are created through pair production from photons in the shower~\cite{Tsai:1966js}.
We use the expressions for $I_e^{(1,2)}$ derived in ref.~\cite{Tsai:1966js}. However, we do not use the simplified distributions, which assume that $E_0 - E \ll E_0$, but instead use the full expressions for the distributions, see appendix~\ref{Ie_details}. Since the electron energy distribution is the only function of $t$ in the integral~\eqref{eq:Implementation:N_a_original}, we define the integrated energy distribution
\begin{align}
    I\lo{int}(E_0,E,T) \equiv \int_0^T \dd t \, I_e(E_0,E,t) \,.
\end{align}
The integrated distributions for first and second-generation electrons/positrons in E137 are shown in figure~\ref{fig:Beam_Dumps:Iint}.
In addition, we indicate the distribution of first-generation electrons obtained when using the approximation $E_0-E \ll E_0$. The approximate expression differs substantially from the exact first-generation distribution. However, it coincidentally matches the more complete result, which includes second-generation effects, quite well.

The absorption of ALPs by electrons in the target is described by $l_\lambda$, which leads to exponential attenuation. This is relevant for thick targets but can be neglected here~\cite{Liu:2023bby}. The ALP decay length $l_a$ depends on the couplings as well as the energy and mass of the ALP. It can be written as
\begin{align}
    l_a = \frac{p_a}{m_a} \frac{1}{\Gamma_a} = \frac{\sqrt{E_a^2-m_a^2}}{m_a} \frac{1}{\Gamma_a} = \frac{\sqrt{x^2 E^2-m_a^2}}{m_a} \frac{1}{\Gamma_a}  \,.
\end{align}
The first exponential function in eq.~\eqref{eq:Implementation:N_a_original} can be understood as the probability that the ALP only decays after leaving the beam dump, such that the decay products do not get absorbed. Conversely, the last term describes the probability that the ALP does not decay behind the detector, i.e.\ that it decays inside the decay volume and can be detected.

With the simplifications discussed above, the general formula we use for the expected number of detected ALPs is
\begin{align}
    N_a = \frac{N_e X}{M\lo{target} / N\lo{avo}} \int_{E\lo{min}}^{E_0}\dd E \int_{x\lo{min}}^{x\lo{max}}\dd x \, I\lo{int}(E_0,E,T) \dv{\sigma}{x} e^{-\frac{L\lo{sh}}{l_a}} \qty(1-e^{-\frac{L\lo{dec}}{l_a}}) \,.
    \label{eq:Implementation:N_a_simplified}
\end{align}
The only missing ingredient is the differential cross section $\mathrm{d}\sigma/\mathrm{d}x$ for the production of ALPs, which will be discussed next.

\subsection{Dark Bremsstrahlung and Primakoff Production}
\label{sec:Implementation:DB_P}

\begin{figure}[tbp]
     \centering
     \includegraphics[width=0.32\textwidth]{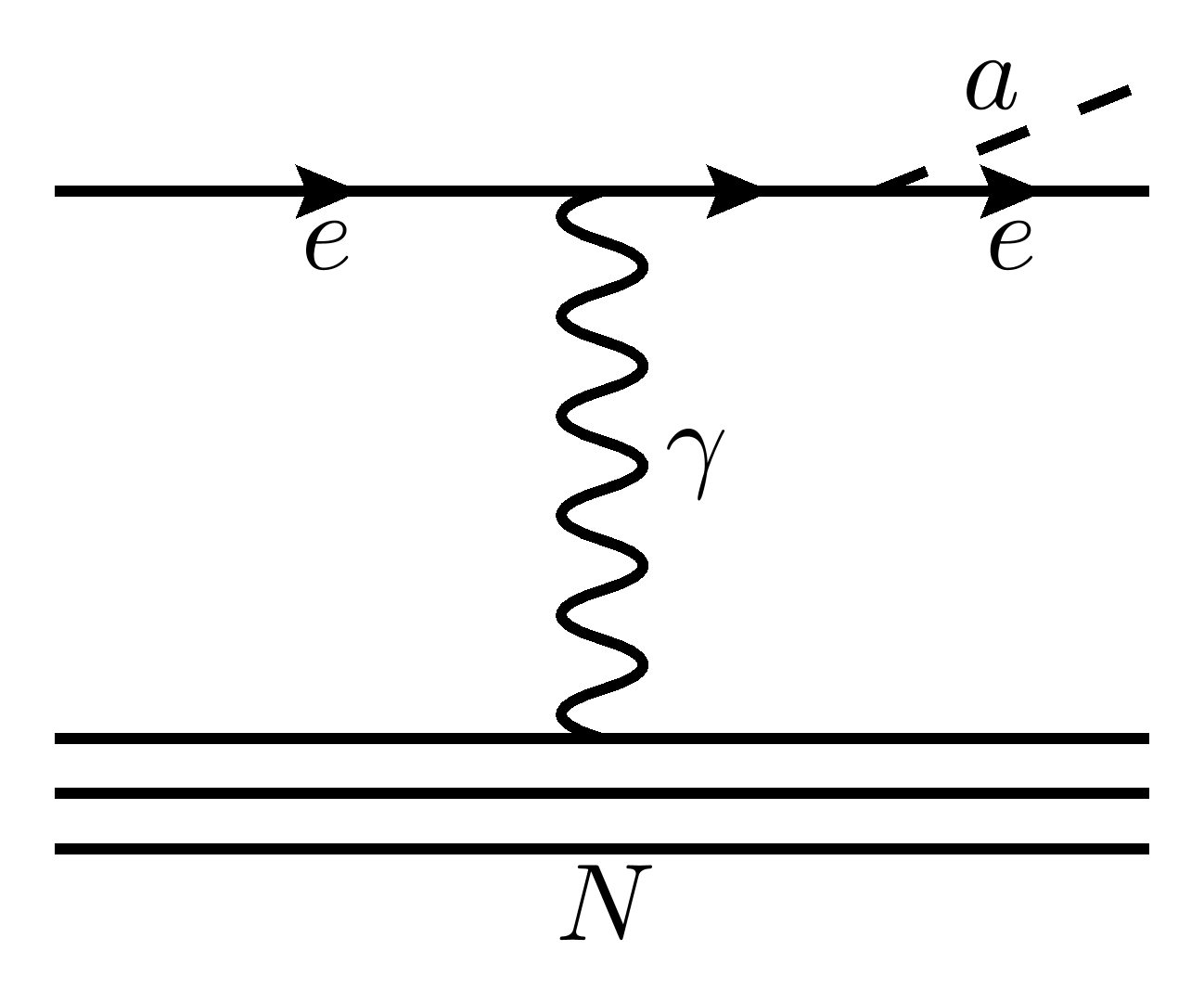}
     \hfill
     \includegraphics[width=0.32\textwidth]{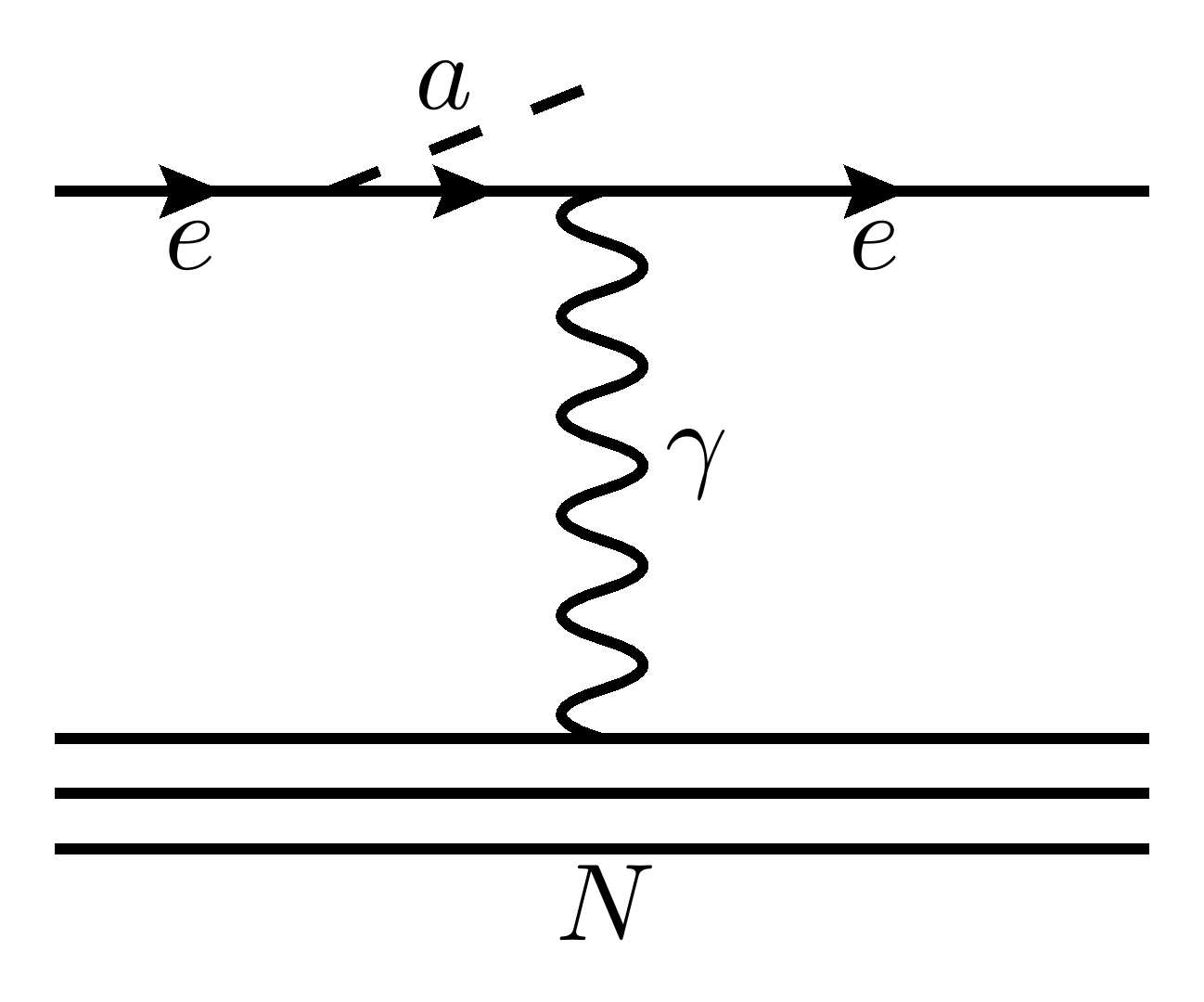}
     \hfill
     \includegraphics[width=0.32\textwidth]{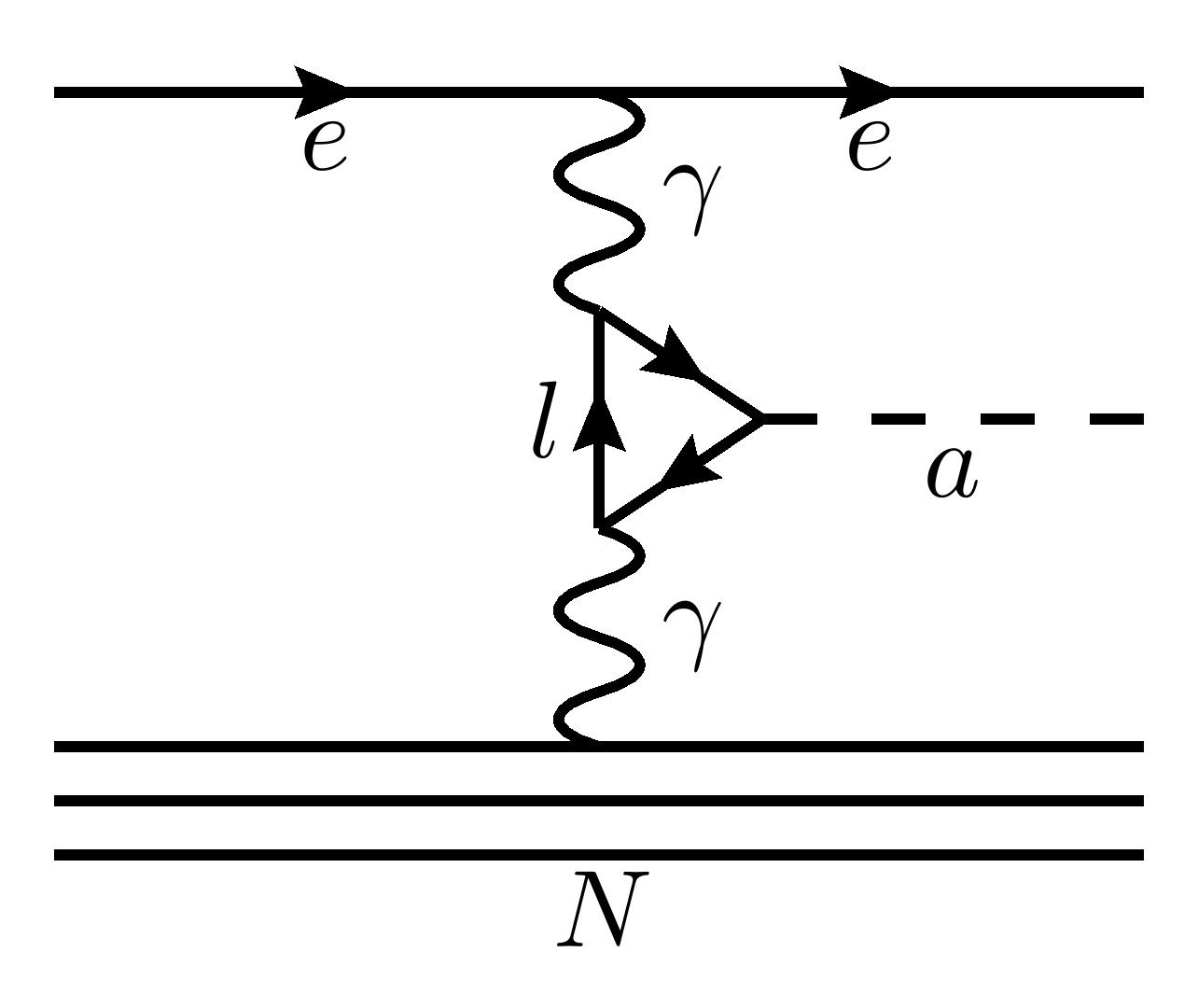}
     \includegraphics[width=0.32\textwidth]{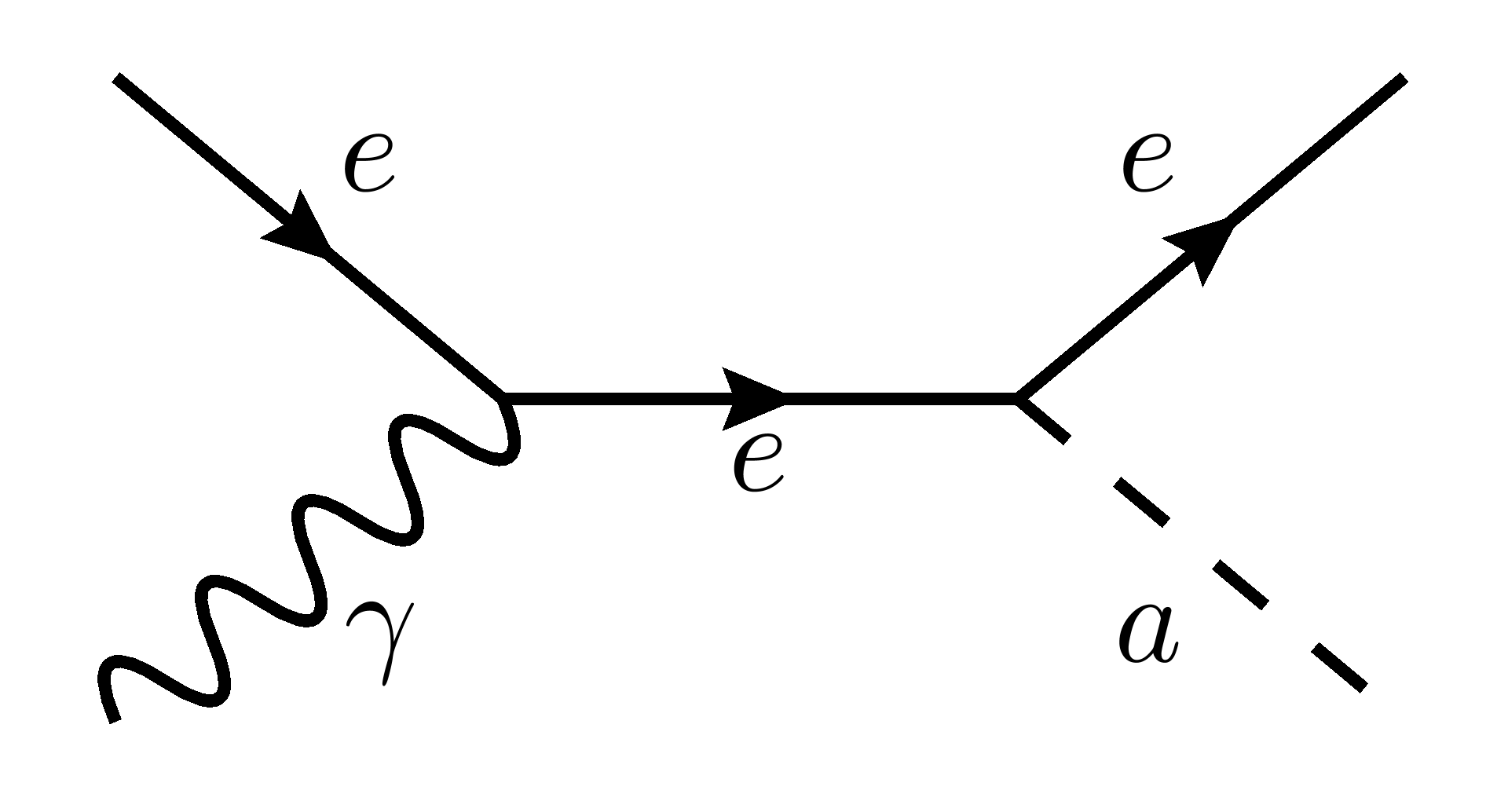}
     \hfill
     \includegraphics[width=0.32\textwidth]{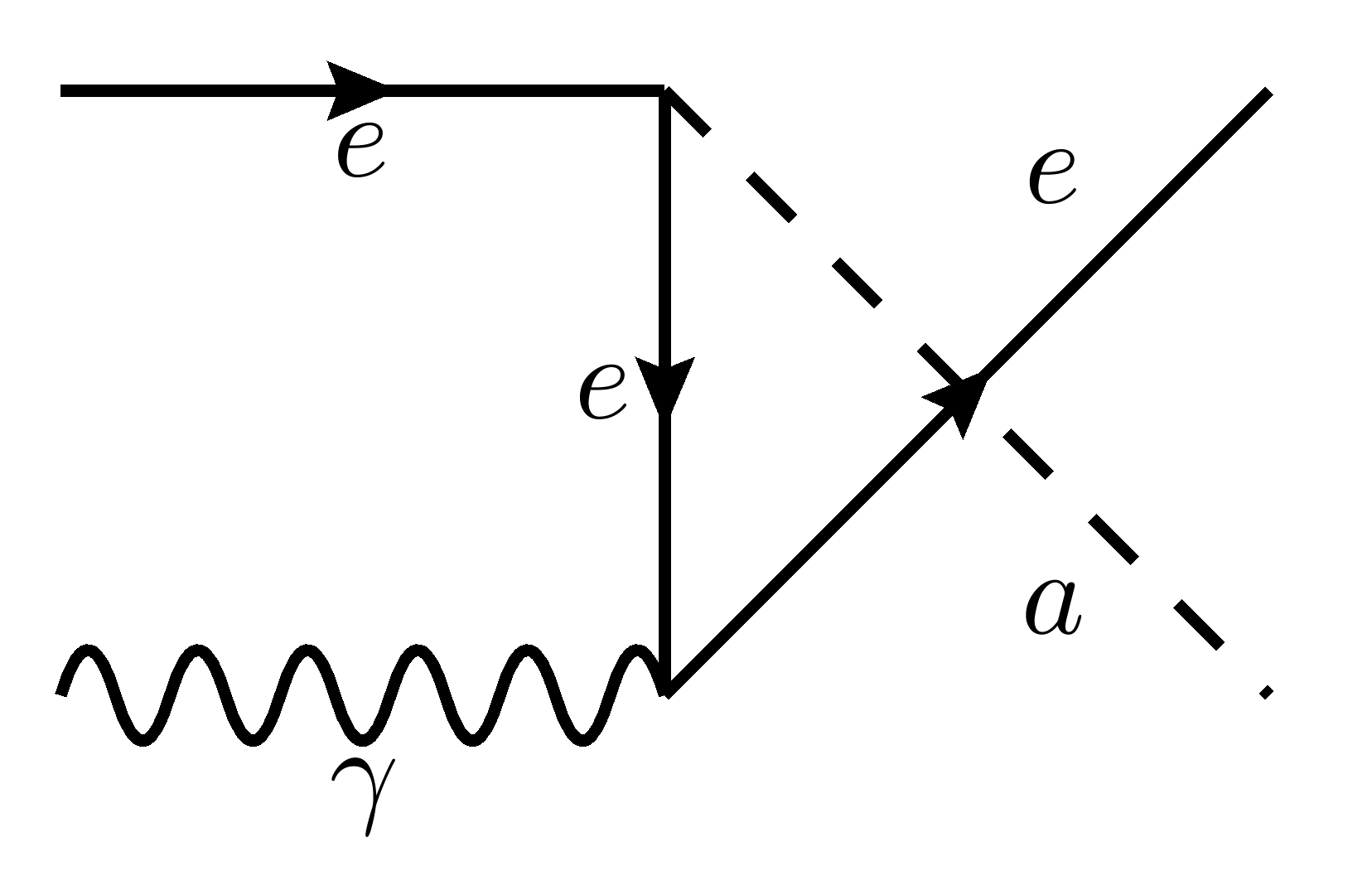}
     \hfill
     \includegraphics[width=0.32\textwidth]{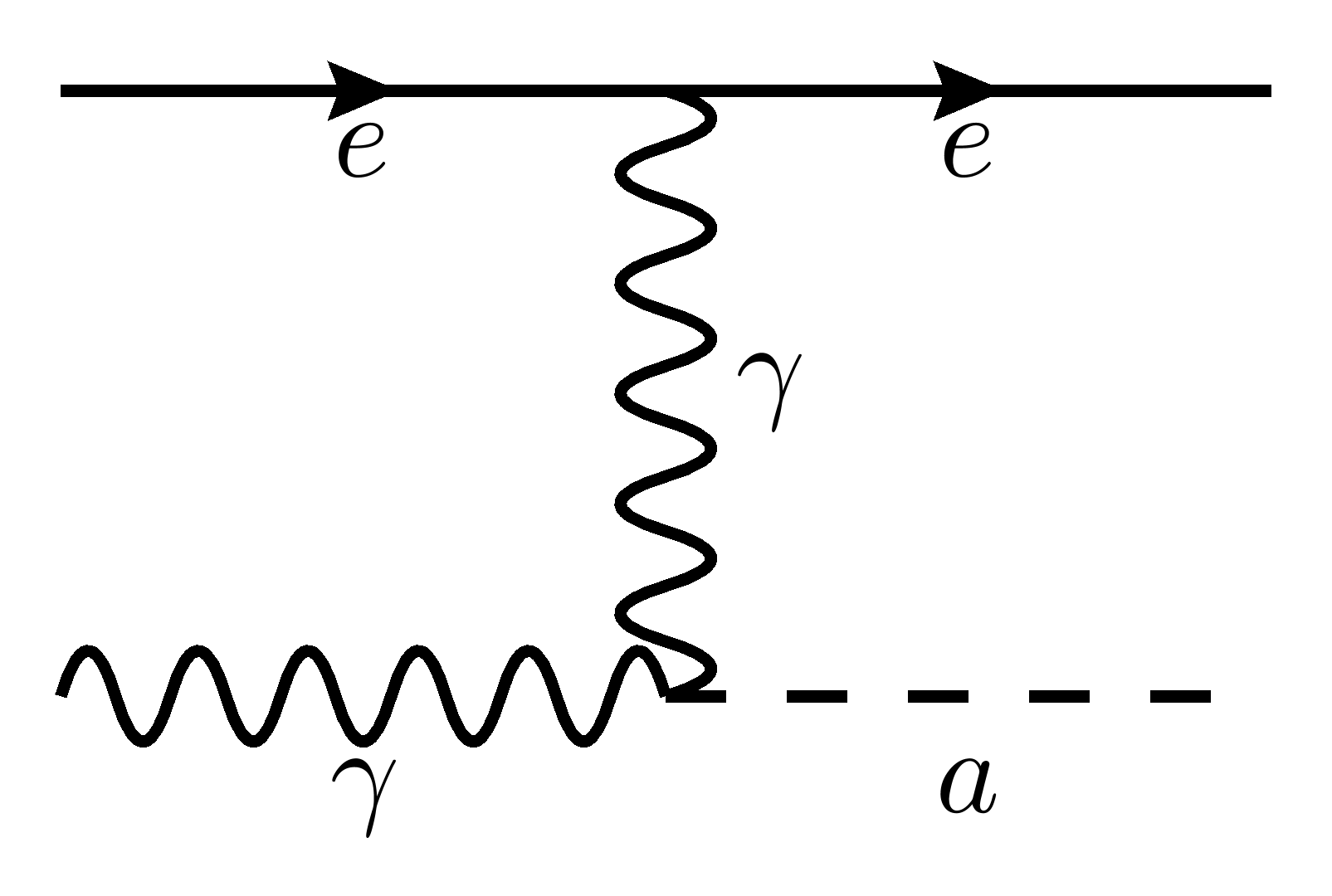}
\caption{Top row: Feynman diagrams contributing to the process $e N \rightarrow e N a$. The diagrams correspond from left to right to the DB $s$-channel process, the DB $u$-channel process and the Primakoff process. Bottom row: Corresponding Feynman diagrams contributing to the process $e \gamma \rightarrow e a$. Note that the lepton loop has been replaced with the effective Primakoff coupling.}
\label{fig:Implementation:Diagrams}
\end{figure}

The main ALP production mechanisms in beam-dump experiments are Dark Bremsstrahlung (DB) and Primakoff production, shown in the top row of figure~\ref{fig:Implementation:Diagrams}.
In the DB diagrams, the leptophilic ALP couples to the beam electron/muon at tree level, while for the Primakoff process, it only couples to the photons via the effective loop coupling.

Due to the high energy of the beam compared to the ALP and lepton masses, the photon that is exchanged between the nucleus and the electron/muon is approximately on-shell~\cite{Kim:1973he}. As a result, the nucleus can be replaced by an effective flux of photons, which is described by the Improved Weizs\"acker-Williams (IWW) approximation. The IWW approximation allows the cross section to be calculated using the reduced $2 \to 2$ scattering amplitude $\overline{\left|\mathcal{M}\right|}^2$. Thus, for a lepton scattering with a nucleus
\begin{align}
    \ell^-(p) + N(P_i) \rightarrow \ell^-(p') + N(P_f) + a(k) \,,
\end{align}
the simplified photoproduction processes are
\begin{align}
    \ell^-(p) + \gamma(q) \rightarrow \ell^-(p') + a(k) \,.
\end{align}
The simplified DB and Primakoff diagrams are shown in the bottom row of  figure~\ref{fig:Implementation:Diagrams}.

The differential cross section for the tree-level DB process can be calculated analytically (see appendix~\ref{app:DB_Primakoff}) and is found to be
\begin{align}
    \begin{split}
        \dv{\sigma\lo{DB}}{x} &= \frac{\alpha}{(4 \pi)^2} \sqrt{x^2-\frac{m_a^2}{E^2}} \chi \frac{1-x}{x} \\
        &\times\frac{e^2 E^2 g_{a \ell}^2 \theta_{a,\text{max}}^2 x^5}{3 (1-x) \left(m_a^2 (1-x)+m_\ell^2 x^2\right)^2 \left(m_a^2 (1-x)+x^2 \left(E^2 \theta_{a,\text{max}}^2+m_\ell^2\right)\right)^3} \\
        &\times \Bigl[3 m_a^4 (1-x)^2 x^2 \left(E^2 \theta_{a,\text{max}}^2+3 m_\ell^2\right)+3 x^6 \left(E^2 \theta_{a,\text{max}}^2 m_\ell+m_\ell^3\right)^2 \Bigr.\\
       \Bigl. &\quad +m_a^2 (1-x) x^4 \left(2 E^4 \theta_{a,\text{max}}^4+9 E^2 \theta_{a,\text{max}}^2 m_\ell^2+9 m_\ell^4\right)+3 m_a^6 (1-x)^3\Bigr] \,,
    \end{split}
    \label{eq:Beam_Dumps:DB_cross_section}
\end{align}
where $\theta_{a,\text{max}}$ denotes the largest angle relative to the beam axis that the ALP can have in order to be detected.
The Primakoff process and the interference term involve the momentum-dependent effective coupling to photons, so we do not obtain a simple analytical expression for the corresponding differential cross sections. Instead, we evaluate them numerically.

\begin{figure}[tbp]
     \centering
     \includegraphics[width=0.495\textwidth]{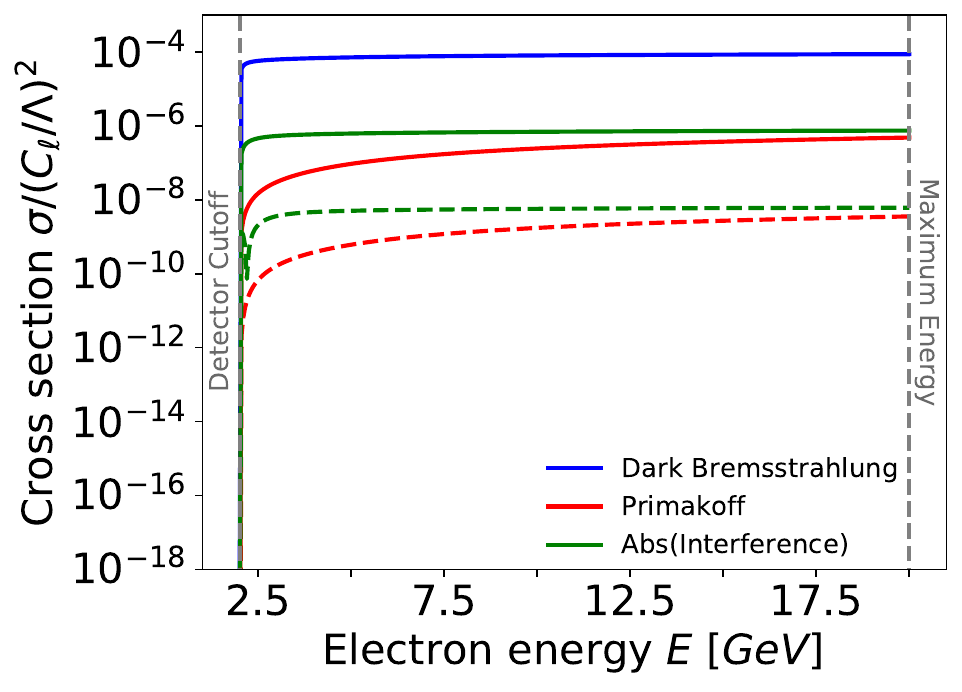}
     \hfill
     \includegraphics[width=0.495\textwidth]{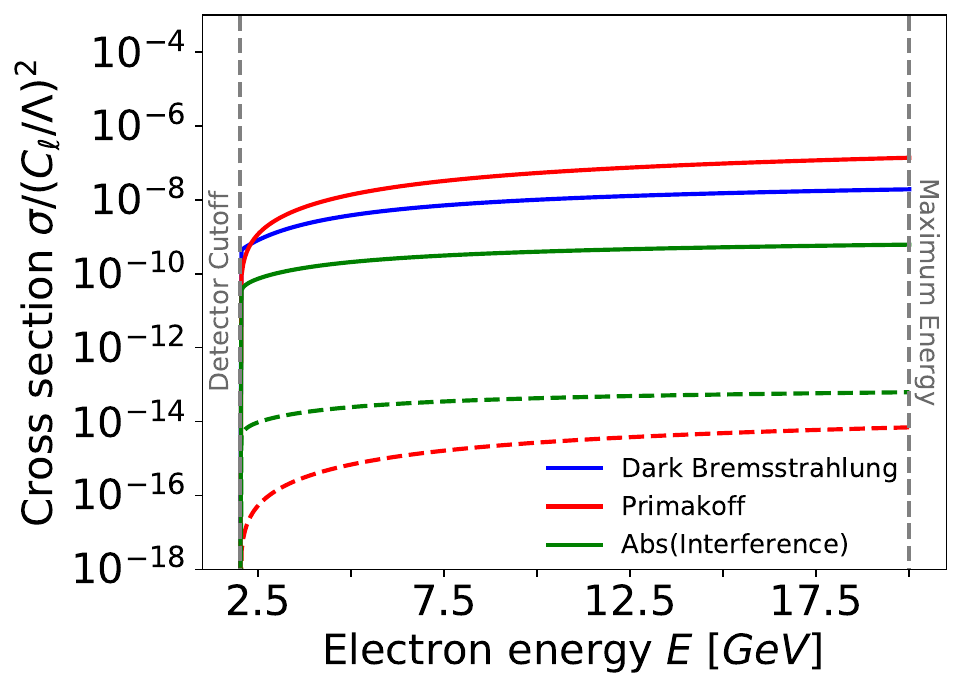}
        \caption{Normalised cross sections for non-vanishing ALP coupling to electrons as a function of the electron energy $E$. The ALP mass is $m_a = \SI{10}{MeV}$ on the left, $m_a = \SI{450}{MeV}$ on the right. The solid lines represent derivative couplings, while the dashed lines represent pseudoscalar couplings.}
        \label{fig:Implementation:CS_El_Total_Comparison}
\end{figure}

In order to analyse the mass and energy dependence of the cross sections, we plot the contributions to $\sigma$ (normalised by the coupling) coming from DB, Primakoff, and interference processes in E137 in figure~\ref{fig:Implementation:CS_El_Total_Comparison}. We assume that the ALP couples exclusively to electrons and compare the normalised cross sections for two masses: $m_a = \SI{10}{MeV}$ and $m_a = \SI{450}{MeV}$.

For $m_a = \SI{10}{MeV}$, the DB cross section is clearly the main contributor to the total cross section. This is to be expected, considering these interactions are mediated by tree-level interactions between the ALPs and electrons. However, it would be wrong to conclude that the Primakoff process does not give important contributions elsewhere, as can be seen for $m_a = \SI{450}{MeV}$. For this mass, Primakoff production outweighs DB production, assuming the ALP exhibits a derivative coupling to electrons (while for pseudoscalar coupling, the Primakoff contribution is again negligible). This mainly happens because, while it is true that all cross sections decrease for an increasing ALP mass, the Primakoff one for derivative couplings decreases only very slowly with respect to the other ones, hence eventually becoming dominant. This difference arises due to the fact that scale-invariant contribution to the effective photon coupling remains constant, while the loop factor is suppressed.

It should be noted that comparing $\sigma$ is only illustrative, as the actual integration over $x$ weighs $\dd \sigma/\dd x$ with the exponential functions given at the end of eq.~\eqref{eq:Implementation:N_a_simplified}. However, it is clear that while the Primakoff production of leptophilic ALPs occurs only via an effective loop coupling, it can still have an impact on constraints that can be set on the ALP mass and couplings.

Finally, there are still the contributions arising from secondary positrons created in the shower. Both the DB and Primakoff processes give exactly the same contributions when the electron line in the Feynman diagrams is replaced with a positron line. The only difference between the contributions are the electron and positron energy distribution functions $I_{e,p}$. This means that instead of considering the electron and positron production processes separately, we summarise them together by using the total electron-positron distribution function
\begin{align}
    I\lo{int}(E_0,E,T) &= \int_0^T \dd t \qty(I_e^{(1)}(E_0,E,t)+I_e^{(2)}(E_0,E,t)) + \int_0^T \dd t I_p^{(2)}(E_0,E,t) \\
    &= \int_0^T \dd t \qty(I_e^{(1)}(E_0,E,t)+2 I_e^{(2)}(E_0,E,t)) \,.
\end{align}

For muonphilic ALPs, there can also be ALP bremsstrahlung from secondary muons produced in the electromagnetic shower~\cite{Marsicano:2018vin}. Comparing our results to the ones in ref.~\cite{Marsicano:2018vin}, we find that the loop-induced Primakoff process dominates ALP production for muonphilic ALPs. Therefore, we do not include the contributions of secondary muons in the present work.

\subsection{Annihilation Processes}
\label{sec:Implementation:Annihilation}

In addition to the contributions from DB and Primakoff production, we also consider ALP production from processes in which a positron created in a secondary shower annihilates with an electron inside the beam dump. Since only positrons from the shower contribute to these processes, the first important change compared to DB and Primakoff production is that the integrated energy distribution appearing in eq.~\eqref{eq:Implementation:N_a_simplified} only contains the contributions from positrons, i.e.
\begin{align}
    I\lo{int}(E_0,E,T) = \int_0^T \dd t \, I_p^{(2)}(E_0,E,t) \,.
\end{align}

\begin{figure}[tbp]
    \centering
    \includegraphics[width=0.25\linewidth]{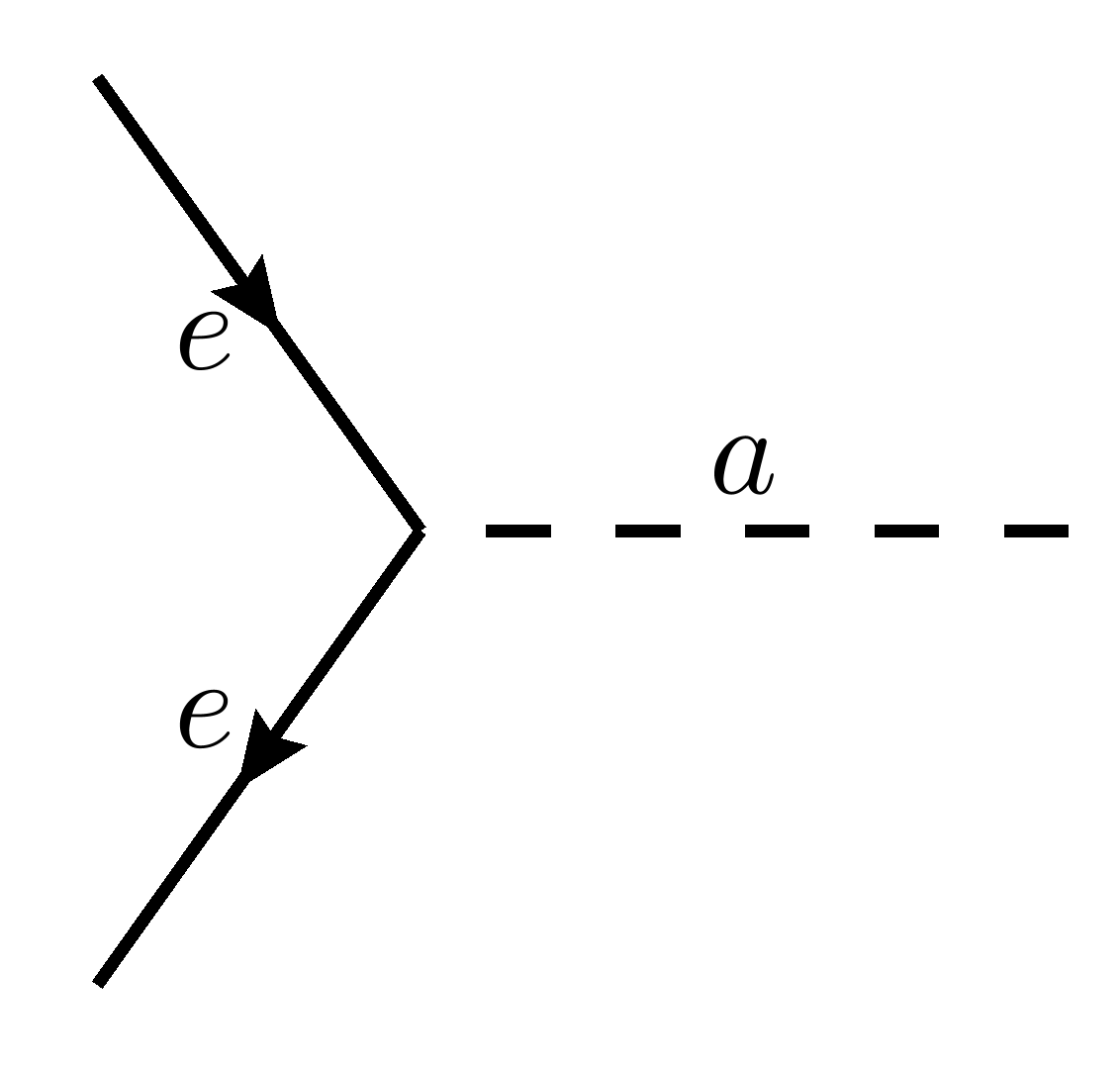}
    \caption{Resonant ALP production through the process $e^- e^+ \rightarrow a$.}
    \label{fig:Implementation:Annihilation_Resonant}
\end{figure}

If the annihilation of the positron and electron has a centre-of-mass energy that is exactly $\sqrt{s} = m_a$, an ALP can be produced resonantly:
\begin{align}
    e^-(p_1) + e^+(p_2) \rightarrow a(k) \,.
\end{align}
The Feynman diagram for this process is shown in figure~\ref{fig:Implementation:Annihilation_Resonant}.
The matrix element for this process is simply
\begin{align}
    \overline{\abs{\mathcal{M}\lo{R}}}^2 &= \frac{g_{a e}^2 s}{2} \, ,
\end{align}
and the differential cross section is given by
\begin{align}
    \dv{\sigma\lo{R}}{x} &= \frac{\pi g_{a e}^2}{4 m_e \sqrt{1-\frac{4 m_e^2}{m_a^2}}} \,\delta\qty(E - \qty(\frac{m_a^2}{2 m_e} - m_e)) \delta\qty(x-\qty(1+\frac{m_e}{E})) \,.
    \label{eq:Implementation:Resonant_CS}
\end{align}
The first delta distribution in eq.~\eqref{eq:Implementation:Resonant_CS} is equivalent to the condition $s = m_a^2$. After inserting this cross section into eq.~\eqref{eq:Implementation:N_a_simplified}, the two delta distributions cancel out the integrations, leaving behind
\begin{align}
    N_{a,\text{res}} = \frac{N_e X}{M\lo{target} / N\lo{avo}} \frac{\pi g_{a e}^2}{4 m_e \sqrt{1-\frac{4 m_e^2}{m_a^2}}} I\lo{int}(E_0,E_*,T) e^{-\frac{L\lo{sh}}{{l_a}_*}} \qty(1-e^{-\frac{L\lo{dec}}{{l_a}_*}}) \,.
\end{align}
Here, $E_*$ and ${l_a}_*$ are the positron energy and decay length, respectively, after applying the conditions given by the two delta distributions in eq.~\eqref{eq:Implementation:Resonant_CS}. The mass range in which resonant production can be detected is bounded from below by the fact that the ALP must have an energy larger than the cutoff energy of the detector and from above by the maximum possible positron energy, i.e.
\begin{align}
    E_a = E + m_e = \frac{m_a^2}{2 m_e} &> E\lo{cut} \,, \\
    E = \frac{m_a^2}{2 m_e} - m_e &< E_0 \,.
\end{align}
This means resonantly produced ALPs can only be detected when the ALP mass is in the interval $\sqrt{2 m_e E\lo{cut}} < m_a < \sqrt{2 m_e (E_0+m_e)}$.

In addition to the resonant production discussed above, electron-positron annihilation can also proceed non-resonantly. The corresponding cross sections are discussed in appendix~\ref{app:Non_Resonant}.

\subsection{ALP Yields in NA64}
\label{sec:Implementation:Modifications}

While eq.~\eqref{eq:Implementation:N_a_simplified} holds for E137 and the visible mode setup of NA64, there are some relevant changes when calculating the expected yield of ALPs for the invisible setup as well as the muon setup of NA64.

\subsubsection*{NA64 in invisible mode}

Since the invisible setup of NA64 is an active beam dump that measures the missing energy of an event instead of detecting particles downstream of the dump, one can consider the decay volume to be infinitely large, i.e. one can set $L\lo{dec} \rightarrow \infty$. This simplifies the equation for $N_a$ to
\begin{align}
    N_a = \frac{N_e X}{M\lo{target} / N\lo{avo}} \int_{E\lo{min}}^{E_0}\dd E \int_{x\lo{min}}^{x\lo{max}}\dd x I\lo{int}(E_0,E,T) \dv{\sigma}{x} e^{-\frac{L\lo{sh}}{l_a}} \,,
    \label{eq:Implementation:N_a_invisible}
\end{align}
where the various quantities that enter this equation are given in appendix~\ref{app:experimental_parameters}.
In principle, it would also be possible to simplify the expression for $\mathrm{d}\sigma/\mathrm{d}x$, since there is no longer a restriction on the emission angle $\theta_a$ of the ALP. Since the main contribution to the cross section comes from small angles~\cite{Dusaev:2020gxi}, one can safely extend the integration to large angles of order unity. Concretely, in analogy to E137, we introduce an arbitrary cutoff for the angle based on the geometry of the detector given by
\begin{align}
    \theta_{a,\text{max}} = \frac{R\lo{sh}}{L\lo{sh}} \,,
\end{align}
where $R\lo{sh} \approx \SI{60}{\centi\m}$ is the approximate radius of the detector, in this case of the hadronic calorimeters (HCALs).

\subsubsection*{NA64\texorpdfstring{\boldmath $\mu$}{mu}}

The general setup of the NA64$\mu$ experiment is similar to that of NA64 run in invisible mode, with it being an active beam dump. This means that the formula~\eqref{eq:Implementation:N_a_invisible} applies to NA64$\mu$ as well, with the caveat that the electron lines in the DB and Primakoff diagrams in figure~\ref{fig:Implementation:Diagrams} need to be replaced with muon lines, and the EOT $N_e$ should be replaced with the muons on target (MOT) $N_\mu$. Since there are no muons for the anti-muons in the beam to annihilate with, the contribution discussed in section~\ref{sec:Implementation:Annihilation} is absent. Finally, an important difference is the energy distribution function of the muons compared to the electrons. Since muons are much more massive than electrons, they lose significantly less energy due to bremsstrahlung compared to electrons, allowing them to easily penetrate the beam dump~\cite{NA64:2024klw}. This means the muon's energy distribution function is independent of the depth $t$. Therefore, we can use the thin-target approximation to calculate the distribution function for muons, which is given by~\cite{Tsai:1966js}
\begin{align}
    I_\mu(E_0,E,t) = \delta(E_0-E) \quad \Rightarrow \quad I\lo{int}(E_0,E,T) = T \delta(E_0-E) \,.
\end{align}
This means the expected yield of ALPs simplifies to
\begin{align}
    N_a = \frac{N_\mu X}{M\lo{target} / N\lo{avo}} T \int_{x\lo{min}}^{x\lo{max}}\dd x \dv{\sigma}{x} e^{-\frac{L\lo{sh}}{l_a}} \,,
\end{align}
where it is implied that $E=E_0$ in the expressions of $x\lo{min,max}$, $l_a$, and the cross section. It is important to note that this makes $N_a$ very sensitive to the length $T$ of the beam dump. This is in contrast with the other electron beam dumps, in which the electron shower attenuates after several radiation lengths.

\section{Other Constraints}
\label{sec:other_constraints}

In this section we review other experimental and theoretical constraints of relevance for leptophilic ALPs.

\subsection{Perturbative unitarity}
\label{sec:PertUni}

A simple but important constraint on the ALP coupling to leptons can be obtained by considering the scattering process $\ell^+ \ell^- \to \ell^+ \ell^-$ via $s$-channel ALP exchange. Even though the ALP interaction is non-renormalisable, the matrix element for this process becomes independent of the centre-of-mass energy for $\sqrt{s} \gg m_a$. The requirement of unitarity of the $J=0$ partial wave $a_0$, i.e.\ $|\text{Re}\ a_0| < 1/2$, implies~\cite{Cornella:2019uxs}
\begin{align}
    \frac{C_\ell}{\Lambda} = \frac{g_{a \ell}}{m_\ell} < \frac{1}{m_\ell} \sqrt{\frac{8\pi}{3}} \,.
\end{align}

\subsection[LEP bounds on exotic $Z$ boson decays]{LEP bounds on exotic \boldmath $Z$ boson decays}
\label{sec:LEP_bounds}

Ref.~\cite{Jaeckel:2015jla} studied constraints on ALP couplings to photons from LEP measurements on the $Z$ pole. Here, we apply this analysis to the present setup, where photon couplings are induced via lepton loops. As we will see, these loops are strongly suppressed by $m_\ell^2/M_Z^2$ for the case of pseudoscalar ALP couplings, meaning relevant constraints from LEP only arise for derivative couplings.  

There are two processes of interest: $e^+ e^- \to Z \to a \gamma$ via on-shell production of a $Z$ boson, and $e^+ e^- \to \gamma^* \to a \gamma$ via an off-shell photon at the $Z$ peak. For light ALPs with mass below $m_\pi$, the photons from the radiative ALP decay are very collimated, so that both processes look like the forbidden SM decay $Z \to 2 \gamma$~\cite{Chala:2015cev}. 
For $Z$-boson decays we follow ref.~\cite{Jaeckel:2015jla} and use the bound
\begin{align}
{\rm BR} (Z \to a \gamma) \times  {\rm BR} (a \to \gamma \gamma) \times P(l_a) \le 5.2 \times 10^{-5} \, ,
\label{eq:BRbound}
\end{align}
which is inferred from the  95\% confidence level LEP limit ${\rm BR} (Z \to \pi^0 \gamma ) \le 5.2 \times 10^{-5}$~\cite{L3:1995nbq}. Here, we conservatively include only the fraction ${\rm BR} (a \to \gamma \gamma)$ of ALPs decaying into two photons since it is unclear whether a leptonically decaying ALP could mimic a $\pi^0$ decay when strongly boosted. Moreover, the factor
\begin{align}
P(l_a) = 1 - e^{- \tfrac{L} {l_a}} \, ,
\end{align}
gives the probability that the ALP decays inside the decay volume and can be detected. Here, $l_a$ is the ALP decay length in the laboratory frame (see section~\ref{sec:Implementation:Yield}), and we take $L = 10\,\mathrm{cm}$. The above limits are used for $m_a \le m_{\pi^0} = 135 \MeV$. For larger ALP masses up to 1 GeV we use the limits obtained in ref.~\cite{Jaeckel:2015jla} from simulating angular distributions for the three-photon decay ($e^+ e^- \to Z \to a + \gamma \to 3 \gamma$), which were compared bin by bin to the distributions given in ref.~\cite{L3:1995nbq}. The resulting bounds are weaker than those for $m_a \le m_{\pi^0}$ by about 30\%.

These constraints can be directly interpreted as a bound on lepton couplings,  using the decay rates for derivative (D) and pseudoscalar (P) couplings~\cite{Bauer:2021mvw} 
\begin{align}
\Gamma (Z \to a \gamma)_{\rm D} & = \frac{m_Z^3}{96 \pi^3} \frac{\alpha^2}{s_{\rm w}^2 c_{\rm w}^2 } \frac{|C_\ell|^2}{\Lambda^2} \left( s_{\rm w}^2 - 1/4 \right)^2 \left| 1 + B (\tau_\ell, \tau_Z)\right|^2 \left( 1 - \frac{m_a^2}{m_Z^2} \right)^3  \, , \nonumber \\
\Gamma (Z \to a \gamma)_{\rm P} & = \frac{m_Z^3}{96 \pi^3} \frac{\alpha^2}{s_{\rm w}^2 c_{\rm w}^2 } \frac{|g_{a\ell}|^2 }{m_\ell^2} \left( s_{\rm w}^2 - 1/4 \right)^2 \left| B (\tau_\ell, \tau_Z) \right|^2 \left( 1 - \frac{m_a^2}{m_Z^2} \right)^3  \, , 
\end{align}
with $s_\text{w}$ denoting the sine of the Weinberg angle, $\tau_Z  = 4 m_\ell^2/M_Z^2$, $\tau_\ell$ is defined below Eq.~\eqref{eq:Theory:C_0_Decay}, and $B (\tau_1, \tau_2)$ given in eq.~\eqref{eq:Bt1t2}. Since here we consider $m_a \ll m_Z$, one has $\tau_Z \ll \tau_\ell$, so that to good approximation
\begin{align}
B (\tau_\ell, \tau_Z) & =   \tau_Z \left(  f^2(\tau_\ell) -   f^2(\tau_Z) \right) \ll 1 \, .
\end{align}
Constraints on the pseudoscalar scenario are thus much weaker than those in the derivative scenario. In fact, they are weaker than the perturbative unitarity bounds for all leptons and, therefore, irrelevant. 

For the derivative scenario, the constraints are more relevant but still rather weak due to the small $Z$ boson couplings to SM leptons resulting from the accidental cancellation $s_w^2 \approx 1/4$. As we will see below, the LEP limits from $Z$ decays are coincidentally of the same magnitude as the limits from  ALP production via off-shell photons. 

Indeed, the same signature as $Z \to a \gamma$ is generated from ALP production via off-shell photons $e^+ e^- \to \gamma^* \to a \gamma$ with $s = M_Z^2$. Therefore, the cross-section is limited by
\begin{align}
\sigma ( {e^+ e^- \to \gamma^* \to a \gamma})|_{s = M_Z^2} & \le \sigma_{\rm limit} ( {e^+ e^- \to Z \to a \gamma})|_{s = M_Z^2} \nonumber \\
& = \frac{12 \pi}{m_Z^2} {\rm BR} (Z \to e^+ e^-) {\rm BR}_{\rm limit} (Z \to a \gamma) \nonumber \\
& = \frac{7.9 \times 10^{-9}}{\GeV^2} \times  \frac{1}{{\rm BR} (a \to \gamma \gamma)} \times \frac{1}{P(l_a)} 
\, ,
\end{align}
where we have used the narrow-width approximation in the second line and eq.~\eqref{eq:BRbound} in the third line.
The cross section at the $Z$ pole can be calculated using the general vertex function with a single off-shell photon in the derivative and pseudoscalar scenario, respectively (cf. eq.~\eqref{eq:Theory:Effective_Coupling})
\begin{align}
\Gamma^{\mu \nu, D}_{a \gamma^* \gamma}    & = i \frac{ \alpha C_\ell}{\pi \Lambda} \eps^{\mu \nu \alpha \beta} q_\alpha k_\beta    \left[ 1 + B (\tau_\ell, \tau_Z) \right] \, , \nonumber \\
\Gamma^{\mu \nu, P}_{a  \gamma^* \gamma}    & =  i \frac{ \alpha g_{a\ell}}{\pi m_\ell} \eps^{\mu \nu \alpha \beta} q_\alpha k_\beta  B (\tau_\ell, \tau_Z)   \, ,
\label{off1}
\end{align}
 where the external on-shell photon has outgoing momentum $q$ and polarisation index $\mu$, $k$ is the outgoing momentum of the axion, and $p = k + q$ is the inflowing momentum of the off-shell photon with polarisation index $\nu$, with $p^2 = s = M_Z^2$. The total cross-section is given by 
\begin{align}
\sigma ({e^+ e^- \to \gamma^* \to a \gamma})_{\rm D} & = \frac{\alpha^3}{24 \pi^2 } \frac{|C_\ell|^2}{\Lambda^2} \frac{\left(1 - \frac{m_a^2}{m_Z^2}\right)^3 \left(1+ \frac{2 m_\ell^2}{m_Z^2}\right)}{\sqrt{1-4 m_\ell^2/M_Z^2}} \left| 1 + B (\tau_\ell, \tau_Z) \right|^2 \, ,  \nonumber \\
\sigma ({e^+ e^- \to \gamma^* \to a \gamma})_{\rm P} & = \frac{\alpha^3}{24 \pi^2 } \frac{|g_{a\ell}|^2}{m_\ell^2} \frac{\left(1 - \frac{m_a^2}{m_Z^2}\right)^3 \left(1+ \frac{2 m_\ell^2}{m_Z^2}\right)}{\sqrt{1-4 m_\ell^2/m_Z^2}} \left|  B (\tau_\ell, \tau_Z) \right|^2 \, , 
\end{align}
Again, constraints on the lepton couplings in the pseudoscalar scenario are weaker than those from perturbative unitarity, such that constraints from ALP production at LEP are relevant only in the derivative scenario. They are accidentally of the same size as the constraints from $Z$ boson decays, since 
\begin{align}
\frac{ \Gamma_Z}{12 \pi M_Z {\rm BR}(Z \to e^+ e^-) } \frac{e^2 s_{\rm w}^2 c_{\rm w}^2}{(s_{\rm w}^2 - 1/4)^2} \simeq  0.94\, .
\end{align}
\begin{figure}[!t!]
	\centering
	\includegraphics[width=.495\textwidth]{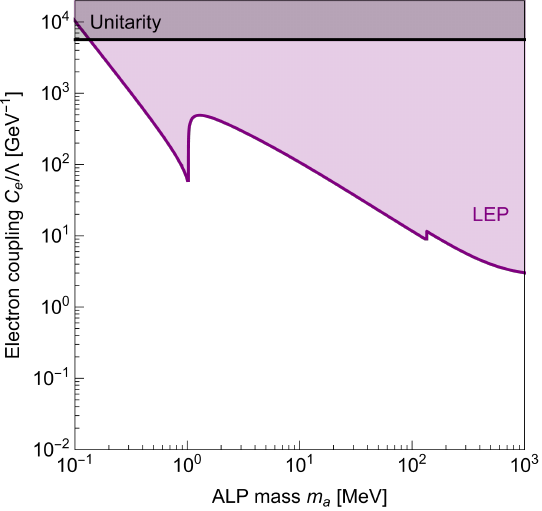}
    \includegraphics[width=.495\textwidth]{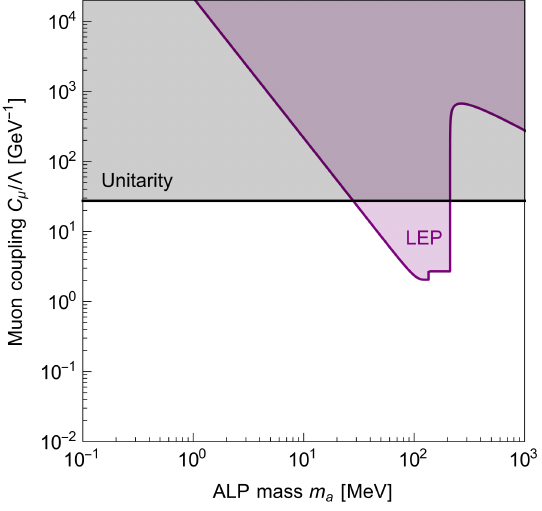}
    \includegraphics[width=.495\textwidth]{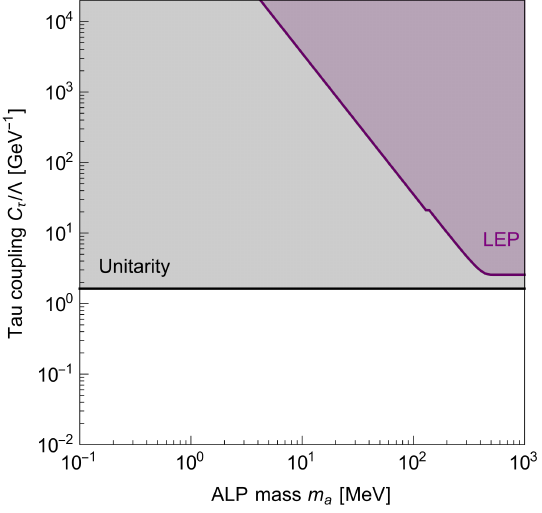}
    \includegraphics[width=.495\textwidth]{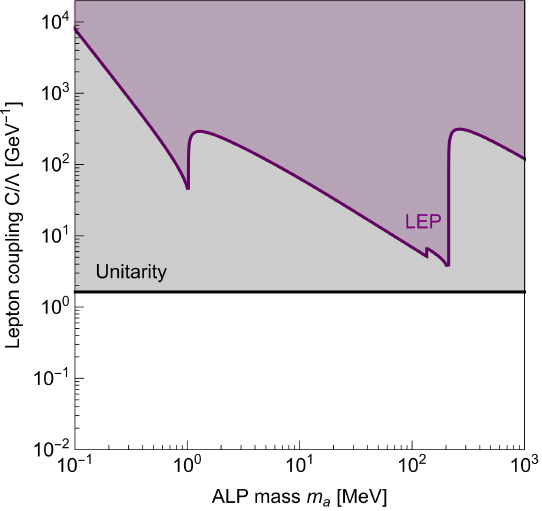}
	 \caption{  Constraints on ALP derivative couplings from LEP  searches for $Z \to 3 \gamma$ \label{ZdecayD} from direct ALP production either via on-shell or off-shell photons (purple), see text for details. Also shown are the limits from perturbative unitarity (gray), which always dominate over analogous constraints on pseudoscalar couplings.} 
\end{figure}

The resulting constraints on the lepton couplings in the derivative scenario are shown in the ($m_a-C_\ell/\Lambda$ plane) in figure~\ref{ZdecayD} for all three individual lepton couplings as well as the universal case, with the limits from perturbative unitarity shown in grey. Note that, as explained above, for $m_a \le m_{\pi^0}$ the bounds are slightly stronger than above, due to highly collimated photons looking like the forbidden $Z \to \gamma \gamma$ decay~\cite{Jaeckel:2015jla}; moreover, the spikes in correspondence of $2m_e$ and $2m_\mu$ originate from the leptonic ALP decay becoming allowed, hence lowering the corresponding ${\rm BR} (a \to \gamma \gamma)$; finally, the plateau observed for the muon case between $\sim100$ MeV and $2m_\mu$, and for the tau case above $\sim400$ MeV, correspond to the ALP mass values where $P(l_a)\simeq 1$.

\subsection{B factories}
\label{sec:Bfactories}

In ref.~\cite{BaBar}, the BaBar collaboration reported results from a search for dark photons $A'$ produced in the reaction $e^+e^- \to \gamma A'$, $A'\to\ell^+\ell^-$ with $\ell = e,\mu$ in terms of a 90\% CL upper bound on the kinetic mixing parameter $\epsilon$ as a function of the dark photon mass $m_{A'}$. We reinterpret these results in terms of the couplings of leptophilic ALPs by comparing the cross section for dark photon production with the one for non-resonant production of ALPs, see appendix~\ref{app:Non_Resonant}.

Since the BaBar search focuses on dark photon decays to light charged leptons, the bound does not apply to models of ALPs coupled exclusively to tau leptons. For ALPs coupled exclusively to electrons, we only reinterpret the bounds up to the muon production threshold, $m_a < 2 m_\mu$. Similarly, for ALPs coupled exclusively to muons, we only reinterpret the BaBar bound for $m_a > 2 m_\mu$. Note that even though there is no tree-level coupling to electrons in this case, the ALP can still be produced through the s-channel process $e^-e^+ \to \gamma^* \to a \gamma$ via the effective photon coupling induced through the muon loop.

In ref.~\cite{Belle-II:2024wtd}, the Belle II collaboration reported results from their search for a muonphilic scalar $S$ via the process $e^- e^+ \to \mu^- \mu^+ S$ with $S \to \mu^-\mu^+$. The Belle II collaboration assumed an interaction Lagrangian of the form $\mathcal{L}_{\text{int}} \supset - i g_{a\mu} a \overline{\psi}_\mu \psi_\mu$, which allows us to directly use their results, neglecting the small differences of scalars compared to pseudoscalars. Moreover, this constraint applies to both the derivative and pseudoscalar coupling cases since it involves a tree-level coupling between the muon and the ALP.

\subsection[Meson and $W^+$ boson decays]{Meson and {\boldmath $W^+$} boson decays}
\label{sec:Raremesondec}

Ref.~\cite{Altmannshofer:2022ckw} pointed out the possibility of searching for leptophilic ALPs using rare meson decays such as $\pi^+ \to e^+ \nu a$~\cite{SINDRUM:1986klz} or $K^+ \to e^+ \nu a$~\cite{Poblaguev:2002ug}, or $W^+$ boson decays such as $W^+ \to e^+ \nu a$. The resulting constraints can be directly applied to our setup, noting that the derivative scenario corresponds to the ``weak-violating'' ALP in ref.~\cite{Altmannshofer:2022ckw} and the pseudoscalar scenario to the ``weak-conserving'' ALP. These terms are defined above their eq.~(7), from which, along with their eq.~(1), one can directly see that the ``weak-violating'' model is identical with our derivative coupling case in eq.~\eqref{eq:Theory:Lagrangian_Der}. On the other hand, the ``weak-conserving'' model is better mapped to our scenarios using the basis in their eq.~(3), which for the ``weak-conserving'' model contains the pseudoscalar coupling, but also the anomalous coupling between ALP and the photon, so it is not identical with the pseudoscalar scenario in eq.~\eqref{eq:LagP}. Nevertheless, as can also be seen from their fig.~1, the limits from rare meson decays in ref.~\cite{Altmannshofer:2022ckw} only constrain the first and last terms in their eq.~(3), and the last term vanishes for the ``weak-conserving'' ALP. Thus, in practice, this case constrains our pseudoscalar scenario.


The decay $\pi^+ \to e^+ \nu a$ can also occur in air showers, in which case the subsequent ALP decays may be observed with Super-Kamiokande~\cite{Cheung:2022umw}. For the case of ALPs coupling only to tau leptons, ref.~\cite{Alda:2024cxn} derived constraints from BESIII searches for $J/\psi \to \gamma + \text{inv}$~\cite{BESIII:2020sdo} and $J/\psi \to 3 \gamma$~\cite{BESIII:2024hdv}.

\subsection{Supernova bounds}
\label{sec:SN1987A}
A burst of ALPs can be produced by the electrons, muons or photons that are present in large numbers in the hot and dense plasmas of core-collapse supernovae. In 1987, one of these explosions, named SN 1987A, was observed in the Large Magellanic Cloud, around 50 kpc away from Earth. This is the only supernova to date for which the accompanying neutrino emission was measured by experiments here on Earth. The duration of this neutrino burst is in overall agreement with simulations~\cite{Fiorillo:2023frv}, which can be used to constrain ALP models. If any exotic particle would be produced in the central neutrinosphere of the nascent proto-neutron star in the supernova core and was coupled so weakly that it escaped without re-depositing its energy, it would provide an additional cooling channel that is not present in simulations. Following approximate, order-of-magnitude physical arguments supported by early simulations of the explosion, the additional cooling power cannot be larger than the neutrino luminosity --- otherwise the duration of the neutrino burst would be significantly shortened, contrary to observations~\cite{Raffelt:1996wa,Raffelt:2006cw}.

Furthermore, one can derive a bound from the ALPs transferring energy into the mantle of the progenitor star, which would contribute to the explosion energy of the circumstellar material \cite{Falk:1978kf,Sung:2019xie}. For typical supernovae such as SN 1987A, this explosion energy is observed to be around $10^{51}$~erg, about one percent of the total neutron star binding energy released in the explosion \cite{Caputo:2021rux}. There are SNe with explosion energies lower by about an order of magnitude, which might also be used to constrain ALP energy deposition \cite{Caputo:2022mah}. Finally, we point out that recently refs.~\cite{Fiorillo:2025sln,Fiorillo:2025yzf} have argued for modifications of both the cooling and energy deposition constraint in the strong coupling regime, which were not included in some of the older references we cite below.

For ALPs coupled to electrons, ref.~\cite{Fiorillo:2025sln} computed a collection of SN constraints, including the above mentioned cooling and explosion energy bounds, which are the relevant ones for the regions of parameter space studied in this work. The loop-induced ALP-photon interactions in the SN plasma are claimed to be negligible, in contrast to the finding of ref.~\cite{Ferreira:2022xlw},\footnote{We point our here that even following the prescription that ref.~\cite{Fiorillo:2025sln} argues for, namely to evaluate the effective loop-induced ALP-photon coupling as in vacuum but with the thermal electron mass instead of $m_e$, would leave a nearly maximal $g_{a\gamma^{*}\gamma}^{\rm eff, \, P,D}\left(|t| \sim T^2 \gtrsim m_{e,{\rm thermal}}^2 \right) \lesssim \alpha C_e/ \pi \Lambda$ for both the pseudoscalar as well as the derivative-coupling case. This is because the thermal electron mass is set by temperature and chemical potential, both of which also determine the typical energy in the Primakoff process --- see section~\ref{sec:Theory:Effective_Coupling} and ref.~\cite{Ferreira:2022xlw}. And since photon coalescence becomes important for $m_a \gtrsim \mathcal{O}(10 \MeV) > m_{e,{\rm thermal}}$, at least in the derivative-coupling case, also that process (or its inverse) is not significantly suppressed as $g_{a\gamma\gamma}^{\rm eff, \, D}\left(m_a > m_{e,{\rm thermal}}\right) \simeq \alpha C_e/\pi \Lambda$. Nevertheless, ref.~\cite{Fiorillo:2025sln} improves over the work in ref.~\cite{Ferreira:2022xlw} in many ways and hence we include their results here.} and thus are not included in either ALP production or reabsorption rates. In this way, the bound could both be underestimated or overestimated at its upper end, depending on which production and reabsorption effects would dominate. However, as long as loop-effects are ignored, pseudoscalar and derivative couplings lead to the same phenomenology and we can show the constraints for both cases in figure~\ref{fig:Results:Total_El}. We show the most conservative constraint shown in ref~\cite{Fiorillo:2025sln}. For ALP masses $m_a < 1$~MeV, ref.~\cite{Fiorillo:2025sln} does not show any constraints. Since scattering processes in the plasma should be independent of the ALP mass for such light ALPs, we extrapolate the cooling constraint to lower masses but choose to cut off the energy deposition bound since in principle it depends on the decay length of the ALP, which of course changes significantly with $m_a$ especially at the threshold $m_a = 2 m_e$.

For interactions with muons, the cooling constraint on the derivative coupling was presented in ref.~\cite{Croon:2020lrf}. Note, however, that these results have been criticized in Refs.~\cite{Caputo:2021rux} and \cite{Caputo:2022rca}, mainly pointing out that since absorption is only considered in the radial direction, i.e., is underestimated, and hence the upper end of the bound overestimated, and furthermore that the muons are assumed to be non-relativistic even though typical thermal energies are $3T \simeq m_\mu$. The latter effect also tends to overestimate the constraint in the strong coupling regime. Hence, we show the bound as a dashed line in the left panel of figure~\ref{fig:Results:Total_Mu} to indicate where SN phenomenology should become relevant, but caution that there are potentially large numerical uncertainties on the exact value of the bound. The constraint is different for pseudoscalar couplings since the differing loop-induced photon coupling significantly affects the upper end of the bound for $5 \MeV \lesssim m_a \lesssim 2 m_\mu \simeq 211 \MeV$, and has been calculated in Ref.~\cite{Caputo:2021rux}. 
More importantly, for $m_a \lesssim 10$~MeV, ref.~\cite{Caputo:2021rux} furthermore estimates that the explosion energy constraint in this case is $g_{a\mu} / m_\mu \gtrsim 2.26 \cdot 10^{-2}~{\rm GeV}^{-1}$. The bound is approximated by demanding that the total energy emitted in ALPs, in the ``Stefan-Boltzmann approximation'' discussed and shown to be a good estimate in simple models of SN-ALP emission \cite{Caputo:2022rca}, does not exceed the SN explosion energy. We show this constraint, which is stronger than the cooling bound in the relevant parameter region, in the right panel of figure~\ref{fig:Results:Total_Mu}. Also at larger ALP masses there should be a constraint on the muon coupling. However, since $m_a \gtrsim T$ for heavier ALPs --- especially near the upper end of the bound that constrains ALPs produced in the colder outer regions of the proto-neutron star --- the estimates of production, absorption, and decay rates in ref.~\cite{Caputo:2021rux} do not hold anymore, and the bound is not simply extrapolated to larger masses.

For ALPs coupled only to tau leptons, one can safely integrate them out, given that the tau mass is much larger than typical temperatures reached in the proto-neutron star. This leaves only a constant on-shell coupling to photons for the pseudoscalar case, which could be constrained by the usual analyses, while for the derivative case, the coupling is suppressed by $m_a^2/m_\tau^2$ or $E_a^2/m_\tau^2$ with the ALP energy $E_a$ (for ALP production or reabsorption via the Primakoff process).
In ref.~\cite{Alda:2024cxn}, SN constraints on $g_{a\gamma}$ are translated into constraints on the derivative coupling $C_{\tau} / \Lambda$. However, the fact that the effective photon coupling is energy dependent if one of the photons is off-shell, as in the Primakoff process, is not taken into account (not many details are given, but the loop-function seems only to be evaluated using the ALP mass). We also note that due to the energy-dependence of the effective ALP-photon coupling, the constraints on $g_{a\gamma}$ cannot simply be rescaled but the energy integrals have to be reevaluated using the respective loop-functions. Thus, we choose not the show these constraints, but remark that they should be present in some form in this region of parameter space.

In the universal coupling case, production from electrons, muons, and the loop-induced photon interaction (also receiving a contribution from $g_{a\tau}$) have to be taken into account, and the same holds true for reabsorption and decay events. Thus, neither of the bounds on electron or muon couplings are accurate estimates for this case, and we thus choose not to show any SN constraints for this scenario either, but again remark that they should be present somewhere in the parameter space.

\subsection{Proton beam-dump experiments and forward-physics detectors}

Proton beam dump experiments provide sensitive probes of the effective ALP-photon coupling ~\cite{Dobrich:2015jyk,Dobrich:2019dxc}. 
A detailed reinterpretation of these constraints in the context of leptophilic ALPs is not straightforward since the typical momentum transfer $t$ from the target nucleus to the ALP can be comparable to $m_\ell^2$, such that the full momentum dependence of the lepton loop needs to be included for an accurate estimate.

To obtain a first estimate of the expected constraints, we implemented the constant piece of the loop-induced photon coupling $g_{a\gamma \gamma}^\text{eff} = \alpha C_\ell / (\pi \Lambda)$ in the public code ALPINIST~\cite{Jerhot:2022chi}, which allows for a reinterpretation of proton-beam dump data for ALPs with different coupling structures. The resulting constraints are found to be weaker than the ones from E137 and NA64. Given the results from ref.~\cite{Rella:2022len} for muonphilic scalars, we anticipate the same conclusion to hold also for muonphilic ALPs, which can be produced via Bremsstrahlung from secondary muons.
Since a more accurate calculation is not expected to give significantly stronger bounds, we refrain from a detailed study and omit these constraints from our analysis.
For the same reason, we do not attempt to reinterpret the recent search for ALPs coupled to photons at FASER~\cite{FASER:2024bbl}.

\subsection[Lepton $g-2$]{Lepton {\boldmath $g-2$}}
\label{sec:gm2}
\begin{figure}[!t]
     \centering
         \includegraphics[width=0.3\textwidth]{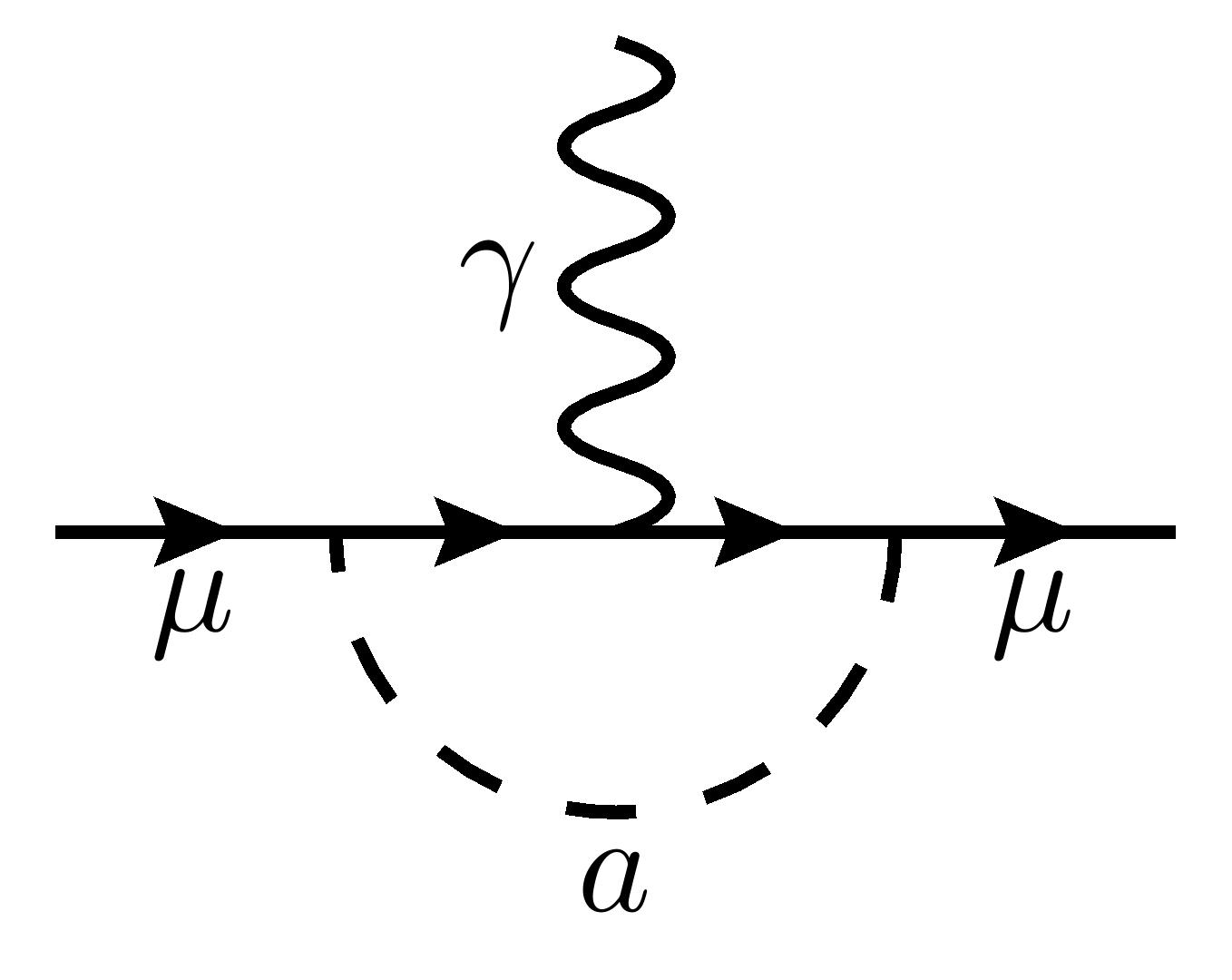}
         \includegraphics[width=0.3\textwidth]{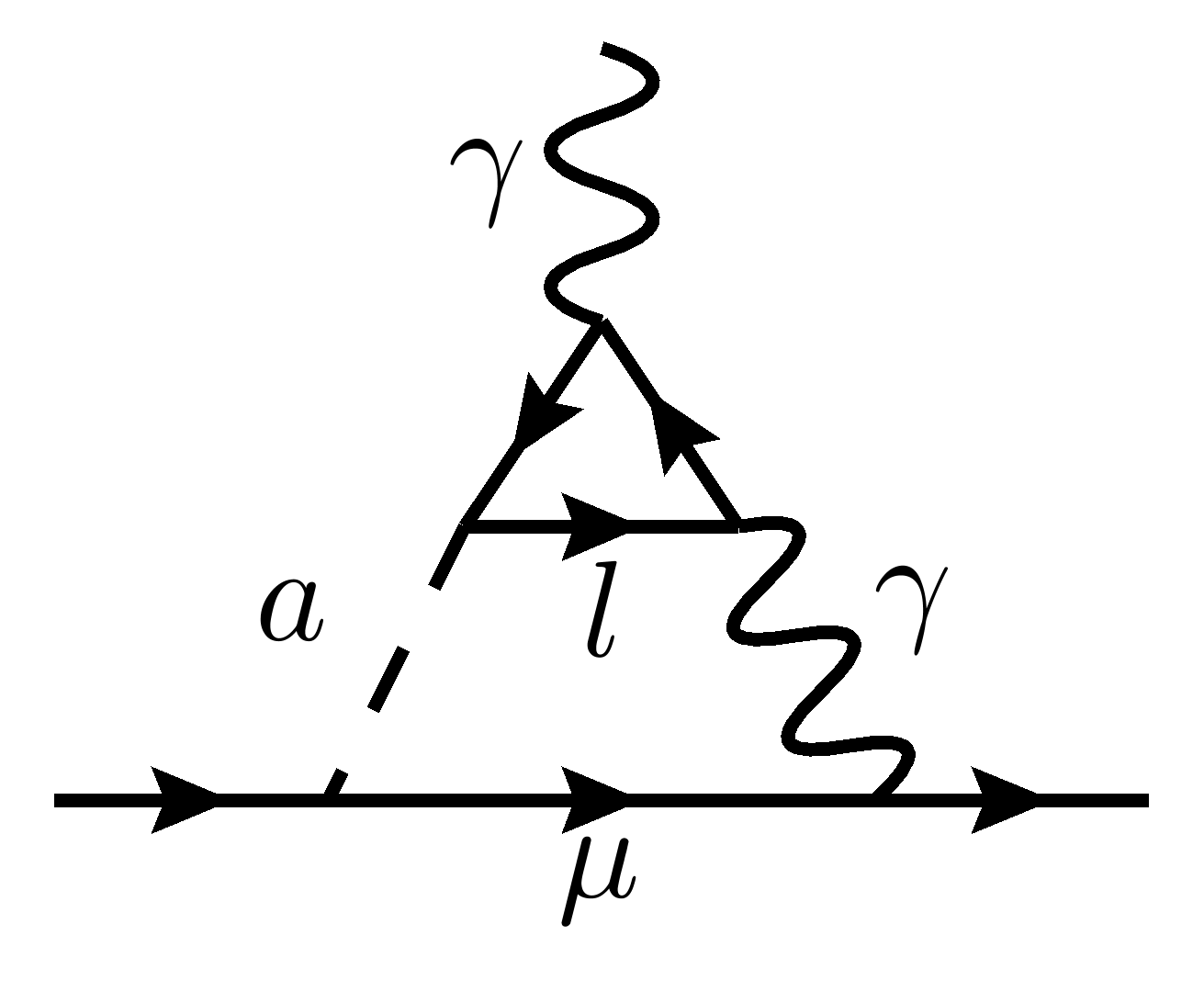}
        \caption{One-loop contributions (left) and two-loop contributions (right) of muonphilic ALPs to $(g-2)_\mu$.}
        \label{fig:gminus2_contributions}
\end{figure}

Leptophilic ALPs contribute to the anomalous magnetic moment of the charged leptons, $a_\ell \equiv (g-2)_\ell/2$ with $\ell=e,\mu,\tau$. The main contributions for the muonic case are shown in figure~\ref{fig:gminus2_contributions}.

At the one-loop level, there is no difference between derivative and pseudoscalar couplings due to the direct coupling of the ALP to the external charged lepton lines. Conversely, the two-loop contribution, which features an effective ALP-photon coupling, distinguishes the two coupling types. The expression for the one-loop contribution is given in ref.~\cite{Bauer:2017ris}, while the expressions for the two-loop contributions can be found in refs.~\cite{Buttazzo:2020vfs, Armando:2023zwz} and~\cite{Buen-Abad:2021fwq} for pseudoscalar and derivative couplings, respectively. Note that, in the latter case, the result depends on the UV cutoff scale $\Lambda$, which we will set here to $\Lambda \sim \SI{1}{TeV}$.

The theoretical situation regarding the muon anomalous magnetic moment is currently under scrutiny, see e.g. ref.~\cite{Davies:2025pmx} and references therein. Indeed, up to a few years ago, the theoretical consensus was a value~\cite{Aoyama:2020ynm} that currently shows a negative 5.2$\sigma$ discrepancy with the current experimental world average~\cite {Muong-2:2023cdq}, a discrepancy that would only be exacerbated by ALPs contributions. However, the latest lattice determinations of the so-called \emph{leading order hadronic vacuum polarisation} (LOHVP) contribution increase the theoretical prediction of $a_\mu$, effectively resulting in a compatibility at the 0.9$\sigma$ level~\cite{Borsanyi:2020mff,Boccaletti:2024guq}. Yet these determinations are incompatible with the previous LOHVP determinations at the 4$\sigma$ level. While waiting for the resolution of this theoretical controversy, we decided to adopt a conservative approach here and include in our results only a dashed line corresponding to the ALP parameter values such that its presence would decrease $a_\mu\hi{theory}$ by $\delta a_\mu=-10^{-9}$, i.e. corresponding to a negative $2\sigma$ shift from the current experimental value, when current theoretical and experimental errors are combined. Nevertheless, this result must not be considered as a binding constraint, as nothing prevents the presence of additional new processes contributing to $a_\mu$ and counteracting the ALP effect.

A similar conservative approach is required as well for the electron anomalous magnetic moment, albeit for different reasons. In this case, the discrepant determinations of the LOHVP do not play a relevant role because this contribution is negligible due to its electron mass suppression. However, at the current level of accuracy of $10^{-13}$, the theoretical prediction is sensitive to different measurements of the fine-structure constant employed in its determination, where a 5.4$\sigma$ difference has been observed in its two latest experimental determinations~\cite{Parker_2018,Morel:2020}. The latest experimental determination of $a_e$, measured with a precision similar to the theoretical uncertainty, sits between the two theoretical determinations~\cite{Fan:2022eto}. Therefore, we adopt a similar approach to the muon case, showing in our results a dashed line corresponding to ALP parameter values such that they would decrease $a_e\hi{theory}$ by $10^{-13}$. Once again, this constraint does not have to be considered binding due to possible additional contributions to this channel.

Finally, we also consider the effect a tauphilic ALP would have on the anomalous magnetic moment of the tau lepton, $a_\tau$. The recent determination of $a_\tau$ from photon fusion at CMS \cite{CMS:2024qjo} provides $a_\tau = \qty(0.9_{-3.1}^{+3.2}) \times 10^{-3}$. Given this experimental uncertainty, we choose once again to mark the parameter space where the presence of a tauphilic ALP decreases $a_\tau\hi{theory}$ by $10^{-3}$ just by a dashed line.

\subsection{ALP lifetime}
\label{sec:ALPtau}

Axion-like particles decaying into leptons and photons in the early universe may change its expansion rate, the temperature of the electron-photon bath relative to the one of neutrinos and the primordial element abundances~\cite{Depta:2020zbh}. While a detailed calculation of these constraints is challenging, in particular for the muonic decay mode, it is expected that ALP decays are harmless for cosmology if the ALP lifetime is shorter than $0.1 \, \mathrm{s}$. To give the reader a feeling for how relevant this consideration is, we indicate a dashed line labelled ``$\tau_a = \SI{0.1}{s}$" in our plots. This line should, however, not be interpreted as a hard constraint since additional effects can change the cosmological evolution, and the resulting constraints may be weaker.

\section{Results}
\label{results}

In this section, we present and discuss the bounds on the leptophilic ALP models obtained from the beam-dump experiments E137 and NA64 discussed in section~\ref{sec:Beam_Dumps} as well as the other constraints discussed in section~\ref{sec:other_constraints}. In order to remain agnostic about the UV completion of this scenario, we analyse different coupling structures independently:
\begin{itemize}
    \item Scenario E: Non-vanishing electron coupling $\hat{g}_{a e} \neq 0$ \,,
    \item Scenario M: Non-vanishing muon coupling $\hat{g}_{a \mu} \neq 0$ \,,
    \item Scenario T: Non-vanishing tau coupling $\hat{g}_{a \tau} \neq 0$ \,,
    \item Scenario U: Universal lepton coupling $\hat{g}_{a e} = \hat{g}_{a \mu} = \hat{g}_{a \tau}$ \,,
\end{itemize}
where we have introduced the definitions $\hat{g}_{a\ell} \equiv g_{a\ell}/m_\ell$ or $\hat{g}_{a\ell} \equiv C_{\ell}/\Lambda$ for the pseudoscalar or the derivative case, respectively.
For each scenario, we consider both derivative and pseudoscalar couplings and set all other ALP couplings to zero. All bounds are calculated at $90 \%$ confidence level, except the ones from LEP and $W^+\to e^+\nu a$, which are at 95\% confidence level.

\subsection{Effect of the coupling type on the E137 bound}
\label{sec:Results:E137}

\begin{figure}[tbp]
    \centering
    \includegraphics[width=0.495\linewidth,clip,trim=7 7 10 7]{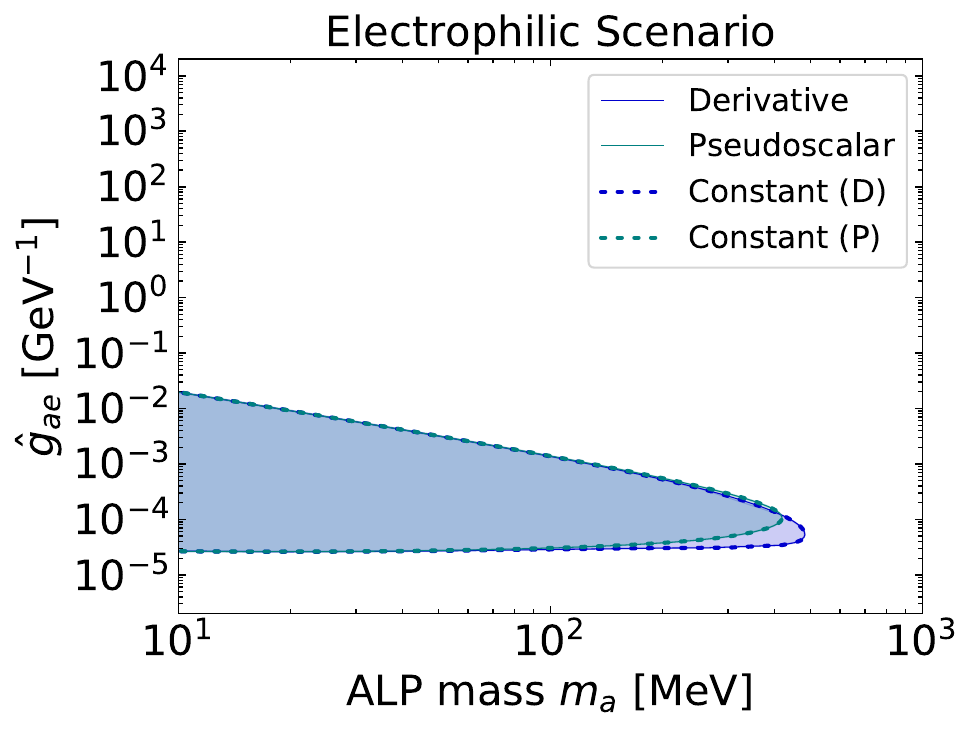}
    \includegraphics[width=0.495\linewidth,clip,trim=7 7 10 7]{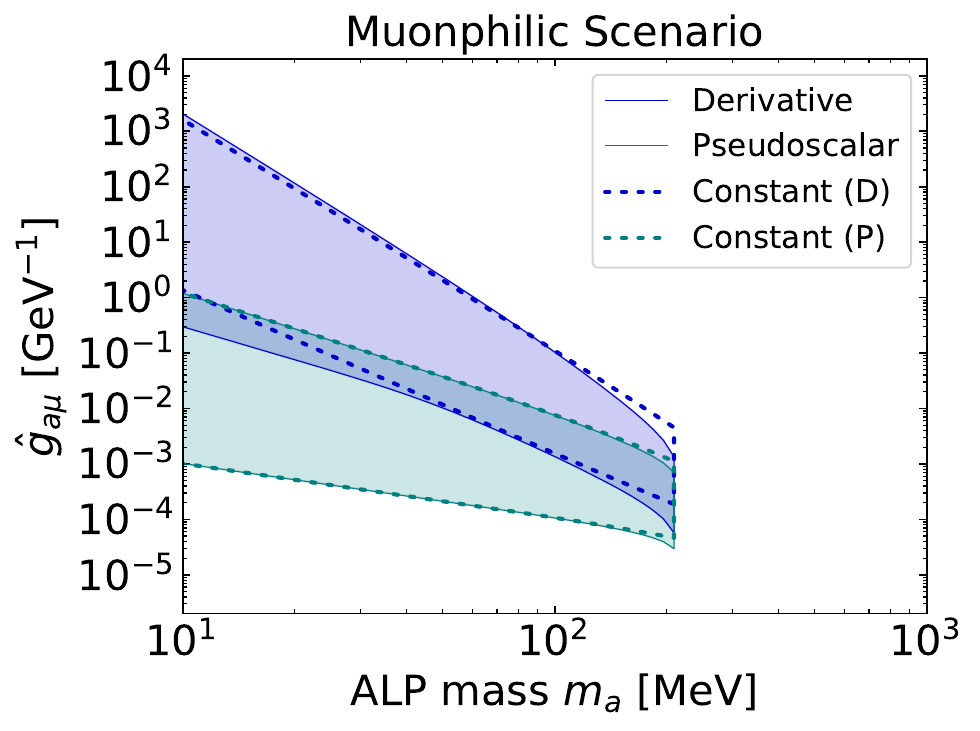}

    \includegraphics[width=0.495\linewidth,clip,trim=7 7 10 7]{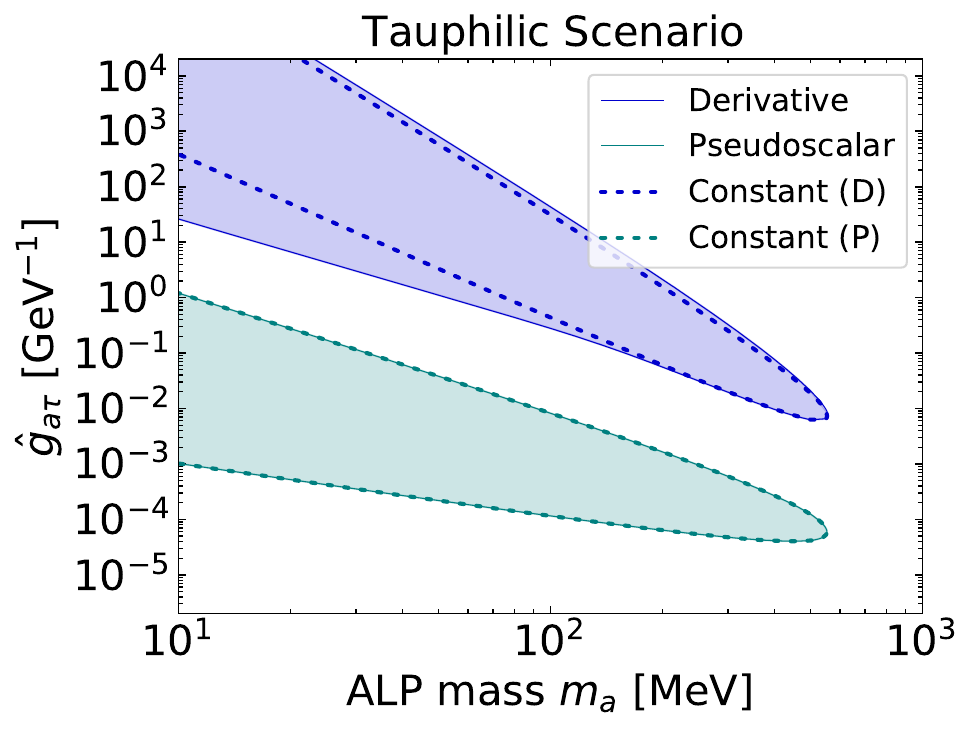}
    \includegraphics[width=0.495\linewidth,clip,trim=7 7 10 7]{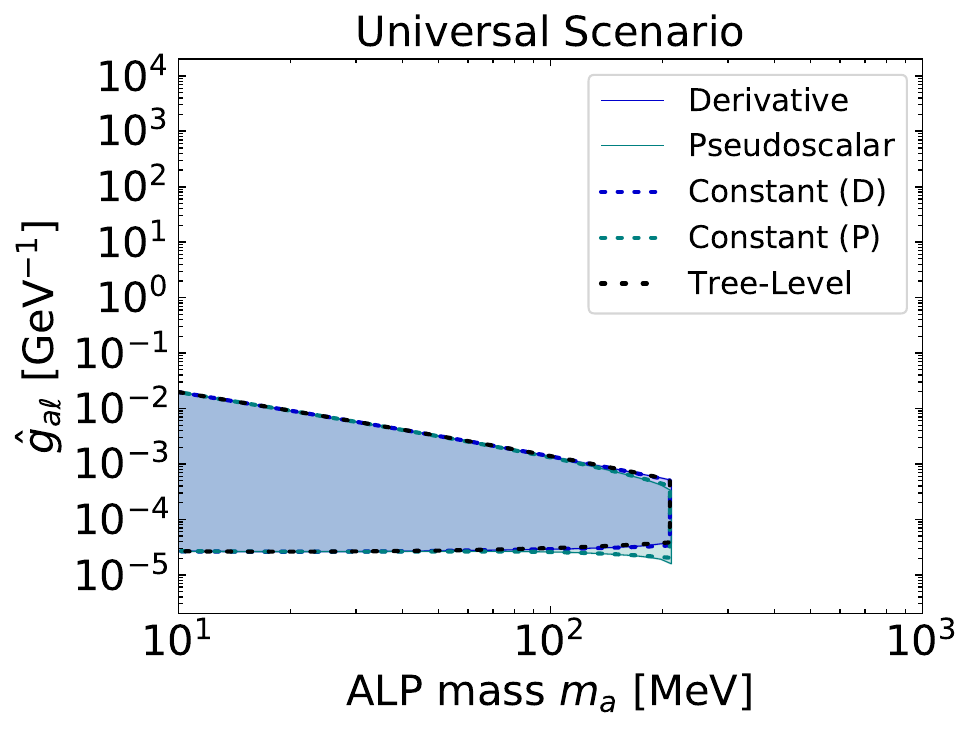}
    \caption{Bounds derived from E137 using different types of couplings to electrons (top-left), muons (top-right), tau leptons (bottom-left) and universal couplings to all leptons (bottom-right).}
    \label{fig:Results:E137_Comparison}
\end{figure}

Before presenting our full results, we compare the effect of the choice of coupling type on the bounds derived from the E137 experiment. The comparisons of the bounds for all scenarios are shown in figure~\ref{fig:Results:E137_Comparison}.

In this figure, aside from comparing the derivative coupling given in eq.~\eqref{eq:Theory:Lagrangian_Der} and the pseudoscalar coupling in eq.~\eqref{eq:LagP}, we also show a comparison with a simplifying approximation, in which the effective off-shell coupling of the ALP to photons is set to the constant on-shell value, in order to judge the relevance of the momentum dependence of the effective photon coupling. We approximate the couplings according to the limits of the derivative and pseudoscalar couplings given in eqs.~\eqref{eq:Theory:C_0_Decay_Der_Cases} and~\eqref{eq:Theory:C_0_Decay_PS_Cases}, respectively. Since the ALP mass is larger than that of the electron in the most relevant regions of parameter space, we assume $\tau_e = 4 m_e^2/m_a^2 \ll 1$. For the muon and tau contribution, on the other hand, we can approximate $\tau_{\mu,\tau} \gg 1$ in the parameter regions probed by E137. In the on-shell approximation, we thus approximate the derivative coupling by the constant values
\begin{align}
    g_{a \gamma^* \gamma,}\hi{eff,D} \to g_{a \gamma \gamma}\hi{eff,D} \approx g_{a \gamma \gamma,\text{const}}\hi{eff,D} = \begin{cases}
        \frac{\alpha C_e}{\pi \Lambda}\,, &\qfor*\text{Scenario E}\,, \\
        -\frac{\alpha C_\mu}{\pi \Lambda}\frac{m_a^2}{12 m_\mu^2}\,, &\qfor*\text{Scenario M}\,, \\
        -\frac{\alpha C_\tau}{\pi \Lambda}\frac{m_a^2}{12 m_\tau^2}\,, &\qfor*\text{Scenario T}\,, \\
       \frac{\alpha C_\ell}{\pi \Lambda}\qty(1-\frac{m_a^2}{12 m_\mu^2}-\frac{m_a^2}{12 m_\tau^2})\,, &\qfor*\text{Scenario U}\,.
    \end{cases}
    \label{eq:Results:g_Der_constant}
\end{align}
Similarly, the approximate pseudoscalar couplings read
\begin{align}
     g_{a \gamma^* \gamma,}\hi{eff,P} \to g_{a \gamma \gamma}\hi{eff,P} \approx g_{a \gamma \gamma,\text{const}}\hi{eff,P} = \begin{cases}
        0 \,, &\qfor*\text{Scenario E}\,, \\
        \frac{\alpha g_{a \mu}}{\pi m_\mu} \qty(-1-\frac{m_a^2}{12 m_\mu^2}) \,, &\qfor*\text{Scenario M}\,, \\
        \frac{\alpha g_{a \tau}}{\pi m_\tau} \qty(-1-\frac{m_a^2}{12 m_\tau^2}) \,, &\qfor*\text{Scenario T}\,, \\
        \frac{\alpha g_{a \ell}}{\pi m_\ell} \qty(-1-\frac{m_a^2}{12 m_\mu^2}-1-\frac{m_a^2}{12 m_\tau^2}) \,, &\qfor*\text{Scenario U}\,.
    \end{cases}
    \label{eq:Results:g_PS_constant}
\end{align}
Specifically, in the approximation for the case of pseudoscalar couplings to electrons only, we set the loop-induced coupling to photons to zero and consider only the tree-level contribution from electrons. 
For Scenario U, we also consider a simplified tree-level model of ALPs, where we set all effective loop-induced photon couplings to zero in order to estimate their relevance for the final exclusion bounds.

Beginning with Scenario E in the top-left panel of figure~\ref{fig:Results:E137_Comparison}, one can see the bounds on the derivative coupling, the pseudoscalar one and their respective constant approximations. For clarity, only the contributions coming from DB and Primakoff production are shown, but not the positron annihilation processes. For ALP masses $m_a \lesssim \SI{100}{MeV}$, the agreement of the bounds between all couplings is close to perfect. However, for larger masses, the differences between the couplings become evident. While the lower bounds on the derivative coupling remain relatively constant until reaching the tip, the pseudoscalar bounds are weaker for the higher ALP mass range.

The derivative and pseudoscalar couplings perfectly match their respective constant approximations. This is unsurprising, especially at the tip of the plot, as the ALP mass is much larger than the electron mass, making eq.~\eqref{eq:Results:g_Der_constant} an excellent approximation for the effective photon coupling in the derivative basis. Similarly, the fact that the bound for the pseudoscalar coupling can be reproduced by setting the loop-induced photon coupling to zero agrees with the expectation from eq.~\eqref{eq:Theory:C_0_Decay_PS_Cases}, as the effective photon coupling in the pseudoscalar basis is strongly suppressed by the electron mass. 

We note that this excellent agreement between the approximated couplings and the exact couplings is still present at the very left side of the plot, where $m_a \approx \SI{10}{MeV}$ and the approximation $\tau \ll 1$ starts to become invalid. This agreement can be explained by the fact that the Primakoff production mechanism, which is the contribution sensitive to the effective coupling, is dwarfed by the tree-level DB production channel, as can be seen in figure~\ref{fig:Implementation:CS_El_Total_Comparison}. No matter which coupling type is used, their differing effects are irrelevant compared to DB production, hence the agreement is good for low masses as well. At the tip of the exclusion, on the other hand, the Primakoff production actually outweighs the DB production (see figure~\ref{fig:Implementation:CS_El_Total_Comparison}), such that the two coupling scenarios give slightly different results.

This difference becomes much more pronounced in the top-right panel of figure~\ref{fig:Results:E137_Comparison}, which shows the comparison of the bounds derived using Scenario M. 
Since in this scenario only the muon coupling is present at tree level, the only production mechanism in E137 is Primakoff production. Below the threshold $m_a = 2 m_\mu \approx \SI{211}{MeV}$, the ALP can only decay into photons, whereas above the muon threshold, the decay into muons proceeds so fast that the ALPs can never leave the dump and reach the detector. As a result, the bounds on this model are very sensitive to the type of coupling used. Indeed, since the effective coupling in the derivative basis is smaller than in the pseudoscalar basis for large lepton masses, the bounds for the derivative coupling are shifted to higher $\hat{g}_{a \mu}$ to compensate.

In addition to the derivative and pseudoscalar couplings, we also show the bounds derived using constant couplings. For the constant approximation of the pseudoscalar coupling, the bounds are generally in very good agreement with the bounds derived from the full pseudoscalar coupling, except close to the threshold $m_a = 2 m_\mu$ where the approximation $\tau \gg 1$ becomes invalid. For the constant approximation of the derivative coupling, the bounds of the full and approximate couplings start to deviate significantly for low masses. In other words, it is essential to consider the full momentum dependence of the effective photon coupling to accurately describe the bounds on the ALP-muon coupling for the derivative coupling case. 

One can observe the same effect in Scenario T, which is shown in the bottom-left panel of figure~\ref{fig:Results:E137_Comparison}. Since the ALP masses are far below the tau threshold, the ALPs can only decay into photons via the effective coupling. As in Scenario M, the ALPs can only be produced via Primakoff production. As a result, the qualitative discussion for Scenarios M and T are almost identical. The only major difference is that, since the relevant ALP mass range remains well below $2 m_\tau$, the constant coupling approximation works very well for $m_a \gtrsim \SI{100}{MeV}$. For lower masses, however, the momentum dependence of the couplings in the derivative case again becomes apparent.

Finally, the bottom-right panel of figure~\ref{fig:Results:E137_Comparison} considers Scenario U, i.e.\ universal couplings to all leptons. The general shape of the bounds is similar to Scenario E for $m_a < 2 m_\mu$, but features the cutoff at $m_a = 2 m_\mu$ seen in Scenario M. Since there is now once again a tree-level coupling to the beam electrons, the DB production of ALPs is the dominant process for lower ALP masses, such that pseudoscalar and derivative couplings agree in this mass range. The difference between the two coupling types only becomes apparent at masses around $m_a \gtrsim \SI{100}{MeV}$, when the loop-induced decays into photons start to compete with the tree-level decays into electrons~\cite{Dolan:2014ska}.

Similarly to Scenario E, both the derivative coupling and the pseudoscalar one are described quite accurately by their relative constant coupling approximation in most of the mass range. However, the additional muon and tau couplings in Scenario U lead to a non-trivial momentum dependence in the effective Primakoff coupling near the threshold $m_a = 2 m_\mu$, particularly in the pseudoscalar case. By coincidence, the derivative case is well approximated by considering only tree-level couplings, i.e.\ setting the effective ALP-photon coupling to zero and neglecting both Primakoff production and decays into photons.

\subsection{Combined Constraints}
\label{sec:Results:Combined}

\begin{figure}[!t]
    \centering
    \includegraphics[width=0.495\linewidth,clip,trim=7 7 10 7]{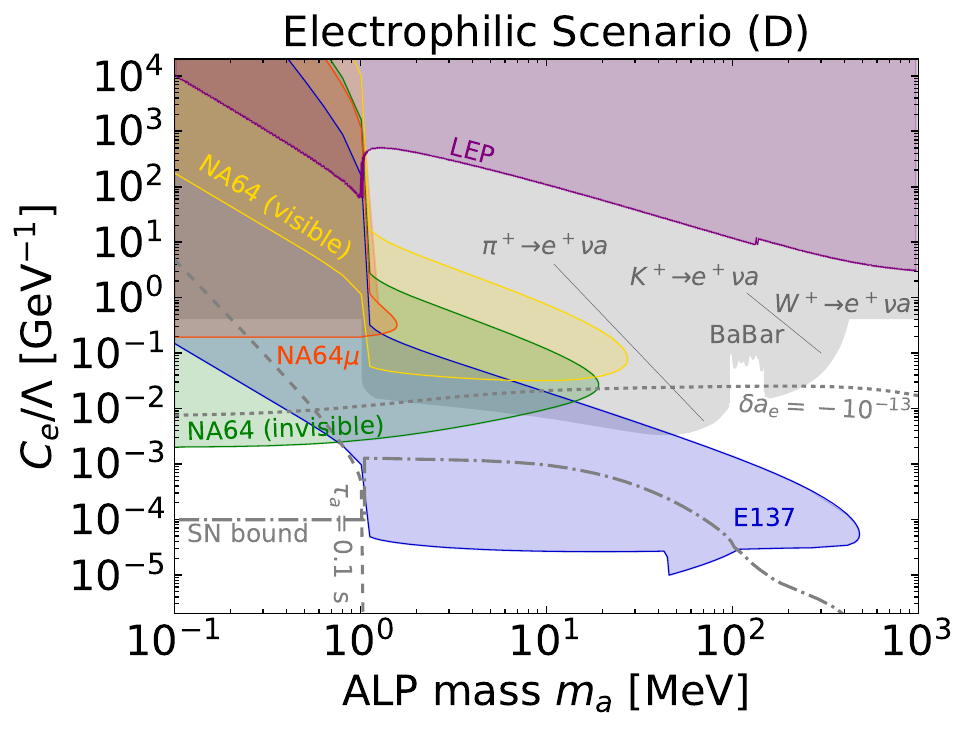} 
    \hfill
    \includegraphics[width=0.495\linewidth,clip,trim=7 7 10 7]{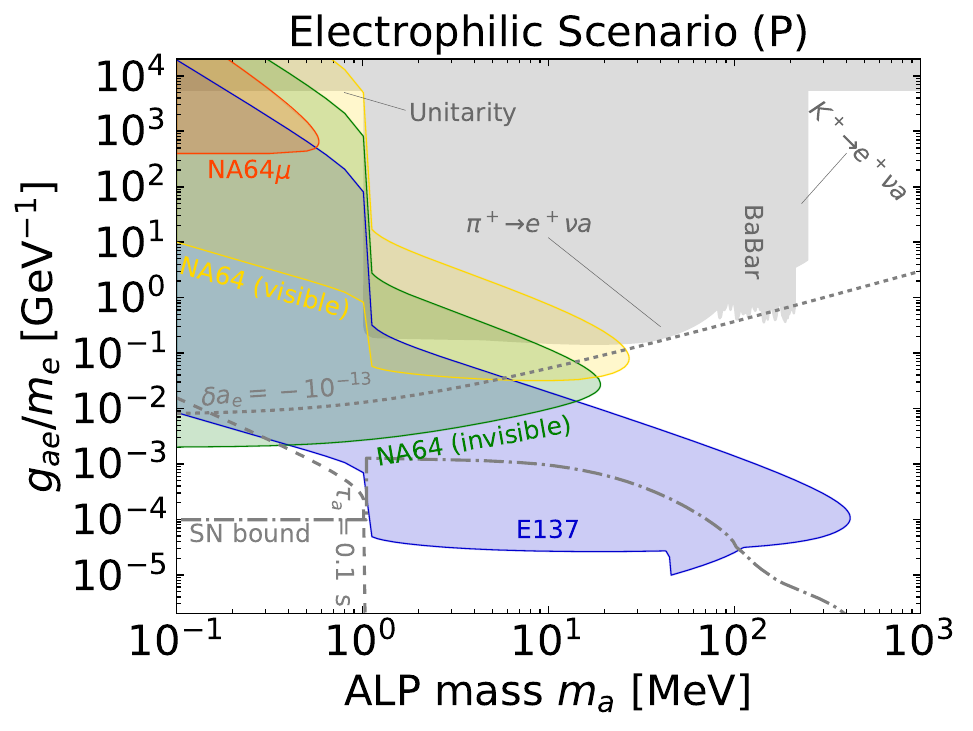}
    \caption{Laboratory bounds on an ALP with derivative (left) and pseudoscalar (right) coupling to electrons.}
    \label{fig:Results:Total_El}
\end{figure}

\begin{figure}[!t]
    \centering
    \includegraphics[width=0.495\linewidth,clip,trim=7 7 10 7]{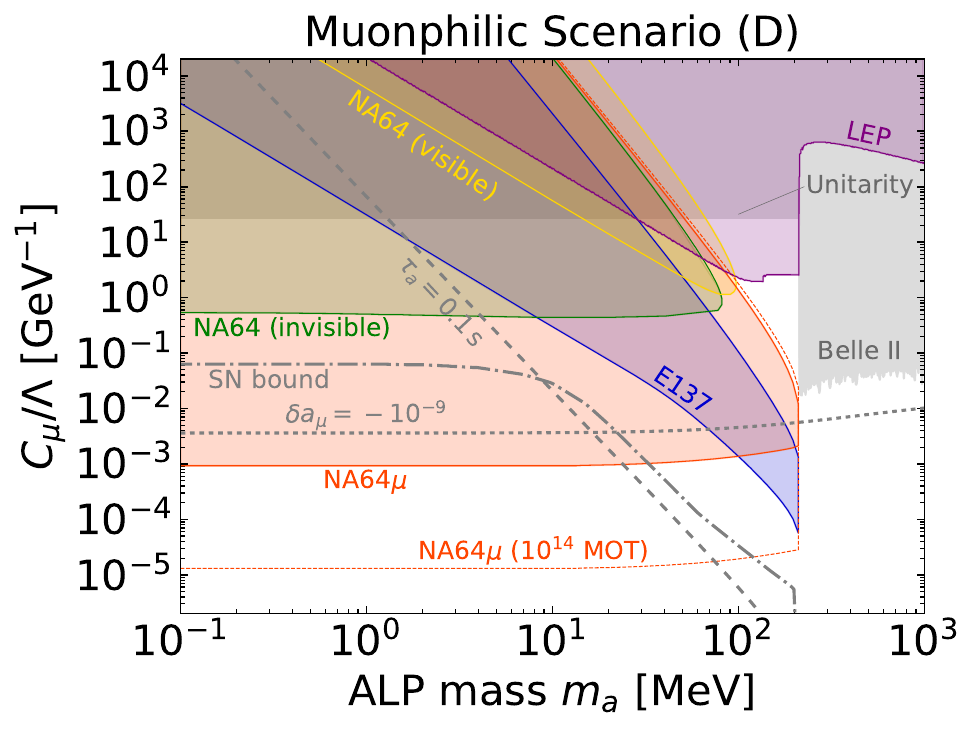}
    \hfill
    \includegraphics[width=0.495\linewidth,clip,trim=7 7 10 7]{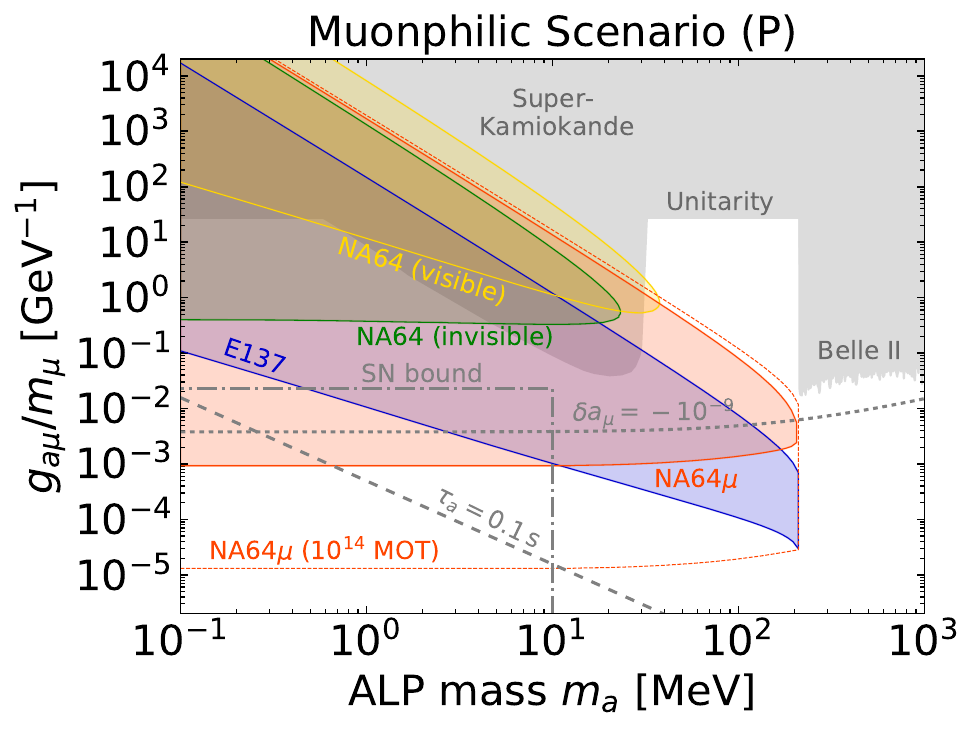}
    \caption{Laboratory bounds on an ALP with derivative (left) and pseudoscalar (right) coupling to muons.
    }
    \label{fig:Results:Total_Mu}
\end{figure}

\begin{figure}[!t]
    \centering
    \includegraphics[width=0.495\linewidth,clip,trim=7 7 10 7]{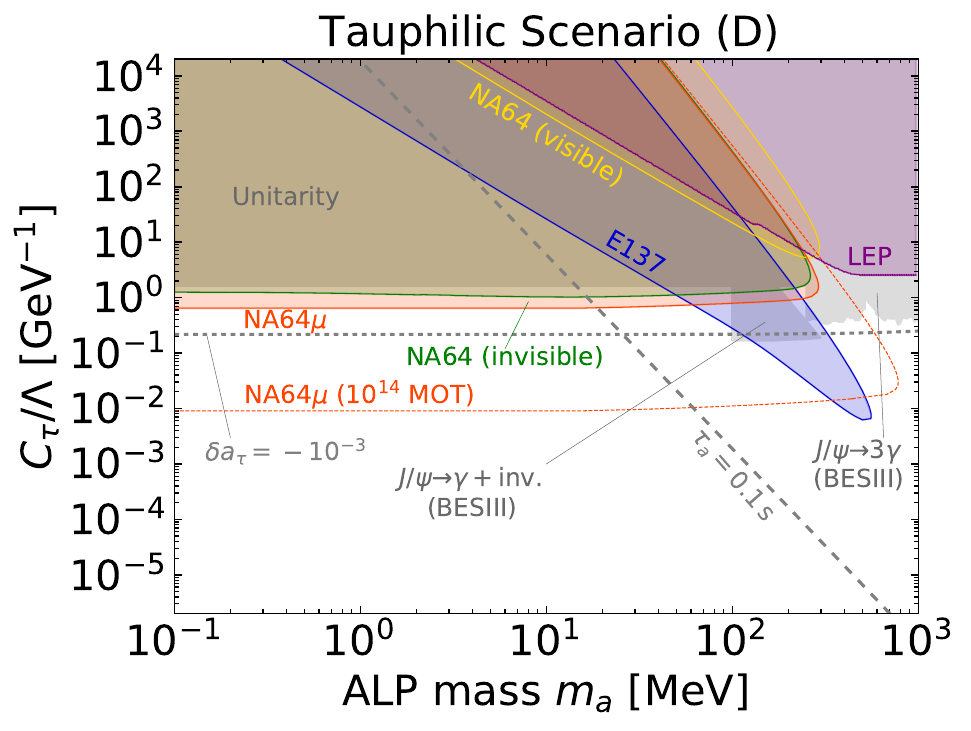}
    \hfill
    \includegraphics[width=0.495\linewidth,clip,trim=7 7 10 7]{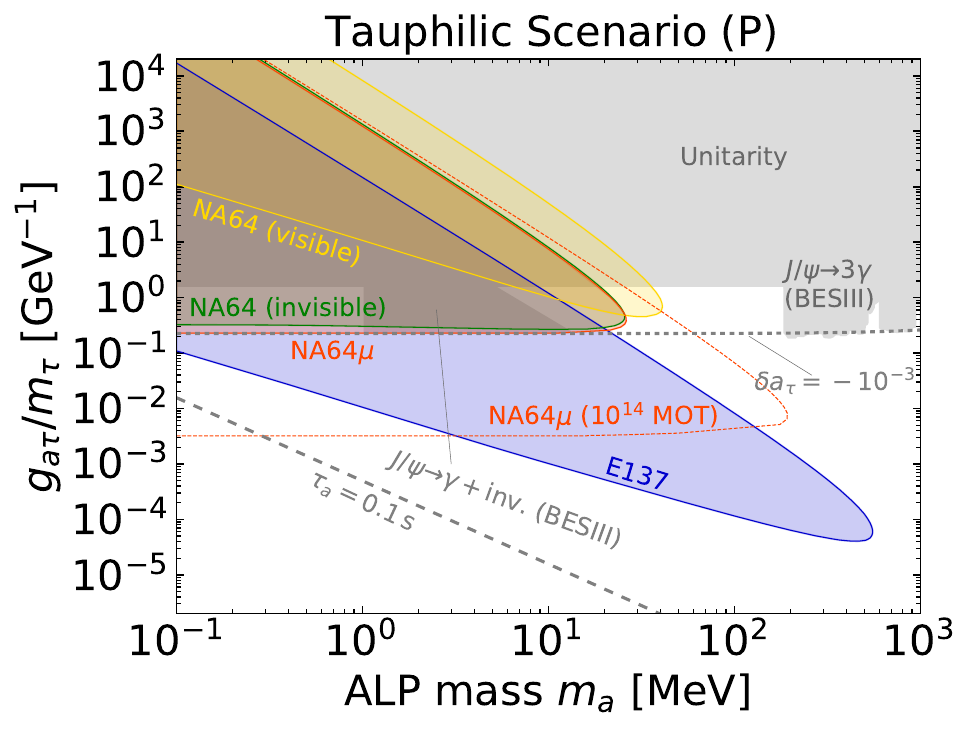}
    \caption{Laboratory bounds on an ALP with derivative (left) and pseudoscalar (right) coupling to tau leptons.
    }
    \label{fig:Results:Total_Tau}
\end{figure}

\begin{figure}[!t]
    \centering
    \includegraphics[width=0.495\linewidth,clip,trim=7 7 10 7]{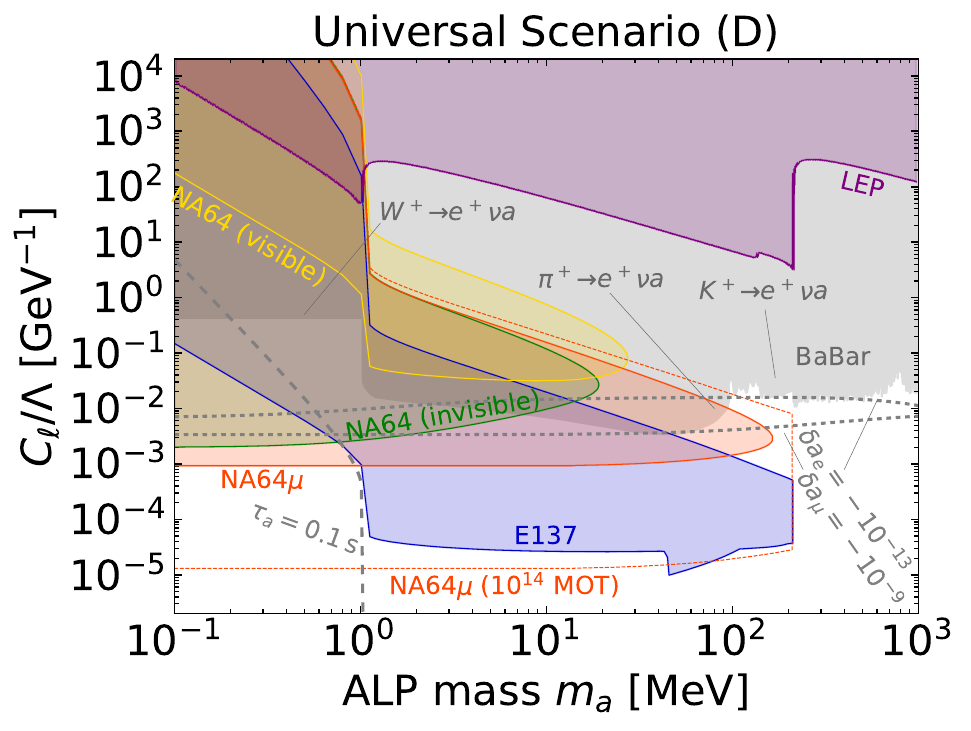}
    \hfill
    \includegraphics[width=0.495\linewidth,clip,trim=7 7 10 7]{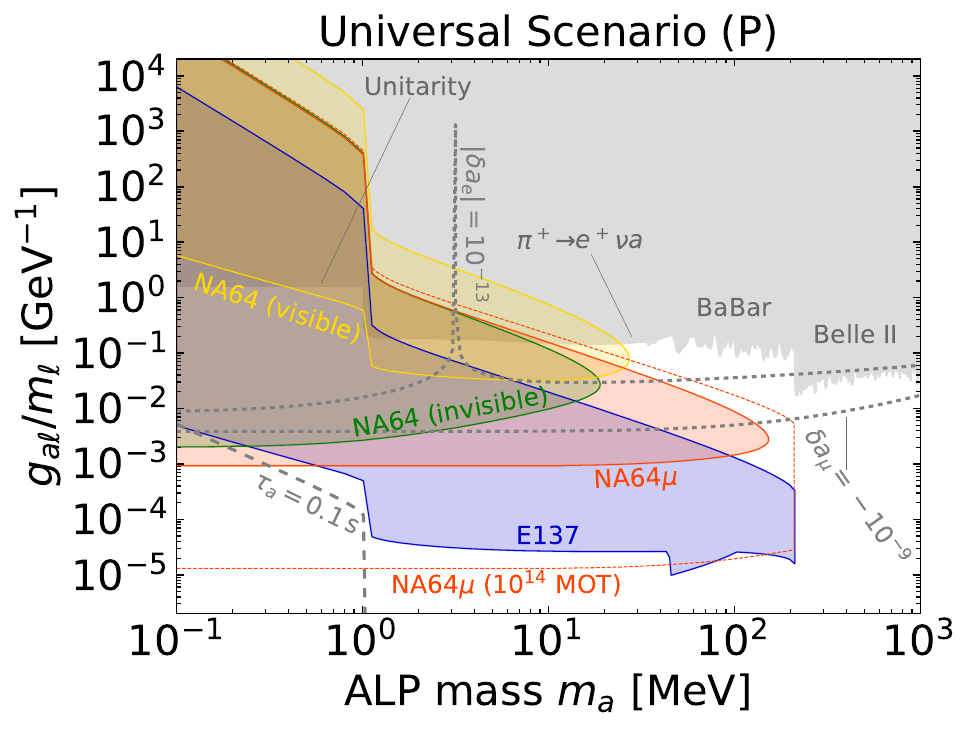}
    \caption{Laboratory bounds on an ALP with universal derivative (left) and pseudoscalar (right) coupling to all charged leptons.
    }
    \label{fig:Results:Total_Uni}
\end{figure}

Having discussed the pertinent differences between pseudoscalar and derivative couplings for the bounds from E137, as well as the relevance of the momentum dependence in the effective couplings, we now turn to the various other constraints on leptophilic ALPs. Our main results are shown in figures~\ref{fig:Results:Total_El} to~\ref{fig:Results:Total_Uni}, corresponding to the four scenarios introduced above. In each case, we show in the left (right) panel the constraints for derivative (pseudoscalar) couplings. The bounds that have been (re)computed by us, namely the ones from sections~\ref{sec:Beam_Dumps} and~\ref{sec:LEP_bounds}, are shown in the foreground, each highlighted with a differently coloured area. Conversely, the laboratory constraints from sections~\ref{sec:PertUni},~\ref{sec:Bfactories},and~\ref{sec:Raremesondec}, which have been taken directly from the literature, are shown in the background as a unified, gray-shaded area. Supernova constraints, where applicable and known, are indicated as dashed lines since they are somewhat approximate (and even contain some errors) in this strong coupling regime, cf.~section~\ref{sec:SN1987A} --- they nevertheless show that SN phenomenology should become relevant in the indicated region of parameter space. Non-binding constraints, namely the ones coming from anomalous magnetic moments described in section~\ref{sec:gm2} and the ALP lifetime discussed in section~\ref{sec:ALPtau}, are indicated for reference as dashed lines. Finally, for the NA64$\mu$ experiment, we also indicate projected bounds for a total of $10^{14}$ MOT, compared to the current $1.98 \cdot 10^{10}$ MOT.

Let us start our discussion from Scenario E, for which bounds are given in figure~\ref{fig:Results:Total_El}. Contrary to the previous section, where only contributions from DB and Primakoff production were considered for estimating E137 constraints, an additional bump is now present because we include the resonant ALP production yields as well.\footnote{While resonant ALP production is not possible in NA64$\mu$, it could in principle play a role in NA64 run in invisible and visible modes. However, due to the high cutoff energy $E\lo{cut} = \SI{20}{GeV}$ of NA64, the condition $m_a > \sqrt{2 m_e E\lo{cut}}$ derived in section~\ref{sec:Implementation:Annihilation} cannot be satisfied in the parameter regions probed by NA64.} We find that E137, which ran over 40 years ago, is still able to set some of the most stringent constraints on the electrophilic ALP models, especially for high ALP masses. However, the bounds set by NA64, both in its visible and invisible modes, complement those by E137, particularly for higher values of the coupling. This is particularly relevant in the pseudoscalar case, where these additional constraints cover all the otherwise unconstrained regions for ALP masses below 10 MeV and couplings above $10^{-2} \, \mathrm{GeV^{-1}}$. While most of the parameter space under consideration is now covered for the derivative case, this is not the case in the pseudoscalar one. In particular, the absence of constraints coming from LEP and $W^+$ decays leave the large coupling region for an ALP mass above $\sim$ 200 MeV unconstrained. To determine the viable parameter regions of smaller couplings, it would be desirable to perform a detailed calculation of the constraints from SNe for pseudoscalar couplings, including the effect of loop-induced couplings to photons.

Moving onto Scenario M, for which bounds are shown in figure~\ref{fig:Results:Total_Mu}, we observe, as expected, that the strongest constraints come from NA64$\mu$, which is exceptionally well suited to probing muonphilic ALPs. Contrary to the previous scenario, E137 is only relevant for couplings between about $10^{-3} \, \mathrm{GeV}^{-1}$ and $10^{-4} \, \mathrm{GeV}^{-1}$, and masses above 100 (10) MeV, for the derivative (pseudoscalar) case. However, if NA64$\mu$ extends its programme to $10^{14}$ MOT, it is expected to push its constraints down to $10^{-5}  \, \mathrm{GeV}^{-1}$, hence totally encompassing E137 ones. Complementary constraints also come from Belle II and from LEP (only in the derivative case) and SNe. We therefore conclude that, as for Scenario E, the parameter space is more constrained in the derivative coupling case than in the pseudoscalar coupling one.

Concerning Scenario T, from figure~\ref{fig:Results:Total_Tau}, it becomes clear that currently not many constraints are stronger than the unitarity bounds. We also observe that, in this scenario, the main additional bound comes from E137, which is stronger in the pseudoscalar coupling case than in the derivative one. As a result, in this scenario, the pseudoscalar coupling case is more tightly constrained than the derivative one. Current bounds from NA64 have marginal relevance in the former case, namely only in the ALP mass range $10\text{--}30$ MeV, while in the latter one, they improve the unitarity bound by a factor of 2. The final NA64$\mu$ results will be able to provide much stronger constraints, improving the current bound by about two orders of magnitudes in the coupling and encompassing almost completely the ones from E137 in the derivative case.

We conclude with Scenario U, for which constraints are reported in figure~\ref{fig:Results:Total_Uni}. As expected, if the ALP couples to all charged leptons, the different constraints from the three previous scenarios become relevant simultaneously. We therefore find that the combination of E137, NA64$\mu$, BaBar and Belle II measurements exclude ALP couplings above about $10^{-2}  \, \mathrm{GeV}^{-1}$. The strongest constraint comes again from E137, which pushes the bound down to almost $10^{-5}  \, \mathrm{GeV}^{-1}$ in the $1\text{--}200$ MeV ALP mass range, while for smaller ALP masses NA64 excludes couplings larger than $10^{-3}  \, \mathrm{GeV}^{-1}$. With $10^{14}$ MOT, NA64$\mu$ is expected to push the coupling bounds down to about $10^{-5} \, \mathrm{GeV}^{-1}$ for ALP masses up to the muon threshold, going beyond the E137 reach and covering the vast majority of the currently unconstrained parameter space studied here. For the case of universal couplings to charged leptons, there is no relevant difference between the derivative and the pseudoscalar coupling cases, except a small unconstrained window for ALP masses in the $30\text{--}200$ MeV mass range for the latter case.

\section{Conclusion}
\label{conclusions}

Axion-like particles (ALPs) arising as pseudo-Goldstone bosons from a spontaneously broken approximate global symmetry are one of the most well-motivated ways to extend the Standard Model with a new state below the GeV scale coupled feebly to known particles. While couplings of ALPs to SM gauge bosons and SM quarks have been studied in great detail, the case of leptophilic ALPs, which couple dominantly to charged leptons, has received much less attention. In the present work, we have filled this gap by providing an exhaustive overview of the relevant constraints from beam-dump, flavour and collider experiments, as well as from theoretical considerations and astrophysics.

First of all, we have pointed out that the precise definition of leptophilic ALPs is rather subtle since the assumed absence of tree-level couplings to photons leads to a distinction between ALPs with pseudoscalar couplings and ALPs with derivative couplings. We take a complete model-independent perspective and discuss both interactions in parallel, highlighting their differences.

In both cases, effective interactions of ALPs to photons are generated through lepton loops, which depend on the momenta of the ALP and the two photons. If all particles are on-shell, one reproduces the standard photon coupling, but for off-shell particles, the loop function depends non-trivially on the momenta, which are specific to the relevant experimental setup. We have shown that this effect plays an important role in the analysis of electron beam-dump experiments such as E137 and collider constraints from, e.g. LEP and have derived suitable approximations that can be used to simplify the analysis.

In our analysis, we have then considered separately the three cases where ALPs couple only to one generation of charged leptons, as well as the case of universal couplings to all charged leptons. In all cases, we find that E137 gives highly relevant constraints, even if both the production and decay of the ALPs proceed exclusively via loop-induced photon couplings. Further relevant constraints come in particular from searches for missing energy and displaced decays at NA64, in particular when using a muon beam. Indeed, future runs of NA64$\mu$ constitute a highly promising way to further probe the parameter space of leptophilic ALPs.

For cases in which the ALPs couple to muons, beam dump experiments are not sensitive to ALP masses above the threshold $m_a > 2 m_\mu$, where ALPs rapidly decay back into muons. In this parameter region, future searches for di-muon resonances at Belle II~\cite{Belle-II:2024wtd} are particularly promising. Below the threshold, relevant sensitivity may come from the planned proton beam-dump experiment SHiP~\cite{Albanese:2878604} as well as from the proposed forward-physics experiment FASER2~\cite{Adhikary:2024nlv}, both of which can exploit the presence of secondary muons in hadron showers to search for leptophilic ALPs. A detailed study of this production mode, including Monte Carlo simulations of the secondary muon distributions, is left for future work. Finally, an improved understanding of the bounds from SN 1987A for all the different coupling scenarios is needed in order to identify the most interesting regions of parameter space at very small couplings. Combining all of these different strategies will enable us to further explore the parameter space of leptophilic ALPs.

\acknowledgments We thank Aoife Bharucha, Paolo Crivelli, Babette D\"obrich, Jeff Dror, Miguel Escudero Abenza, Joerg Jaeckel, Jan Jerhot, Yan Luo, M.C.~David Marsh, Laura Molina Bueno, Uli Nierste, \'{A}lvaro Santos, Thomas Schwetz, Vladyslav Shtabovenko and Joachim Weiss for discussions. This work has received support from the European Union’s Horizon 2020 research and innovation programme under the Marie Sklodowska-Curie grant agreement No 860881-HIDDeN and from the Deutsche Forschungsgemeinschaft (DFG)
through Grant No. 396021762 -- TRR 257 and through the Cluster of Excellence \textit{PRISMA}$^+$ (EXC 2118/1, Project ID 390831469). MF acknowledges funding by the Generalitat Valenciana (Grant PROMETEO/2021/071) and by MCIN/AEI/10.13039 /501100011033  (Grant No. PID2020-114473GB-I00). FK thanks the Instituto de Física Teórica at UAM-CSIC for hospitality during the final stages of this project. ER acknowledges partial funding from Villum Experiment grant no. 00028137, as well as support from Manuel Meyer through guidance and mentorship. This article is based upon work from COST Action COSMIC WISPers (CA21106), supported by COST (European Cooperation in Science and Technology).

\appendix

\section{Experimental details}
\label{app:experiments}

\subsection{E137}
\label{sec:Beam_Dumps:E137}

An electron beam with an energy of $\SI{20}{GeV}$ was directed at an aluminium target. Over the course of the experiment, a total of $\SI{30}{C}$ (Coulombs) of electrons were deposited in the dump, corresponding to $N_e = 1.87 \cdot 10^{20}$ electrons on target (EOT). The target was placed in front of a hill $\SI{179}{m}$ in length, which absorbed all SM particles except neutrinos. The detector was placed behind the hill, providing a decay length of $\SI{204}{m}$. The detector consisted of an 8-radiation-length shower calorimeter, which could detect photons and electrons. For the first phase of the experiment, the detector was arranged in a $2 \times 3$ mosaic of $\SI{1}{m} \times \SI{1}{m}$ chambers, which was later changed to $\SI{3}{m} \times \SI{3}{m}$ chambers. The cutoff energy for the detector, i.e. the minimum ALP energy to be considered a candidate, was set to $\SI{2}{GeV}$. No candidate events were observed.

\subsection{NA64}
\label{sec:Beam_Dumps:NA64}

\subsubsection*{Invisible mode}

The detector consists of one electromagnetic calorimeter (ECAL) and three HCAL. The ECAL consists of lead and scintillator plates, having a total length of $\SI{45}{cm}$ and a width of $\SI{23}{cm} \times \SI{23}{cm}$. The ECAL is the active target and detects the energy deposition of electrons. Each HCAL is made of iron and scintillator plates intended to detect charged and neutral hadrons. The three HCALs have a combined length of approximately $\SI{5}{m}$, and a width of $\SI{60}{cm} \times \SI{60}{cm}$. Using this setup, the calorimeters are able to determine the energy deposited inside the beam dump. If an ALP or other weakly interacting particle were to be produced inside the dump and was long-lived enough, it would escape the detector and carry away a fraction of the original energy of the beam. This missing energy can be measured and registered as an event~\cite{Gninenko:2719646}.

The main beam has an energy of $\SI{100}{GeV}$. The energy cutoff for the detector is set at $\SI{50}{GeV}$. In the runs from 2016--2022, NA64 had a total of $N_e = 9.37 \cdot 10^{11}$ EOT. In the 2016 run, a fourth HCAL was placed in line with the other three HCALs, which was moved downstream to be a zero-degree HCAL to help with background detection in later runs. A detailed description of the layout of the experiment can be found in ref.~\cite{MolinaBueno:2868332}. No candidate events were observed~\cite{NA64:2023wbi}.

\subsubsection*{Visible mode}

The setup used for the visible mode is described in ref.~\cite{NA64:2019auh}. The beam is directed toward an electromagnetic calorimeter composed of tungsten and scintillator plates (WCAL). The WCAL serves as an active beam dump, with the first layers being the main regions in which BSM particles would be produced. The rest of the WCAL serves as a dump for the SM particles created in other processes, while weakly interacting particles would pass through the dump. After a decay length of approximately $\SI{3.5}{m}$, the WCAL is followed by the ECAL and HCALs discussed above. If an ALP or any other particle was produced and subsequently decayed inside the decay volume, the ECAL would register the energy of the decay products. If the energy that is deposited in the ECAL is the same as the missing energy in the WCAL, i.e. $E\lo{ECAL} = E\lo{beam} - E\lo{WCAL}$, then this is a candidate event~\cite{Gninenko:2320630}.

In 2017, the beam was run at an energy of $\SI{100}{GeV}$. The length of the WCAL was varied from $\SI{290}{mm}$, with $2.4 \cdot 10^{10}$ EOT, to $\SI{220}{mm}$, with $3 \cdot 10^{10}$ EOT. The cutoff energy for the detector was set to $\SI{20}{GeV}$. In 2018, the beam energy was increased to $\SI{150}{GeV}$, with $3 \cdot 10^{10}$ EOT, with the cutoff energy being set to $\SI{30}{GeV}$. No candidate events were observed~\cite{NA64:2019auh,Gninenko:2320630}.

\subsubsection*{Muon mode}

The experimental setup for NA64$\mu$ is illustrated in ref.~\cite{NA64:2024klw}. As with the original NA64 run with an electron beam, the experiment is built to detect the invisible decay of BSM particles by measuring the missing energy of an event. A muon beam with an energy of $\SI{160}{GeV}$ is directed at an ECAL, similar to the one in previous runs. The ECAL has a total length of $\SI{49}{cm}$. Further downstream, a magnetic spectrometer is placed, which uses a bending magnet (MS2) to deflect the muon beam, as well as two HCALs to remove residuals from the upstream detectors~\cite{NA64:2024klw}. The muon momentum can then be reconstructed. Candidate events are defined as events in which the reconstructed muon momentum is smaller than $\SI{80}{GeV}$ and the energy deposition in the calorimeters is compatible with a minimum ionising particle, such that the missing energy is larger than $\SI{80}{GeV}$. This setup results in a shielding of approximately $\SI{11}{\m}$, which the ALP must traverse before decaying. After the 2021 test run and the 2022-2023 runs, the experiment accumulated a total of $1.98 \cdot 10^{10}$ muons on target (MOT). No candidate events were observed~\cite{Crivelli:2907892}.

\subsection{Summary of Experimental Parameters}
\label{app:experimental_parameters}

For ease of reference, the values of all experimental parameters we use to calculate $N_a$ are listed in Tables~\ref{tab:Implementation:Material_Parameters} and~\ref{tab:Implementation:Experimental_Parameters}.
\begin{table}[t]
    \centering
    \begin{tabular}{c|cccccc}
        \toprule
        Experiment & Target & Z & A & $X$ & $\rho$ & $M\lo{target}$ \\
        & & & & $\qty[\SI{}{\g\per\square\cm}]$ & $\qty[\SI{}{\g\per\cm\cubed}]$ & $ [\SI{}{\g\per\mole}]$ \\
        \midrule
        E137 & Al & $13$ & $27$ & $24.01$ & $2.7$ & $26.98$ \\
        NA64 (invisible,$\mu$) & Pb & $82$ & $207$ & $6.37$ & $11.35$ & $207.2$ \\
        NA64 (visible) & W & $74$ & $184$ & $6.76$ & $19.3$ & $183.84$ \\
        \bottomrule
    \end{tabular}
    \caption{Material parameters for E137 and NA64~\cite{ParticleDataGroup:2024cfk}.}
    \label{tab:Implementation:Material_Parameters}
\end{table}
\begin{table}[t]
    \centering
    \begin{tabular}{c|ccccccc}
        \toprule
        Experiment & EOT/MOT & $E_0$ & $E\lo{cut}$ & $L\lo{sh}$ & $L\lo{dec}$ & $T$ & $\theta_{a,\text{max}}$ \\
        & & [GeV] & [GeV] & [m] & [m] & & \\
        \midrule
        E137 & $1.87 \cdot 10^{20}$ & $20$ & $2$ & $179$ & $204$ & $2012.9$ & $4.4 \cdot 10^{-3}$ \\
        NA64 (invisible) & $9.37 \cdot 10^{11}$ & $100$ & $50$ & $5.5$ & $\infty$ & $979.8$ & $0.11$ \\
        NA64 (visible): & & & & & & & \\
        Run 1 (2017) & $2.4 \cdot 10^{10}$ & $100$ & $20$ & $0.29$ & $3.5$ & $82.8$ & $6.9 \cdot 10^{-2}$ \\
        Run 2 (2017) & $3 \cdot 10^{10}$ & $100$ & $20$ & $0.22$ & $3.5$ & $62.8$ & $6.9 \cdot 10^{-2}$ \\
        Run 3 (2018) & $3 \cdot 10^{10}$ & $150$ & $30$ & $0.22$ & $3.5$ & $62.8$ & $6.9 \cdot 10^{-2}$ \\
        NA64$\mu$ & $1.98 \cdot 10^{10}$ & $160$ & $80$ & 11 & $\infty$ & $87.3$ & $5.5 \cdot 10^{-2}$ \\
        \bottomrule
    \end{tabular}
    \caption{Experimental parameters for E137 and NA64.}
    \label{tab:Implementation:Experimental_Parameters}
\end{table}
Since the experimental parameters were varied for the different runs of NA64 operated in visible mode, the sum over all runs is taken when determining the total number $N_a$
\begin{align}
    N_a = \sum_{\text{runs}} N_{a,i} \,.
\end{align}

The invisible mode of the NA64 experiment has three HCALs downstream of the ECAL, giving a total length of approximately $\SI{5.5}{\m}$. In the first run in 2016, the experiment had a fourth HCAL, which was later moved to a zero-degree HCAL. This means the length of the shielding of the 2016 run was greater compared to the later runs. Nevertheless, we approximate $L\lo{sh} = \SI{5.5}{\m}$ for all runs, as the error would be minimal on account of the relatively low contribution of the $4.5 \cdot 10^{10}$ EOT from this run to the total $9.37 \cdot 10^{11}$ EOT.

\section{Details on Electron Energy Distribution Functions}
\label{Ie_details}

For the calculation of the electron and positron energy distributions, we use the expressions that were derived in ref.~\cite{Tsai:1966js}, namely, eq. (22) for the first generation electrons, eq. (24) for the first generation photons, and eq. (12) for the second generation electrons and positrons:
\begin{align}
    I_e^{(1)}(t,E) &= \frac{1}{E_0} \frac{\ln(E_0/E)^{b t - 1}}{\Gamma(b t)} \,, \\
    \begin{split}
        I_\gamma^{(1)}(t,k) &= \frac{1}{k} \int_0^t \dd t' \frac{e^{-c (t-t')}}{\Gamma(b t' + 1)} \qty(\ln\frac{1}{u})^{b t'} \\
        &\times\qty[u + \sum_{n=0}^\infty \frac{b (-1)^n - u^2}{n! (n+b t'+1)} \qty(\ln\frac{1}{u})^{n+1}] \,,
    \end{split} \\
    \begin{split}
        I_{e,p}^{(2)}(t,E) &= \int_0^t \dd t' \int_E^{E_0} \dd E' \frac{1}{E'} \frac{\ln(E'/E)^{b (t-t')-1}}{\Gamma(b (t-t'))} \\
        &\times \int_{E'}^{E_0} \dd k \, 2 I_\gamma^{(1)}(t',k) \qty[b \qty(1-\frac{k}{E'})+\qty(\frac{k}{E'})^2] \frac{E'^2}{k^3} \,,
    \end{split}
\end{align}
where $E_0$ is the initial beam energy, $u = k / E_0$, $b = 4/3$, and $c = 7/9$.

\section{Details on Cross Section Calculations}
\label{Xsecs}

\subsection{Dark Bremsstrahlung and Primakoff Production}
\label{app:DB_Primakoff}

It is convenient to define modified Mandelstam variables $\tilde{s}$, $\tilde{u}$, and $t_2$ in the same way as~\cite{Liu:2017htz}, noting the use of the opposite sign of the metric. Assuming the incoming lepton has an energy that is much greater than the masses of the ALP and lepton, one can make the approximations~\cite{Liu:2017htz}
\begin{align}
    \Tilde{s} = -\frac{\Tilde{u}}{1-x} \,\qc \Tilde{u} = - x E^2 \theta_a^2 - m_a^2 \frac{1-x}{x} - m_\ell^2 x \,\qc t_2 = \frac{x}{1-x} \Tilde{u} + m_a^2 \,.
    \label{eq:Beam_Dumps:Substitutions}
\end{align}

In the IWW approximation, the cross section is given by~\cite{Liu:2017htz}
\begin{align}
    \qty(\dv{\sigma}{x})\lo{IWW} = \frac{\alpha}{(4 \pi)^2} \sqrt{x^2-\frac{m_a^2}{E^2}} \chi \frac{1-x}{x} \int_{\Tilde{u}\lo{min}}^{\Tilde{u}\lo{max}} \dd \Tilde{u} \frac{\overline{\left|\mathcal{M}\right|}^2}{\Tilde{u}^2} \,,
    \label{eq:Beam_Dumps:Cross_Section}
\end{align}
with the boundaries
\begin{align}
    \Tilde{u}\lo{min} = - x E^2 \theta_{a,\text{max}}^2 - m_a^2 \frac{1-x}{x} - m_\ell^2 x \qc \Tilde{u}\lo{max} = - m_a^2 \frac{1-x}{x} - m_\ell^2 x \,,
\end{align}
where the maximum scattering angle $\theta_{a,\text{max}}$ depends on the geometry of the detector. The effective flux of photons $\chi$ is given by
\begin{align}
    \chi = \int_{t\lo{min}}^{t\lo{max}} \dd t \frac{t-t\lo{min}}{t^2} F^2(t) \,.
    \label{eq:Beam_Dumps:Photon_Flux}
\end{align}
Here, $F$ is the form factor. The inelastic form factor can be neglected here as its contribution is negligible compared to the elastic form factor~\cite{Liu:2017htz}, given by
\begin{align}
    F(t) = \qty(\frac{a^2 t}{1+a^2 t}) \qty(\frac{1}{1+t/d}) Z \,.
\end{align}
The first part accounts for the atomic form factor, while the second part accounts for the nuclear form factor~\cite{Liu:2023bby,Fayet:2024ddk}. Here, $a = 111 Z^{1/3} / m_e$ and $d = \SI{0.164}{\square {GeV}} A^{-2/3}$. The integers $Z$ and $A$ are the atomic number and mass number of the target atom, respectively.

Using \texttt{FeynCalc}~\cite{Shtabovenko:2023idz}, we calculate that the spin-averaged, squared amplitude for the simplified DB processes is
\begin{align}
    \overline{\abs{\mathcal{M}\lo{DB}}}^2 = \frac{e^2 g_{a \ell}^2 \Bigl[2 m_a^2 (\Tilde{s}+\Tilde{u}) \left(m_\ell^2 (\Tilde{s}+\Tilde{u})+\Tilde{s} \Tilde{u}\right)-\Tilde{s} \Tilde{u} (\Tilde{s}+\Tilde{u})^2 -2 m_a^4 \Tilde{s} \Tilde{u} \Bigr]}{\Tilde{s}^2 \Tilde{u}^2} \,.
\end{align}
Using eq.~\eqref{eq:Beam_Dumps:Cross_Section} and the substitutions in eq.~\eqref{eq:Beam_Dumps:Substitutions}, we then obtain eq.~\eqref{eq:Beam_Dumps:DB_cross_section}. Since the DB diagrams have tree-level couplings to the ALP, the choice between a pseudoscalar or derivative ALP-lepton coupling has no effect on the amplitude. 

For the Primakoff process, the amplitude is
\begin{align}
    \overline{\abs{\mathcal{M}\lo{P}}}^2 = \frac{e^2 \abs{g_{a \gamma^* \gamma}\hi{eff} (t_2)}^2 \Bigl[-t_2 \left(\Tilde{s}^2+\Tilde{u}^2\right)-2 m_\ell^2 (\Tilde{s}+\Tilde{u})^2\Bigr]}{4 t_2^2} \,,
\end{align}
where the effective off-shell coupling $g_{a \gamma^* \gamma}\hi{eff} (t_2)$ is defined in eq.~\eqref{eq:Theory:Effective_Coupling} and 
should be understood as the sum over the three effective couplings coming from the couplings to the leptons, i.e.
\begin{align}
    g_{a \gamma^* \gamma}\hi{eff} (t_2) = \sum_{\ell=e,\mu,\tau} g_{a \gamma^* \gamma}\hi{eff} (t_2,m_\ell) \,.
\end{align}
Additionally, the interference term between the two processes gives the contribution
\begin{align}
    \overline{\abs{\mathcal{M}\lo{I}}}^2 = \frac{e^2 \Re\qty(g_{a \gamma^* \gamma}\hi{eff} (t_2)) g_{a \ell} m_\ell (\Tilde{s}+\Tilde{u})^3}{\Tilde{s} t_2
   \Tilde{u}} \,.
\end{align}

\subsection{Non-Resonant Production}
\label{app:Non_Resonant}

\begin{figure}[tbp]
     \centering
     \includegraphics[width=0.32\textwidth]{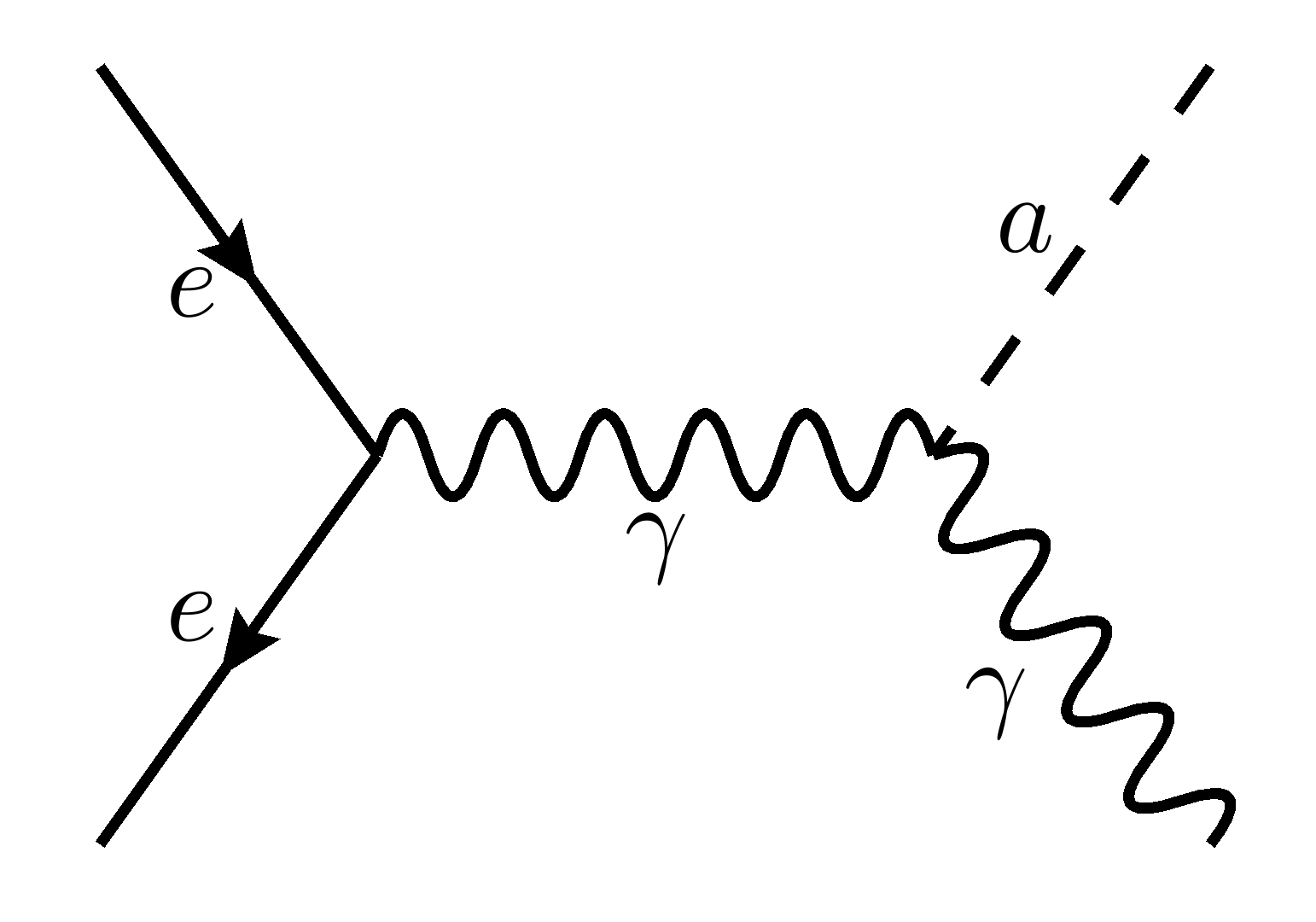}
     \hfill
     \includegraphics[width=0.32\textwidth]{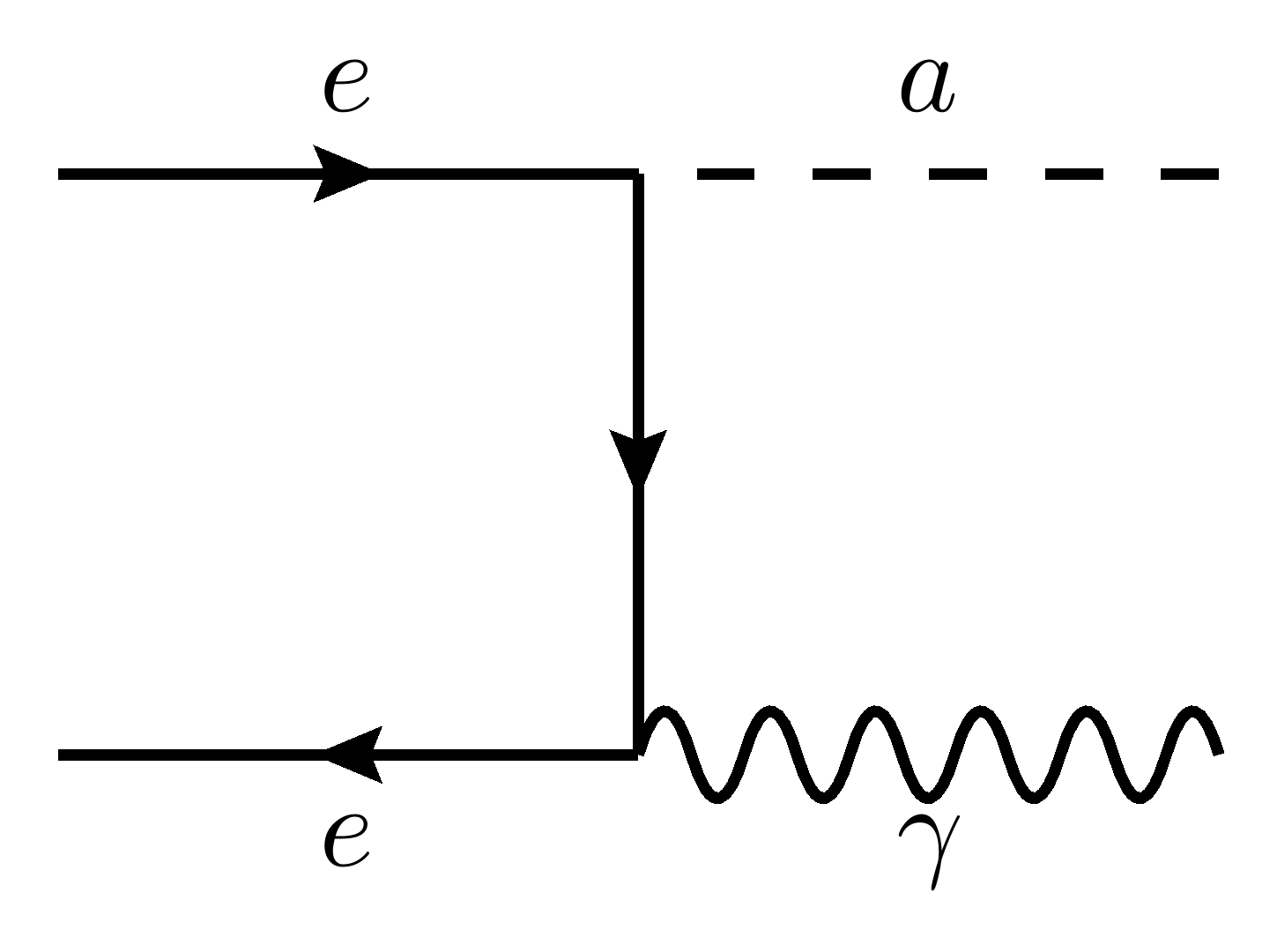}
     \hfill
     \includegraphics[width=0.32\textwidth]{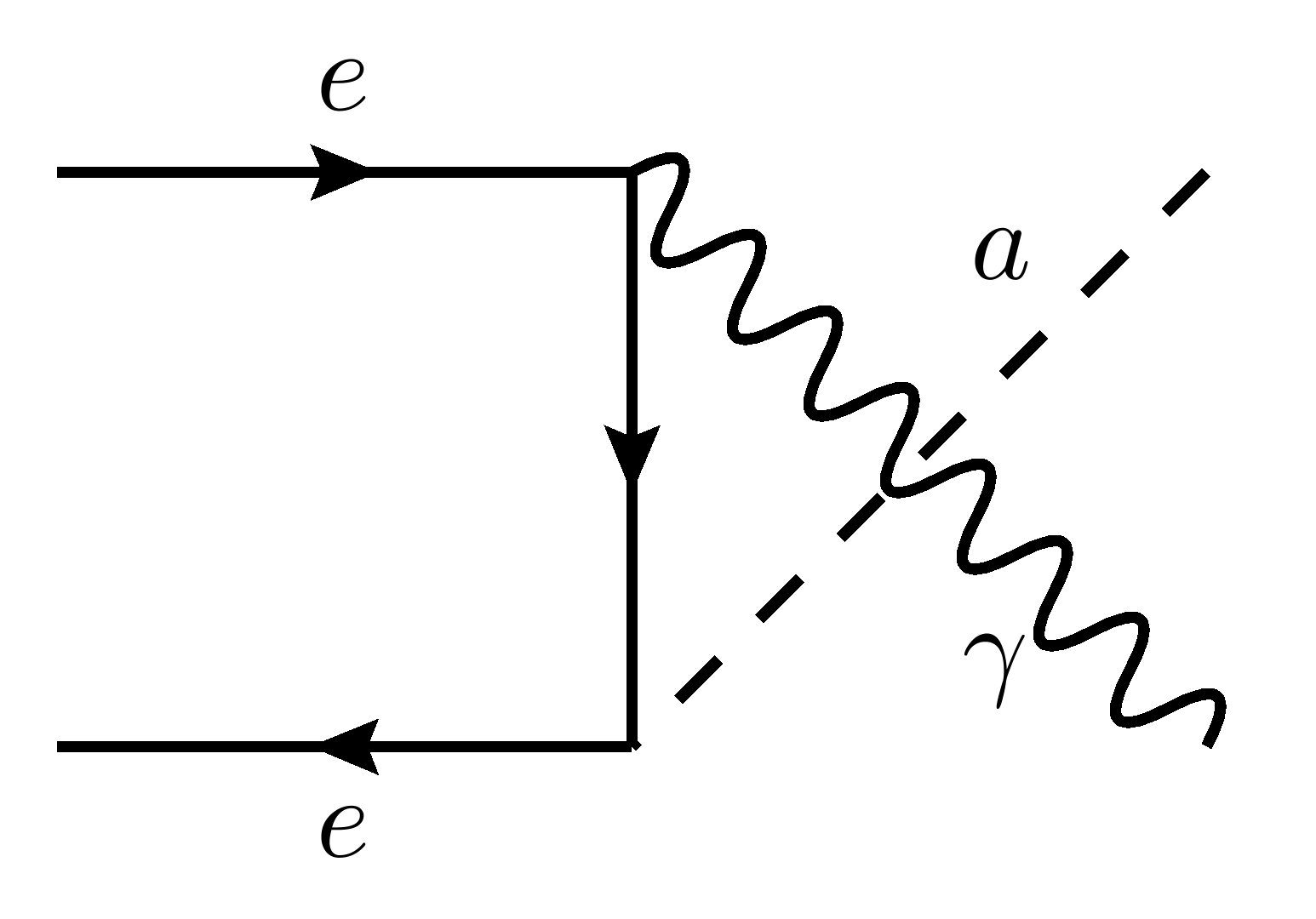}
     \caption{Feynman diagrams contributing to the process $e^- e^+ \rightarrow a \gamma$.}
        \label{fig:Implementation:Annihilation_Diagrams}
\end{figure}

The Feynman diagrams for the non-resonant production of an ALP
\begin{align}
    e^-(p_1) + e^+(p_2) \rightarrow a(k_1) + \gamma(k_2)
\end{align}
are shown in figure~\ref{fig:Implementation:Annihilation_Diagrams}.
The positron can annihilate with the electron through either an s-channel, t-channel or u-channel diagram, of which the latter two are tree-level diagrams. The squared, spin-averaged matrix elements for the tree-level, loop-level, and interference terms are
\begin{align}
       \overline{\abs{\mathcal{M}\lo{NR,tree}}}^2 = & \frac{e^2 g_{a e}^2}{\left(t-m_e^2\right)^2 \left(u-m_e^2\right)^2} \nonumber \\
        &\times\Bigl[\left(t-m_e^2\right) \left(u-m_e^2\right) \left(2 m_a^4 + (s-m_a^2)^2 \right) +2 m_a^2 \left(s-m_a^2\right) \left(t u - m_e^4\right)\Bigr] \,, \nonumber \\
        \overline{\abs{\mathcal{M}\lo{NR,loop}}}^2 =&  \frac{e^2 \abs{g_{a \gamma^* \gamma}\hi{eff}(s)}^2}{4 s^2}  \\
        &\times\Bigl[m_a^4 \left(2 m_e^2+s\right)-2 m_a^2 s \left(m_e^2+s+t\right)+2 s \left(m_e^4-2 m_e^2 t+t (s+t)\right)+s^3\Bigr] \,,
        \nonumber \\
        \overline{\abs{\mathcal{M}\lo{NR,inter}}}^2 = & \frac{e^2 \Re\qty(g_{a \gamma^* \gamma}\hi{eff}(s)) g_{a e} m_e \left(s-m_a^2\right)^3}{s \left(t-m_e^2\right) \left(u-m_e^2\right)} \,, \nonumber
\end{align}
respectively. The Mandelstam variables are defined in the standard way
\begin{align}
    s = (p_1 + p_2)^2 \qc t = (p_1-k_1)^2 \qc u = (p_1-k_2)^2 \,.
\end{align}
The cross section for non-resonant production of ALPs is
\begin{align}
    \dv{\sigma\lo{NR}}{x} &= \frac{1}{16 \pi s} \frac{1}{1-\frac{m_e}{E}} \overline{\abs{\mathcal{M}\lo{NR}}}^2 \theta(\theta_{a,\text{max}}-\theta_a) \,,
    \label{eq:Implementation:Non_Resonant_CS}
\end{align}
with the angle $\theta_a$ defined via
\begin{align}
    \cos(\theta_a) \equiv \frac{2(E+m_e) E_a-s-m_a^2}{2\abs{\vb{p}} \abs{\vb{k_a}}} = \frac{2(E+m_e)(E_a-m_e)-m_a^2}{2\abs{\vb{p}} \abs{\vb{k_a}}} \,.
\end{align}
The final Heaviside step function ensures that the trajectory of the ALP is within the angular coverage of the detector. For non-resonant annihilation, the integration boundaries in eqs.~\eqref{eq:Implementation:Boundaries_x} and~\eqref{eq:Implementation:Boundaries_E} are not applicable. Instead, using the Mandelstam variable $s = 2 m_e (E+m_e)$, the integration boundaries are
\begin{align}
    x\lo{min} &= \frac{1}{E}\cdot\max\qty[E\lo{cut} \,,\, \frac{1}{4 m_e} \qty(s+m_a^2 - \sqrt{1-\frac{4 m_e^2}{s}} (s-m_a^2))] \,, \\
    x\lo{max} &= \frac{1}{E}\cdot\frac{1}{4 m_e} \qty(s+m_a^2 + \sqrt{1-\frac{4 m_e^2}{s}} (s-m_a^2)) \,, \\
    E\lo{min} &= \frac{s\lo{min}}{2 m_e} - m_e \,\qc E\lo{max} = E_0 \,, \\
    s\lo{min} &= \max\qty[m_a^2 \,,\, m_e \frac{(m_a^2-2 m_e E\lo{cut})\qty(\sqrt{E\lo{cut}^2-m_a^2}+E\lo{cut})+m_a^2 m_e}{m_a^2 - 2 m_e E\lo{cut} + m_e^2}] \,.
\end{align}

It is important to note that for $m_a > \sqrt{2 m_e E\lo{cut}}$, the diagrams contributing to the non-resonant production of ALPs exhibit an infrared divergence at the lower integration bound $s\lo{min} = m_a^2$ due to the low-energy photon. This could be remedied by including one-loop corrections to the ALP-lepton vertex, which are, however, beyond the scope of the present work. We therefore neglect the sub-leading non-resonant production in this region of parameter space and focus solely on the dominant contribution from resonant production.

Non-resonant production of ALPs can also be compared to the production of Dark Photons in BaBar to convert bounds on Dark Photons into bounds on ALPs. The differential production cross section for Dark Photons for high lepton beam energies is approximately~\cite{Essig:2009nc}
\begin{align}
    \dv{\sigma_{A'}}{\cos(\theta)} = \frac{2 \pi \epsilon^2 \alpha^2}{s} \qty(1-\frac{m_{A'}^2}{s}) \frac{1+\cos[2](\theta) + \frac{m_{A'}^2/s}{(1-m_{A'}^2/s)^2}}{1-\cos[2](\theta)} \,,
\end{align}
with the Dark Photon mass $m_{A'}$ and the kinetic mixing parameter $\epsilon$. In order for the Dark Photon to be detected, we require that the angles satisfy $\cos(\theta\lo{min}) < \cos(\theta) < \cos(\theta\lo{max})$ in the centre-of-mass system, where $\cos(\theta\lo{min}) \approx -0.92$ and $\cos(\theta\lo{max}) \approx 0.87$~\cite{Essig:2009nc}. By integrating and matching the cross sections of Dark Photons and ALPs
\begin{align}
    \int_{\cos(\theta\lo{min})}^{\cos(\theta\lo{max})} \dd \cos(\theta) \, \dv{\sigma\lo{NR}}{\cos(\theta)} = \int_{\cos(\theta\lo{min})}^{\cos(\theta\lo{max})} \dd \cos(\theta) \, \dv{\sigma_{A'}}{\cos(\theta)} \,,
\end{align}
we can derive constraints on our leptophilic ALP model using the constraints on $\sigma_{A'}$ in ref.~\cite{BaBar}.

\section{Comparison of E137 with other implementations}
\label{comparison}

\begin{figure}[t]
    \centering
    \includegraphics[width=0.6\linewidth]{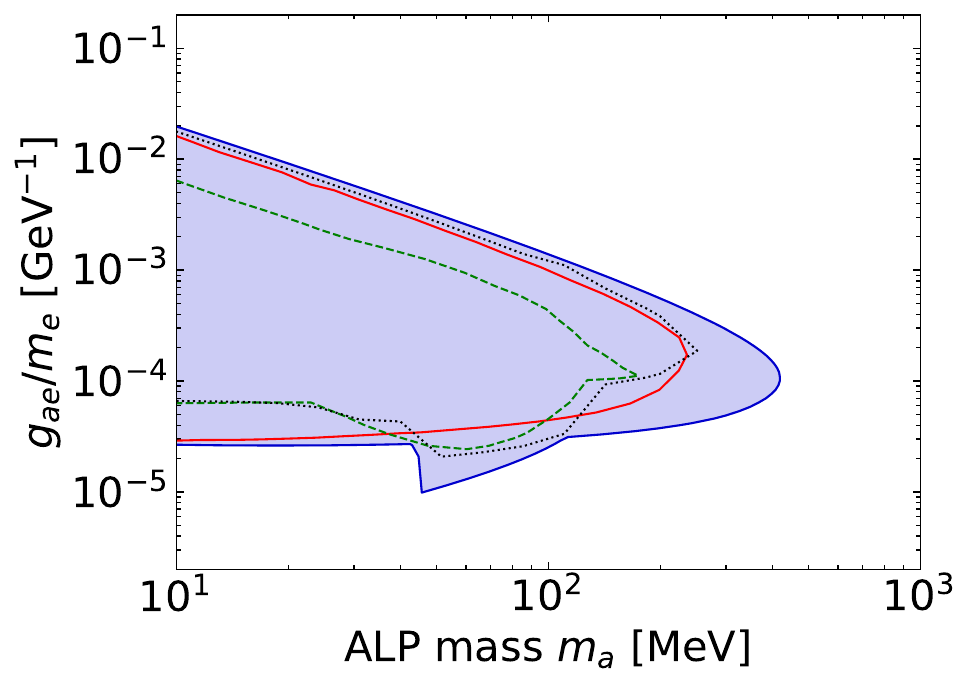}
    \caption{Constraints on the tree-level electron coupling from E137. The result from this work is shown as the blue-shaded region, while the results from refs.~\cite{Liu:2023bby}, \cite{Altmannshofer:2022ckw}, and~\cite{Bharucha:2022lty} are shown as solid red, dashed green, and dotted black lines, respectively.}
    \label{fig:Results:Tree_Level_Comparison}
\end{figure}

In this appendix, we compare our implementation of E137 and the resulting constraints on the ALP-lepton couplings with previous studies in the literature. In ref.~\cite{Liu:2023bby}, the concurrent effects of an ALP-photon and ALP-electron coupling were analysed. The authors give the constraints on a derivative coupling $g_{a\overline{e}e}\hi{eff}$ for various values of $g_{a\gamma\gamma}\hi{eff}$, which are however taken to be constant. Our results can be compared by taking the constraints in ref.~\cite{Liu:2023bby} for $g_{a\gamma\gamma}\hi{eff} = 0$ and comparing them to our tree-level model constraints from figure~\ref{fig:Results:E137_Comparison} (``Constant (P)'' for Scenario E). This comparison is shown in figure~\ref{fig:Results:Tree_Level_Comparison}. For low ALP masses, our constraints are in good agreement with those in ref.~\cite{Liu:2023bby}, but we obtain substantially stronger bounds for high masses. 

It should be noted that ref.~\cite{Liu:2023bby} does not consider the contributions to ALP production arising from electron-positron annihilation. Therefore, we also compare our constraints with those given in ref.~\cite{Altmannshofer:2022ckw}, based on ref.~\cite{Bjorken:1988as}, where this contribution is included.
The original analysis from ref.~\cite{Bjorken:1988as} did not discuss the bounds for large couplings, which were estimated in ref.~\cite{Altmannshofer:2022ckw} assuming the ALPs to be monoenergetic. Clearly, this approximation is not very accurate and leads to a very different upper bound. A more accurate derivation of the upper bound was performed in ref.~\cite{Bharucha:2022lty} using a formalism similar to ours. Indeed, the upper bound obtained in this case\footnote{Note that ref.~\cite{Bharucha:2022lty} only considers ALPs with universal couplings to all leptons. The constraint on ALPs coupled only to electrons shown in figure~\ref{fig:Results:Tree_Level_Comparison} was kindly provided by Sophie Mutzel.} agrees well with our results.

The shape of the lower bounds from refs.~\cite{Bjorken:1988as,Bharucha:2022lty}, which include the contribution from positron annihilation, matches ours relatively well, but the bound is shifted to larger couplings. This shift may be attributed to the different implementations of the track length. Indeed, the track length distribution that we use (see figure~\ref{fig:Beam_Dumps:Iint}) is larger for low energies than the one given in figure~14 of ref.~\cite{Bjorken:1988as}, which is based on the (unpublished) EGS shower code. We note that our implementation agrees very well with the results from ref.~\cite{Marsicano:2018krp}, which has run GEANT4 simulations for E137. Another potential difference between our analysis and the one from ref.~\cite{Bjorken:1988as} is that the E137 analysis uses a much more sophisticated implementation of the detector acceptance, tracking the ALP decay products and their interactions with the detector, which is beyond the scope of our analysis.

As discussed in appendix~\ref{app:Non_Resonant}, our bounds have a discontinuity at $m_a = \sqrt{2 m_e E\lo{cut}} \approx \SI{45.2}{MeV}$, because we conservatively neglect the non-resonant production channels. Including these channels is expected to further improve the agreement with ref.~\cite{Bjorken:1988as}, which finds a continuous bound.

\bibliographystyle{JHEP}
\bibliography{bib}

\end{document}